\documentclass{article}
\usepackage{amsmath,amssymb,epsfig}
\usepackage{natbib}
\usepackage{a4}
\usepackage{makeidx}

\newcommand{\binko}[2]{
\left(\!\!\!\begin{array}{c}{#1}\\{#2}\end{array}\!\!\!\right) }  

\def\d{{\rm d}}
\def\e{{\rm e}}
\def\bg{{\mathbf{g}}}
\def\bk{{\mathbf{k}}}

\def\bv{{\mathbf{v}}}
\def\bx{{\mathbf{x}}}
\def\by{{\mathbf{y}}}
\def\bz{{\mathbf{z}}}

\def\BR{{\mathbb{R}}}

\def\CR{{\cal R}}

\def\hMpc{\ifmmode{h^{-1}{\rm Mpc}}\else{$h^{-1}$Mpc}\fi}

\makeindex
\index{intensity|see{number density}}
\index{pair correlation function|see{two--point correlation function}}
\index{void probability|see{spherical contact distribution}}
\index{Querma\ss{} integral|see{Minkowski functional}}
\index{intensities of Minkowski functionals|see{volume densities of 
Minkowski functionals}}
\index{planarity|see{shape--finder}}
\index{filamentarity|see{shape--finder}}
\index{form factor|see{shape--finder}}

\newlength{\myhalfpage}
\divide\myhalfpage by 2
\newlength{\mythirdpage}
\divide\mythirdpage by 3
\newlength{\mypage}
\newlength{\mymediumpage}
\begin{document}

\title{Statistical analysis of large--scale structure in the Universe}
\author{
Martin  Kerscher\thanks{Ludwig--Maximilians--Universit\"{a}t,  Sektion
Physik,  Theresienstra{\ss}e 37,  80333  M\"{u}nchen, Germany,  email:
kerscher@theorie.physik.uni-muenchen.de}
}

%\date{December 10, 1999}

\maketitle

%%%%%%%%%%%
\begin{abstract}
  Methods  for the  statistical characterization  of  the large--scale
  structure  in the Universe  will be  the main  topic of  the present
  text.   The  focus  is  on  geometrical  methods,  mainly  Minkowski
  functionals  and  the $J$--function.   Their  relations to  standard
  methods  used   in  cosmology  and  spatial   statistics  and  their
  application  to cosmological  datasets will  be discussed.   A short
  introduction to the standard picture of cosmology is given.
\end{abstract}

%%%%%%%
\section{Introduction}

A fundamental problem  in cosmology is to understand  the formation of
the  large--scale  structure in  the  Universe.  Normally  theoretical
models  of   large--scale  structure,  whether   involving  analytical
predictions or numerical simulations, are based on some form of random
or  stochastic  initial conditions.   This  means  that a  statistical
interpretation of  clustering data  is required, and  that statistical
tools  must be  deployed in  order to  discriminate  between different
cosmological models.
Moreover the identification and characterization of specific geometric
features  in  the  galaxy  distribution  like  walls,  filaments,  and
clusters will deepen our  understanding of structure formation, assist
in  the construction  of  approximations and  also  help to  constrain
cosmological models.

During the  past two  decades enormous progress  has been made  in the
mapping of  the distribution of  galaxies in the Universe.   Using the
measured  redshifts of  galaxies as  distance indicators,  and knowing
their angular positions on the sky, we can obtain a three--dimensional
view of the distribution of luminous matter in the Universe. Presently
available redshift  surveys already permit  the detailed study  of the
statistical  properties  of  the  spatial  distribution  of  galaxies.
Surveys of galaxy redshifts that cover reasonable solid angles and are
significantly deeper than  those presently available present important
challenges, and not  just for the observers.  A  precise definition of
the statistical  methods is needed to  extract most out  of the costly
data, and this is an important goal for theorists.

A  complete review  of  the  variety of  statistical  methods used  in
cosmology is  not attempted.   The focus of  this overview will  be on
methods  of  point process  statistics  using  geometrical ideas  like
Minkowski functionals and the $J$--function; moment based methods will
also be  mentioned.  For reviews  with a different emphasis  see e.g.\ 
{}\citet{peebles:lss},               {}\citet{bertschinger:largescale},
{}\citet{peacock:statistics},                {}\citet{borgani:scaling},
{}\citet{efstathiou:observations}, and {}\citet{martinez:measures}.

This text is  organized as follows:\\ 
In Sect.~\ref{sect:cosmo-models} we will  give a short introduction to
the common  theoretical ``prejudice''  in cosmology and  describe some
observational issues.   We briefly comment  on two--point correlations
(Sect.~\ref{sect:twopoint})      and     moment      based     methods
(Sect.~\ref{sect:higher-moments}), and  focus on Minkowski functionals
(Sect.~\ref{sect:minkowski})  and the  $J$--function, as  well  as its
extensions    the     $J_n$--functions    (Sect.~\ref{sect:J}).     In
Sect.~\ref{sect:summary} we summarize and provide an outlook.

\section{Cosmological models and observations}
\label{sect:cosmo-models}

\begin{figure}
\epsfig{figure=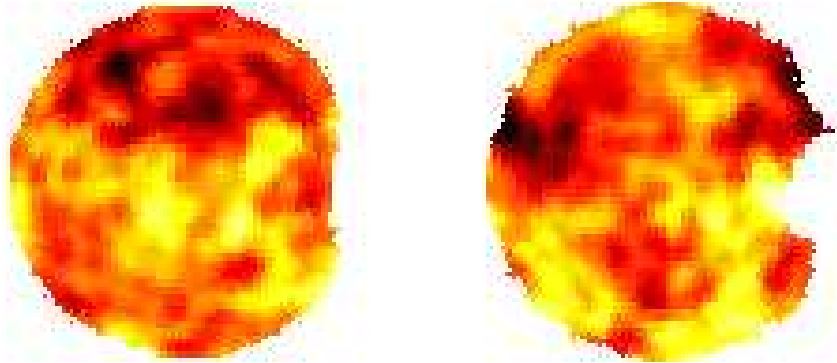,width=\textwidth}
\caption{ Projection of the  temperature fluctuations in the microwave
background  radiation   as  observed  by  the   COBE  satellite  (from
{}\protect\citealt{schmalzing:minkowski_cmb}). The relative fluctuations
are of the order of $10^{-5}$.}
\label{fig:cobe}
\end{figure}

Most cosmological models studied today  are based on the assumption of
homogeneity and isotropy  (see however {}\citealt{buchert:averaging} and
{}\citealt{buchert:onaverage}).   Observationally one can  find evidence
that supports  these assumptions on  very large scales,  the strongest
being the  almost perfect isotropy of the  cosmic microwave background
radiation\index{cosmic microwave background radiation} 
(after assigning  the  whole dipole  to  our proper  motion
relative to  this background).  The  relative temperature fluctuations
over   the  sky  are   of  the   order  of   $10^{-5}$  as   shown  in
Fig.~\ref{fig:cobe}.  This tells us that the Universe {\em was} nearly
isotropic and,  with some  additional assumptions, homogeneous  at the
time of decoupling of approximately 13Gy (Giga years) ago.

\begin{figure}
\begin{center}
\begin{minipage}{\mymediumpage}
    \epsfig{figure=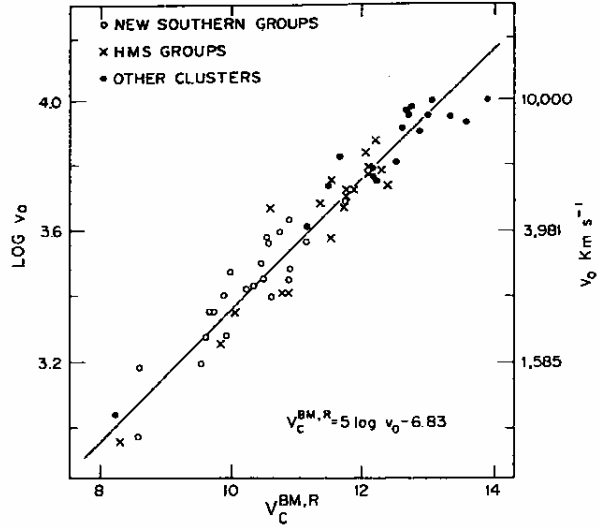,width=\mymediumpage}
\end{minipage}
\end{center}
\caption{\label{fig:hubble} Hubble law  for galaxy clusters and groups
taken  from   {}\protect\citep{sandage:practical}.   The  $x$--axis  is
proportional  to  distance  indicator  obtained  from  the  a  certain
luminosity  of  the clusters  and  groups,  whereas  the $y$--axis  is
proportional to the redshift.}
\end{figure}

For such a  highly symmetric situation the universal  expansion may be
described  by a position  vector $\bx_H(t)$  at time  $t$ that  can be
calculated from the initial position $\bx_i$
\begin{equation}
\bx_H(t) = a(t)\ \bx_i
\end{equation}
using  the  scale  factor   $a(t)$  with  $a(t_i)=1$.   The  dynamical
evolution  of $a(t)$  is determined  by the  Friedmann  equations (see
e.g.~\citealt{padmanabhan:structure}).   As  a  direct  consequence  the
velocities may be approximated by the Hubble law\index{Hubble law},
\begin{equation}
\label{eq:hubble-law}
\bv_H(t) = H(t)\ \bx_H(t)
\end{equation}
relating the  distance vector $\bx_H(t)$ with  the velocity $\bv_H(t)$
by the Hubble parameter $H(t)={\dot a}(t)/a(t)$.  Indeed such a mainly
linear     relationship    is     observed    for     galaxies    (see
Fig.~\ref{fig:hubble}).   The deviations  visible may  be  assigned to
peculiar motions, as caused by mass density perturbations.

\begin{figure}
\begin{center}
\begin{minipage}{\mypage}
\epsfig{file=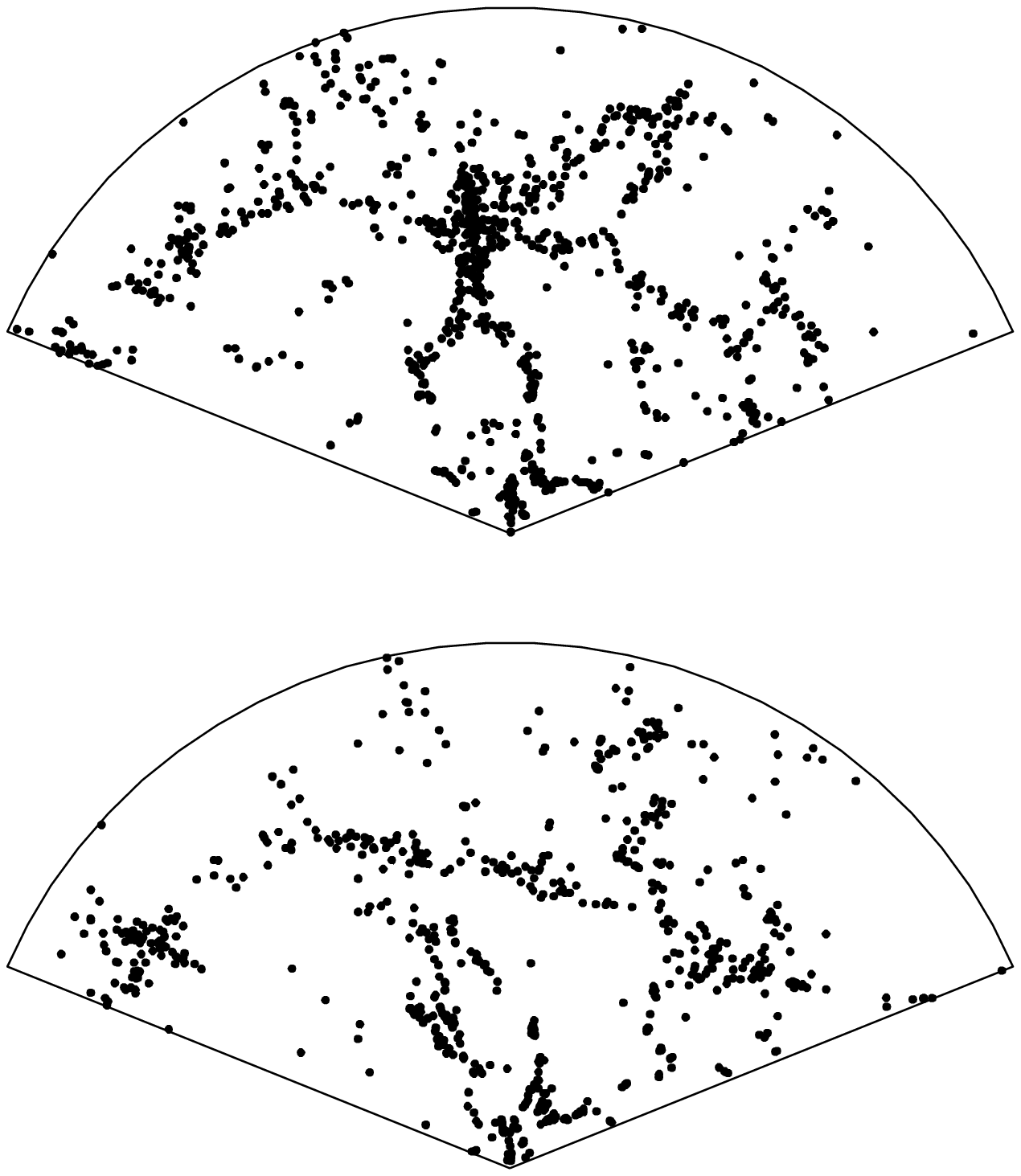,width=\mypage}
\end{minipage} 
\begin{minipage}{\mypage}
\epsfig{file=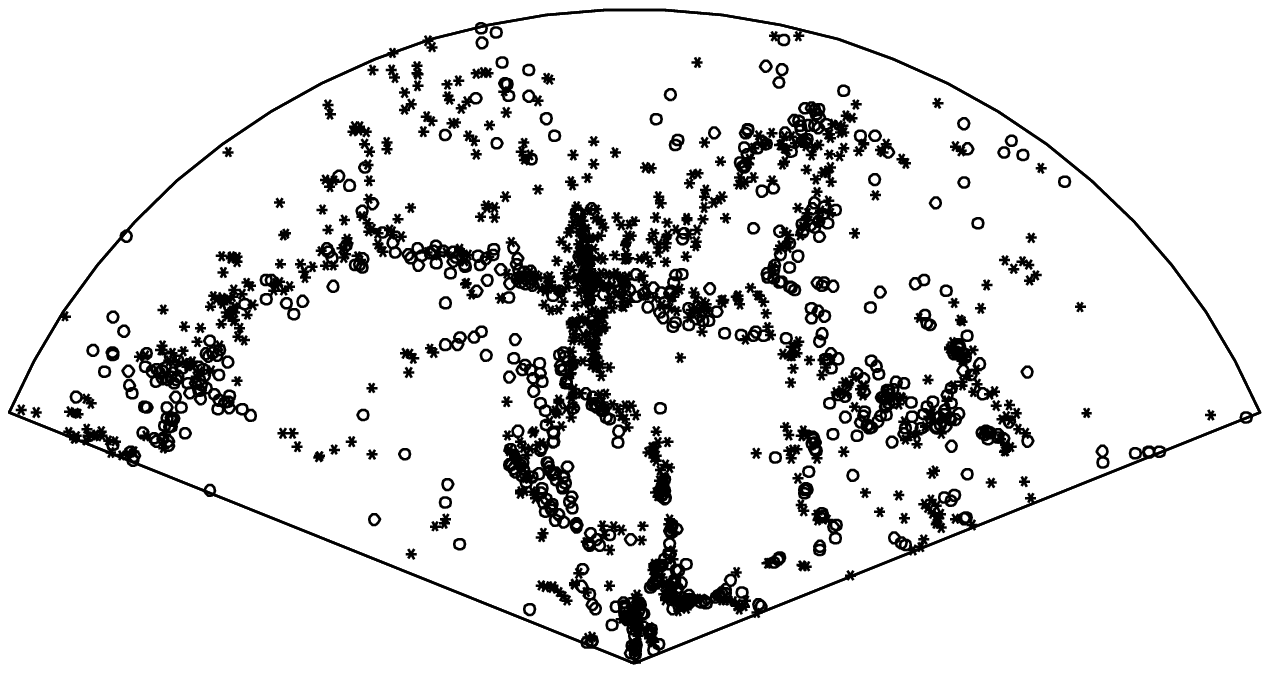,width=\mypage}
\end{minipage} 
\end{center}
\caption{ In the upper two panels, the position of the galaxies in two
neighboring  slices with an  angular extent  of $135\times5$~deg$^{2}$,
and a maximum  distance of 120\hMpc\ from our  galaxy which is located
at the  tip of the  cone. The galaxies  are shown projected  along the
angular coordinate spanning only 5deg.
In the lower plot both slices are shown projected on top of each other
(data         from         {}\protect\citealt{huchra:cfa2s1}         and
{}\protect\citealt{huchra:cfa2s2}).}
\label{fig:cfa2}  
\end{figure}

However, on small and on intermediate scales up to several hundreds of
Mpcs, there  are significant deviations from  homogeneity and isotropy
as visible in the spatial distribution of galaxies. (Mega parsec (Mpc)
is  the  common  unit  of  length in  cosmological  applications  with
1pc=3.26~light years.)  Large holes, filamentary as well as wall--like
structures    are     observed    (Fig.~\ref{fig:cfa2},    see    also
sect.~\ref{sect:fluctuations}).

One of the  goals in cosmology is to understand  how these large scale
structures  form,  given a  nearly  homogeneous  and isotropic  matter
distribution at  some early time.  In the  Newtonian approximation the
process  of structure formation  is modeled  using a  self gravitating
pressure--less fluid,  with the mass density  $\varrho(\bx,t)$ and the
velocity field $\bv(\bx,t)$:
\begin{align}
\partial_t\varrho + \nabla(\varrho\bv) & = 0 ,\nonumber \\
\partial_t\bv + (\bv\cdot\nabla)\bv & = \bg ,\nonumber \\
\label{eq:euler-newton}
\nabla\times\bg & = 0 ,\\ 
\nabla\cdot\bg & = -4\pi G\varrho .\nonumber 
\end{align}
The   first    equation   is   the    continuity   equation,   stating
mass--conservation, the  second comes from  momentum conservation with
the   gravitational   acceleration   $\bg(\bx,t)$   self--consistently
determined  from  the  mass   density.   With  small  fluctuations  in
$\varrho$ and $\bv$  given at some early time,  this system of partial
differential equations constitutes  a highly non--linear initial value
problem.  Up to now no general solution is known.
Approximate  solutions   may  be  constructed   using  a  perturbative
expansion around  the homogeneous background solutions  either for the
fields $\varrho$  and $\bv$ directly or for  the characteristics.  The
first    one   is   called    Eulerian   perturbation    theory   (see
e.g.~\citealt{peebles:lss}),  whereas  the  second is  named  Lagrangian
perturbation   theory  (see   e.g.~\citealt{buchert:lagrangian}).   Also
numerical integration with N--body simulations is used.

The initial conditions  are often chosen as realization  of a Gaussian
random field for the density contrast $(\varrho-\varrho_H)/\varrho_H$.
In principle  a Gaussian random  field model for the  density contrast
allows for  unphysical negative mass  densities, however we  find that
the  initial fluctuations  in  the mass  density  are by  a factor  of
$10^{5}$--times  smaller  than  the  mean  value  of  the  field,  and
therefore negative densities are practically excluded.
Using the  methods mentioned  above we can  follow the  nonlinear time
evolution  of the  density field,  leading to  a  highly non--Gaussian
field.   In this evolved  mass density  field galaxies  are identified
sometimes   also  utilizing   the  velocity   field.    Moreover,  our
understanding  of  the   physical  processes  determining  the  galaxy
formation is still limited.

Two  popular stochastic models  used to  describe the  distribution of
galaxies  are  the Poisson  model\index{Poisson  model}  and the  peak
selection.
In the Poisson model we assume that the mean number of galaxies inside
a region  $C$ is directly proportional  to the total  mass inside this
region  (see  e.g.\  \citealt{peebles:lss},  often also  called  Poisson
sampling).   Hence  the intensity  measure  $\Lambda(C)$  -- the  mean
number of galaxies inside $C$ -- is
\begin{equation}
\label{eq:poisson-sampling}
\Lambda(C) \propto \int_C\d\bx\ \varrho(\bx).
\end{equation}
If the mass density $\varrho$ is modeled as a random field the Poisson
model  results in  a  double--stochastic point  process,  i.e.\ a  Cox
process {}\citep{stoyan:stochgeom}.

\begin{figure}
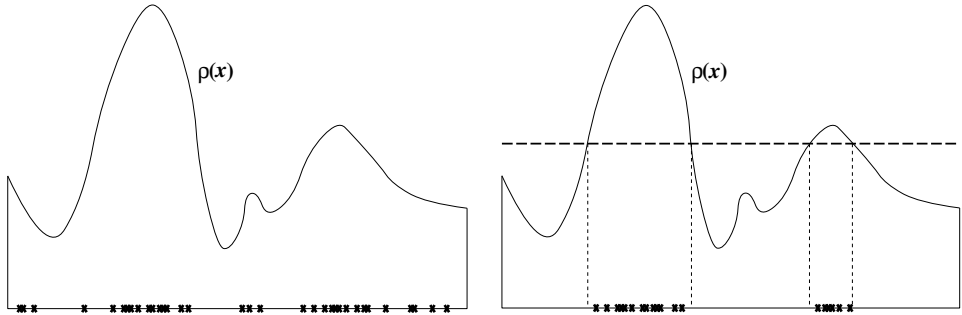

\begin{center}
\begin{minipage}{\myhalfpage}
\epsfig{file=cox.eps,width=\myhalfpage}
\end{minipage} 
\hfill
\begin{minipage}{\myhalfpage}
\epsfig{file=peak.eps,width=\myhalfpage}
\end{minipage} 
\end{center}
\caption{The left figure illustrates the Poisson model, whereas the 
right figure shows the peak selection for the same density field.}
\label{fig:galaxy-identification}
\end{figure}

Within the peak selection model,  galaxies appear only at the peaks of
the density  field above some given  threshold {}\citep{bardeen:gauss}. 
This  model  is  an  example  for  an  ``interrupted  point  process''
{}\citep{stoyan:stochgeom}.   In Figure~\ref{fig:galaxy-identification}
we illustrate both models in the one--dimensional case.
There are also dynamically  and micro--physically motivated models for
the identification  of galaxies  in simulations we  do not  cover here
({}\citealt{kates:highres},                      {}\citealt{weiss:highres},
{}\citealt{kauffmann:galaxy}).

As   we  have   seen   several  ``parameters''   enter  these   partly
deterministic, partly  stochastic models for  the galaxy distribution.
Before  describing the  statistical  methods used  to constrain  these
parameters, typical  observational problems entering  the construction
of galaxy catalogues will be mentioned.

The starting point is the two--dimensional distribution of galaxies on
the celestial  sphere.  Their  angular positions are  known to  a high
precision  compared  to their  radial  distance  $r$.  In most  galaxy
catalogues the radial distance is estimated utilizing the 
redshift\index{redshift}:
\begin{equation}
z=\frac{\lambda_{\rm obs}-\lambda_{\rm lab}}{\lambda_{\rm lab}} ,
\end{equation}
with the  observed wavelength of  a spectral line  $\lambda_{\rm obs}$
and  with the  wavelength  of the  same  spectral line  measured in  a
laboratory $\lambda_{\rm lab}$.  Out  to several hundreds of Mpc's the
relation between the radial distance $r$  and the redshift $z$ is to a
good approximation
\begin{equation}
{\rm c} z \approx |v_{H}| + u = H_0r + u ,
\end{equation}
with  the velocity  of  light c,  and  the Hubble  parameter $H_0$  at
present  time  (see   {}\eqref{eq:hubble-law}).   $u$  is  the  radial
component of the peculiar velocity, i.e.\ the local deviation from the
global  expansion due to  inhomogeneities.  Galaxy  catalogues sampled
homogeneously and with $r$  determined independently from the redshift
are still rare.  Therefore the distance is simply estimated by
\begin{equation}
r=\frac{{\rm c}z}{H_0},
\end{equation}
neglecting  the  peculiar  velocities   $u$.   This  is  often  called
``working in redshift space''.   There is still some controversy about
the actual  value of the  Hubble parameter\index{Hubble law}  which is
parameterized by the number $h$: $H_0=h\ 100$km~sec$^{-1}$~Mpc$^{-1}$.
Likely values are in the range $h=0.5-0.8$.

\begin{figure}
\begin{center}
\begin{minipage}{\myhalfpage}
\epsfig{file=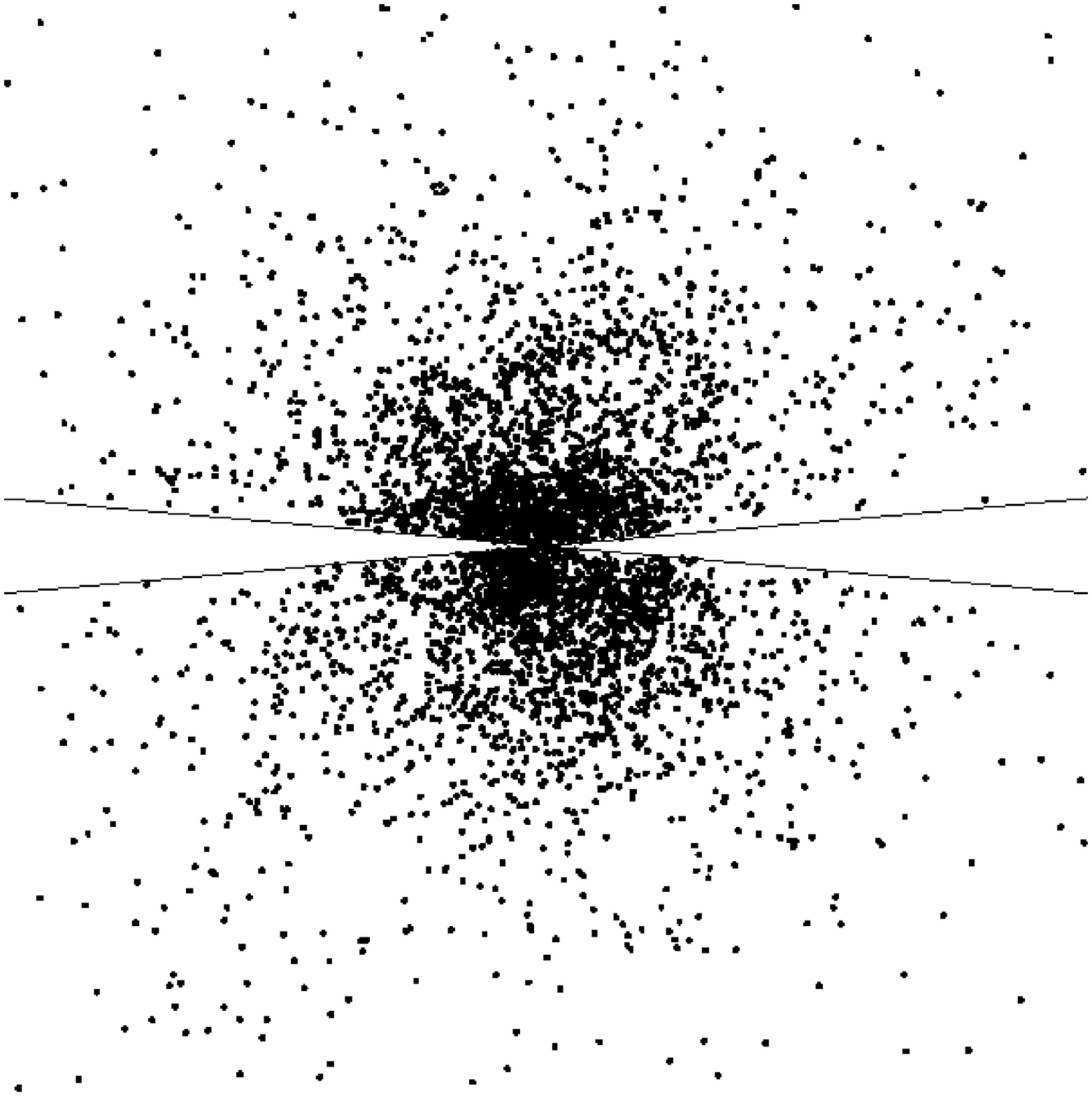,width=\myhalfpage}
\end{minipage} 
\hfill
\begin{minipage}{\myhalfpage}
\epsfig{file=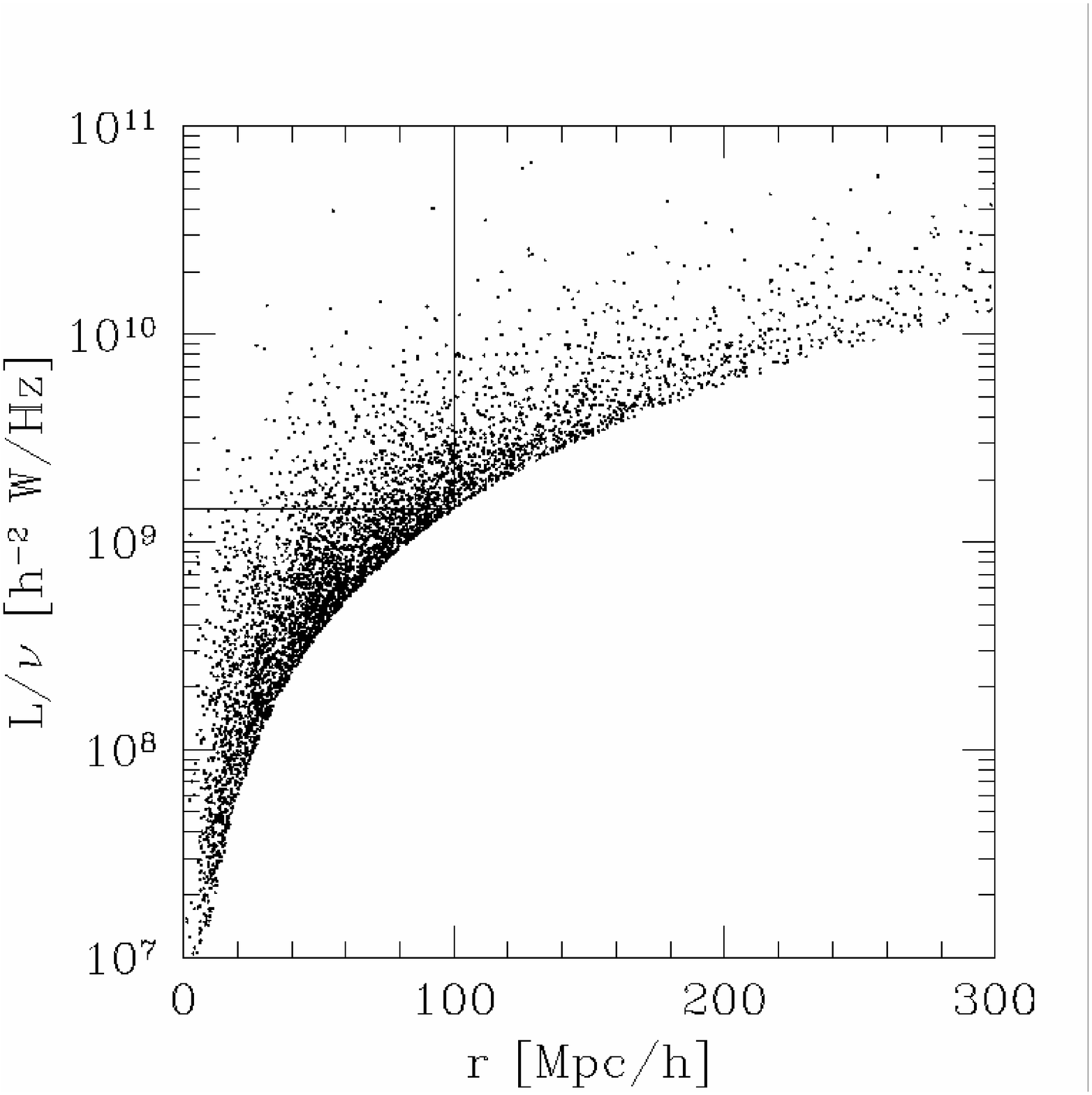,width=\myhalfpage}
\end{minipage} 
\end{center}
\caption{  \label{fig:galaxy-flux-limited}  In  the  left  figure  the
spatial distribution of the  galaxies taken from the IRAS~1.2Jy galaxy
catalogue {}\protect\citep{fisher:irasdata},  projected along one axis.
The horizontally  cones indicate the region where  the observation was
obscured due to  the absorption in our own galaxy.   In the right plot
the absolute  luminosity of  a galaxy against  its radial  distance is
shown, each point represents one galaxy.  The volume limited subsample
with limiting distance of 100\hMpc\  includes only the galaxies in the
marked upper left corner of the figure.}
\end{figure}

\begin{figure}
\begin{center}
\begin{minipage}{\myhalfpage}
\epsfig{file=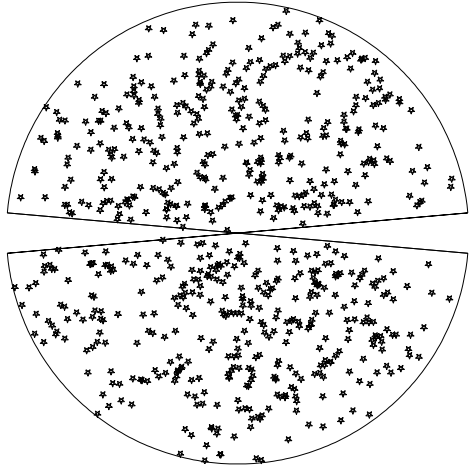,width=\myhalfpage}
\end{minipage} 
\end{center}
\caption{The spatial distribution of IRAS galaxies in a volume limited
sample with a depth of  100\hMpc, projected along one coordinate axis.
This  volume limited sample  is formed  by the  galaxies shown  in the
upper left corner of the  plot with luminosity against radial distance
(Figure~\ref{fig:galaxy-flux-limited}).}
\label{fig:galaxy-volume-limited}
\end{figure}

Furthermore we have  to face another problem.  The  majority of galaxy
catalogues  is flux (i.e.\  magnitude) limited.   This means  that the
catalogue  is complete  for  galaxies  with a  flux  higher than  some
minimum  flux $f_{\rm min}$.   As a  first approximation  the absolute
luminosity  $L$  of a  galaxy  with  observed  flux $f_{\rm  obs}$  at
distance $r$ may be calculated by $L=4\pi r^2\ f_{\rm obs}$.  Hence at
larger distances we observe only the brightest galaxies as can be seen
in Figure~\ref{fig:galaxy-flux-limited}, resulting in a systematically
in--homogeneously sampled point--set in three dimensions.
To  construct a  homogeneously sampled  point set  from such  a galaxy
catalogue we  may restrict ourselves  to galaxies closer  than $r_{\rm
lim}$ with  a absolute luminosity  higher than $L_{\rm  lim}=4\pi r^2\
f_{\rm min}$. This procedure  is called ``volume limitation''.  Such a
set   of   galaxies   for   $r_{\rm  lim}=100\hMpc$   is   marked   in
Figure~\ref{fig:galaxy-flux-limited}  and the spatial  distribution is
shown in Figure~\ref{fig:galaxy-volume-limited}.
Especially in the direction of the disc of our galaxy, in the galactic
plane, we suffer from extinction mainly  due to dust.  To take care of
this we use a cut of 5 to 30 degrees (depending on the catalogue under
consideration)  around the  galactic  plane, resulting  in a  deformed
sampling      window      as      it      can     be      seen      in
Figure~\ref{fig:galaxy-volume-limited}.

The   following   discussion  will   refer   to   a   set  of   points
$X=\{\bx_i\}_{i=1}^N$.  The objects located at these points are either
galaxies, or  galaxy clusters,  and also super--clusters  (clusters of
galaxy clusters).  Galaxies\index{galaxy}  are well defined objects in
space,  with  an extent  of  typically  0.03\hMpc.  Similarly,  galaxy
clusters\index{galaxy  cluster}  are  well  defined  objects,  clearly
visible  in  the two--dimensional  distribution  of  galaxies, with  a
typical  extent  of  1-3\hMpc.   Whether  the  combination  of  galaxy
clusters  to  super--clusters\index{super--cluster}  is  a  reasonable
concept is still some matter of debate {}\citep{kerscher:regular}.

%%%%%%%
\section{Statistics of large scale structure}
\label{sect:stat}

New observations of our Universe  will give us an increasingly precise
mapping  of the  galaxy distribution  around  us ({}\citealt{gunn:sdss},
{}\citealt{maddox:2df}).  But  we will have only  {\em one} realization.
This  makes  a  statistical  analysis  problematic,  especially  model
assumptions like stationarity (homogeneity) and isotropy may be tested
locally  only.  For  an interesting  discussion of  such  problems see
{}\citet{matheron:estimating}.    Still,  global   methods   like  the
Minkowski functionals give us information on the shape and topology of
this point set.

A pragmatic  interpretation is that  with a statistical analysis  of a
galaxy   catalogue,  one   wants  to   constrain  parameters   of  the
cosmological  models.   These   models  incorporate  some  randomness,
quantifying our  ignorance of the  initial conditions, or  our limited
understanding of the exact physical processes leading to the formation
of galaxies.

%%%
\subsection{Two--point statistics}
\label{sect:twopoint}

Second--order statistics, also called two--point statistics, are still
among  the major  tools to  characterize the  spatial  distribution of
galaxies.   With the  mean number  density\index{number density}, or  
intensity,  denoted by
$\rho$, the product density\index{product density}
\begin{equation}
\label{eq:product-density}
\rho_2(\bx_1,\bx_2)\d V(\bx_1)   \d V(\bx_2)
= \rho^2 g(r)\ \d V(\bx_1) \d V(\bx_2)
\end{equation}
describes the  probability to find a  point in the  volume element $\d
V(\bx_1)$  and  another  point  in  $\d  V(\bx_2)$,  at  the  distance
$r=|\bx_1-\bx_2|$;  $|\cdot|$   is  the  Euclidean   norm  (we  assume
stationarity and isotropy).  The product density $\rho_2(\bx_1,\bx_2)$
is  the  Lebesgue  density  of  the second  factorial  moment  measure
(e.g.~\citealt{stoyan:stochgeom}).    Often    the   (full)   two--point
correlation  function\index{two--point   correlation  function},  also
called   pair   correlation    function,   $g(r)$   and   the   normed
cumulant\index{cumulant} $\xi_2(r)=g(r)-1$ are considered.  Throughout
the  cosmological literature  $\xi_2(r)$ is  also  called (two--point)
correlation    function    {}\citep{peebles:lss}.    For   a    Poisson
process\index{Poisson process} one has $g(r)=1$.
Closely   related   is   the  correlation   integral\index{correlation
  integral}    $C(r)$   (e.g.~{}\citealt{grassberger:dimensions}),   the
average number  of points  inside a  ball of radius  $r$ centred  on a
point of the distribution
\begin{equation}
C(r) = \int_0^r\d s\ \rho\ 4\pi s^2 g(s) ,
\end{equation}
which  is related by  $K(r)=C(r)/\rho$ to  Ripley's $K$  
function\index{Ripley's $K$--function}, see
Stoyans's paper  in this volume.   Another common way  to characterize
the second--order  properties is the excess fluctuation  of the number
density inside of $C$ with respect to a Poisson 
process\index{$\sigma^2$}:
\begin{equation}
\label{eq:sigma2}
\sigma^2(C) = \frac{1}{|C|^2}\int_C\d\bx\int_C\d\by\  
\xi_2(|\bx-\by|) .
\end{equation}
Often  the  power spectrum\index{power  spectrum}  $P(k)$  is used  to
quantify  the  second  order   statistical  properties  of  the  point
distribution  {}\citep{peebles:lss}.   $P(k)$  may  be defined  as  the
Fourier transform of $\xi_2(r)=g(r)-1$:
\begin{equation}
\label{eq:power-spectrum}
P(k) = \frac{1}{(2\pi)^3}\ \int\d\bx\ \e^{-i\bk\cdot\bx} \xi_2(|\bx|),
\end{equation}
with $k=|\bk|$.

%%
%\subsubsection{Observed two--point correlations}
\paragraph{Observed two--point correlations}

The  first  analysis  of  a  galaxy  catalogue  using  the  two--point
correlation function  was presented by  {}\citet{totsuji:correlation}. 
Following  the   work  of  {}\citet{peebles:statistical},   today  the
two--point correlation function\index{two--point correlation function}
has  become   {\em  the}  standard  tool,  applied   to  nearly  every
cosmological dataset.  The need  for boundary corrected estimators was
recognized  early.   Several  estimators  have been  introduced,  with
differing   claims  on   their   applicability  ({}\citealt{landy:bias},
{}\citealt{hamilton:towards},                 {}\citealt{stoyan:improving},
{}\citealt{kerscher:twopoint},   {}\citealt{ponsborderia:comparing}).    A
clarification   for   cosmological   applications  is   attempted   in
{}\citet{kerscher:comparison}.

\begin{figure}
\begin{center}
\begin{minipage}{\myhalfpage}
\epsfig{file=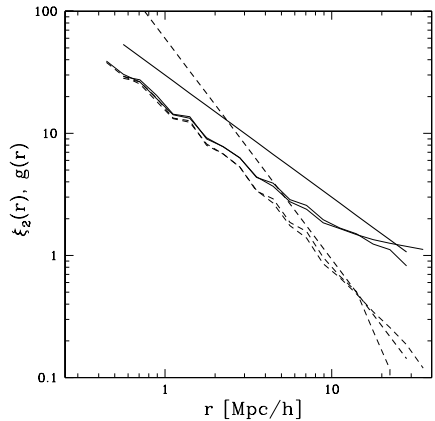,width=\myhalfpage}
\end{minipage} 
\end{center}
\caption{ 
\label{fig:xig-ssrsii}
Estimated  two--point  correlation  function  $g(r)$ (solid)  and  the
normed  cumulant $\xi_2(r)=g(r)-1$  (dashed) in  a  double logarithmic
plot  for the  volume limited  sample  from the  SSRS2 with  100\hMpc\ 
depth.  The  results of the minus (reduced--sample)  estimator and the
{}\protect\citet{fiksel:edge}  estimator are shown,  illustrating that
only on large scales differences occur.
The  straight lines  correspond  to $g(r)\propto  r^{-1}$ (solid)  and
$\xi_2(r)\propto r^{-1.81}$ (dashed).}
\end{figure}

Fig.~\ref{fig:xig-ssrsii} shows the (full) correlation function $g(r)$
and the  normed cumulant $\xi_2(r)$  determined from a  volume limited
sample   of    the   Southern   Sky   Redshift    Survey   2   (SSRS2;
{}\citealt{dacosta:southern}) with 1179 galaxies.  The strong clustering
of galaxies, due to their gravitational interaction, is shown by large
values of $g(r)$ and $\xi_2(r)$ for small $r$.

Of special  physical interest is, whether  the two--point correlations
are  scale--invariant\index{scale   invariance}.   A  scale--invariant
$g(r)\propto r^{D-3}$  is an indication for a  fractal distribution of
the galaxies  (\citealt{mandelbrot:fractal}, {}\citealt{labini:scale}).  A
scale--invariant $\xi_2(r)\propto r^{-\gamma}$ is expected in critical
phenomena (see {}\citealt{goldenfeld:lectures}, \citealt{gaite:fractal}).

Now  lets look at  the log--log  plot in  Figure~\ref{fig:xig-ssrsii}. 
{}\citet{willmer:southern}    give   a    scale--invariant    fit   of
$\xi_2(r)\propto r^{-\gamma}$ with a scaling exponent $\gamma=1.81$ in
the range  of 3-12\hMpc for the  volume limited sample  with 100\hMpc. 
However  on  smaller  scales  the  slope  of  $\xi_2$  is  flattening,
suggesting  that  a  scale--invariant function  $\propto  r^{-\gamma}$
gives  only a  poor description  of  the observed  $\xi_2(r)$ in  this
SSRS2--sample.
If    we    look   at    the    correlation    function   $g(r)$    in
Figure~\ref{fig:xig-ssrsii}, the observed  data may be approximated by
$g(r)\propto  r^{3-D}$   with  $D=2$   over  the  larger   range  from
0.5-20\hMpc.  However the scale--invariance of $g(r)$ is observed over
less  than 2  decades only,  and therefore  an estimate  of  a fractal
dimension $D$  from the scaling  exponent of $g(r)$ may  be misleading
({}\citealt{stoyan:caution},               {}\citealt{mccauley:galaxy},
{}\citealt{mccauley:thegalaxy}, {}\citealt{kerscher:twopoint}).
On large scales the observed  $g(r)$ also deviates from a purely scale
invariant  model, and shows  a tendency  towards unity.   This however
depends on the estimator chosen.
In  this  specific  sample,  a  scale--invariant $g(r)$  seems  to  be
suitable, but  this is not  so clear from  other data sets.   Also the
result on small scales might be  unreliable due to the small number of
pairs with  a short  separation. For a  comprehensive analysis  of the
SSRS2  catalogue focusing  on  two--point properties  and scaling  see
{}\citet{cappi:fractal}.

Hence,  currently  we  cannot  exclude a  scale--invariant  $g(r)$,  a
scale--invariant $\xi_2(r)$, or no  scale--invariance at all, with the
limited    observational    range    provided   by    the    available
three--dimensional  catalogues.   Hopefully  this controversial  issue
will be  clarified in the near  future by the advent  of deeper galaxy
catalogues ({}\citealt{gunn:sdss}, {}\citealt{maddox:2df}).

%%%
\subsection{Higher moments}
\label{sect:higher-moments}

The  two--point  correlation  function  plays  an  important  role  in
cosmology, since  the inflationary paradigm suggests  that the initial
deviations  from the  homogeneous density  field may  be modeled  as a
Gaussian  random  field\index{Gaussian  random field},  stochastically
completely  specified   by  its   mean  density  and   its  two--point
correlation function  (see e.g.~\citealt{boerner:early}).  The analogous
construction  for point  distributions is  the  Gauss--Poisson process
{}\citep{milne:further}, with subtle but important differences from the
Gaussian random field model.   However, the nonlinear evolution of the
mass density  given by {}\eqref{eq:euler-newton}  generates high order
correlations,  not  explainable   within  a  Gaussian  model.   Hence,
assuming  an  initial  Gaussian  density  field,  these  higher  order
correlations  give   us  information  on  the   process  of  structure
formation.

To investigate  these nonlinear structures  several methods are  used. 
In the Sections~\ref{sect:minkowski}  and {}\ref{sect:J} we will focus
on  morphological tools  like  the Minkowski  functionals  and on  the
$J$--function.  A geometrical  method we do not cover  in this text is
percolation    analysis     as    introduced    to     cosmology    by
{}\citet{shandarin:percolation}  (see  also {}\citealt{sahni:probing}).  
Yet another  one is  the analysis based  on the minimal  spanning tree
({}\citealt{barrow:minimalspanning},
{}\citealt{doroshkevich:superlargenbody}).
A  description  of  the  direct,  moment  based  methods  employed  in
cosmology is given now {}\citep{peebles:lss}:

As      a     generalization      of      the     product      density
{}\eqref{eq:product-density}  one  considers  $n$--th order  product
densities\index{$n$--th order  product density}
\begin{equation}
\rho_n(\bx_1,\ldots,\bx_n)\ \d V_1\ldots\d V_n .
\end{equation}
giving the  probability of finding  $n$ points in the  volume elements
$\d  V_{1}$ to  $\d  V_{n}$, respectively.   Again  $\rho_{n}$ is  the
Lebesgue   densities  of   the  $n$--th   factorial   moment  measures
{}\citep{stoyan:stochgeom}.  In  physical applications the (normalized)
cumulants\index{cumulant} are often considered.  As an example we look
at the three--point correlations:
\begin{multline}
\label{eq:rho3-xi3}
\rho_3(\bx_1,\bx_2,\bx_3) = 
\rho^{3} \Big( 1 + 
\xi_2(|\bx_1-\bx_2|)+\xi_2(|\bx_2-\bx_3|)+\xi_2(|\bx_1-\bx_3|) +
\xi_3(\bx_1,\bx_2,\bx_3)\Big) .
\end{multline}
The  three--point correlation  function\index{three--point correlation
  function}, i.e.\  the cumulant $\xi_{3}$,  describes the correlation
of three points in addition  to their correlations determined from the
pairs.  For a Poisson  process\index{Poisson process} all $\xi_n$ with
$n\ge2$  equal   zero.   A   general  definition  of   the  $n$--point
correlation functions\index{$n$--point correlation function} $\xi_{n}$
is         possible         using         generating         functions
(e.g.~\citealt{daley:introduction},      {}\citealt{borgani:scaling}).     
Although  the interpretation  is straightforward,  the  application is
problematic, because a large number of triples etc.\ are needed to get
a stable estimate.  Therefore, one looks for $\xi_{n}$, $n=3,4,\ldots$
mainly     in     angular,     two--dimensional,    surveys     (e.g.\ 
{}\citealt{szapudi:comparison}); for a recent three dimensional analysis
see {}\citet{jing:threepointlcrs}.

More  stable  estimates of  $n$--point  properties,  but with  reduced
informational      content,       may      be      obtained      using
counts--in--cells\index{counts--in--cells}  {}\citep{peebles:lss}.  For
a test volume $C$, typically chosen  as a sphere, we are interested in
the  probability $P_{N}(C)$  of finding  exactly  $N$ points  in $C$.  
These    $P_{N}(C)$   determine   the    one--dimensional   (marginal)
distributions\index{one--dimensional     (marginal)     distributions}
considered  in spatial  statistics  {}\citep{stoyan:stochgeom}.  For  a
Poisson process\index{Poisson process} we have
\begin{equation}
\label{eq:PN-poisson}
P_{N}(C) = \frac{(\rho|C|)^N}{N!}\exp(-\rho|C|) ,
\end{equation}
with the  volume $|C|$  of the  set $C$.  Of  special interest  is the
``void  probability''   $P_{0}(C)$,  which  serves   as  a  generating
functional for all the $P_{N}(C)$, and relates the $P_{N}(C)$ with the
$n$--point     correlation    functions    discussed     above    (see
{}\citealt{stratonovich:topicsI},              {}\citealt{white:hierarchy},
{}\citealt{daley:introduction},  and {}\citealt{balian:I}).  For  a sphere
$B_{r}$  we have  $P_{0}(B_{r})=1-F(r)=1-H_s(r)$,  with the  spherical
contact  distribution\index{spherical  contact  distribution}  $F(r)$,
also denoted by $H_s(r)$ (see Sect.~\ref{sect:J}).

To       facilitate        the       interpretation       of       the
counts--in--cells\index{moments  of  counts--in--cells} one  considers
their $n$--th moments:
\begin{equation}
\sum_{N=0}^{\infty} N^{n} P_{N}(C) .
\end{equation}
They can  be expressed by  the $n$--th moment measures  $\mu_{n}$ (for
their definition see e.g.~\citealt{stoyan:stochgeom}):
\begin{equation}
\mu_{n}(C,\ldots,C) = \sum_{N=0}^{\infty} N^{n} P_{N}(C).
\end{equation}
Especially the centered moments can be related easily to the $n$--point
correlation functions.
As   an   example   consider    the   third   centered   moment   with
$\overline{N}=\rho|C|$ (e.g.\ {}\citealt{coles:cosmology}):
\begin{multline}
\label{eq:centered-moments}
\sum_{N=0}^{\infty} \left(N-\overline{N}\right)^3 P_{N}(C)  = 
\overline{N} + 3\overline{N}^2\sigma^2(C) +
\rho^3\int_C\d\bx\int_C\d\by\int_C\d\bz\ \xi_3(\bx,\by,\bz)
\end{multline}
where  $|C|$  is  the  volume  of  $C$,  and  $\sigma^2(C)$  given  by
{}\eqref{eq:sigma2}.   This centered  moment  incorporates information
from the two--point and  three--point correlations integrated over the
domain $C$.
One may go one step further. The factorial moments\index{factorial 
moments} 
\begin{equation}
\sum_{N=0}^{\infty} N(N-1)\cdots(N-n+1) P_{N}(C) .
\end{equation}
attracted more attention recently,  since they may be estimated easier
with       a        small       variance       ({}\citealt{szapudi:new},
{}\citealt{szapudi:newmethod}), and  offer a concise way  to correct for
typical      observational      problems     (\citealt{colombi:effects},
{}\citealt{szapudi:cosmic}). The  factorial moments may  be expressed by
the      $n$--th     factorial     moment      measures     $\alpha_n$
{}\citep{stoyan:stochgeom} or the $n$--th order product densities:
\begin{multline}
\label{eq:factorial-moment}
\sum_{N=0}^{\infty} N(N-1)\cdots(N-n+1) P_{N}(C) 
= \alpha_n(C,\ldots,C)\\
= \int_C\d\bx_1\ldots\int_C\d\bx_n\ \rho_n(\bx_1,\ldots,\bx_n) ,
\end{multline} 
yielding a simple relation  with the integrated $n$--point correlation
functions by {}\eqref{eq:rho3-xi3}  and its generalizations for higher
$n$.

The moments and the factorial  moments are well defined quantities for
a  stationary   point  process.    Especially  the  relation   of  the
(factorial)  moments  to   the  $n$--point  correlation  functions  in
{}\eqref{eq:centered-moments}   and  {}\eqref{eq:factorial-moment}  is
valid for any stationary point process.  It is worth to note that this
does   not  depend   on  Poisson   sampling  from   a   density  field
{}\eqref{eq:poisson-sampling}.  A lot of work is devoted to relate the
properties of the counts in  cells with the dynamics of the underlying
matter         field         (see         e.g.~\citealt{bouchet:weakly},
{}\citealt{juszkiewicz:weakly},          {}\citealt{padmanabhan:zeldovich},
{}\citealt{bernardeau:properties}).  However, this relation is depending
on   the  galaxy   identification  scheme.    Typically   the  Poisson
model\index{Poisson model} is assumed {}\eqref{eq:poisson-sampling}.

%%%
\subsection{Minkowski functionals}
\label{sect:minkowski}

Minkowski functionals\index{Minkowski functional}, also called 
Querma\ss{} integrals are well known in stochastic and integral 
geometry (see e.g.~\citealt{hadwiger:vorlesung}, 
{}\citealt{weil:stereology}, {}\citealt{schneider:integralgeometrie}, 
{}\citealt{klain:introduction}).
Quantities like  volume, surface  area, and sometimes  also integrated
mean curvature and Euler characteristic were used to describe physical
processes  and  to  construct  models.  Such  models  and  significant
extensions of them were put into the context of integral geometry just
recently  ({}\citealt{mecke:euler}, {}\citealt{mecke:diss}), see  also the
article  by   K.~Mecke  in   this  volume.   The   first  cosmological
application    of    all    Minkowski    functionals   is    due    to
{}\citet{mecke:robust}, marking the advent of Minkowski functionals as
analysis tools for point  processes.  In the following years Minkowski
functionals became more and more common in cosmology.
The interested reader may consider the articles by
% 1995-1996
{}\citet{platzoeder:ringberg}, {}\citet{schmalzing:minkowski},
% 1997
{}\citet{kerscher:abell},                 {}\citet{winitzki:minkowski},
{}\citet{schmalzing:beyond},
% 1998
{}\citet{kerscher:fluctuations},    {}\citet{schmalzing:minkowski_cmb},
{}\citet{novikov:minkowski},         {}\citet{beisbart:characterizing},
{}\citet{sahni:shapefinders},       {}\citet{sathyaprakash:morphology},
{}\citet{hobson:wavelet}, {}\citet{sathyaprakash:filaments},
% 1999
{}\citet{schmalzing:disentanglingI}, {}\citet{schmalzing:quantifying},
{}\citet{dolgov:geometry}, {}\citet{schmalzing:cfa2}.
In the next section a short introduction to Minkowski functionals will be 
given.  See also the articles by K.~Mecke and W.~Weil in this volume.

%%%%%%%
\subsubsection{A short introduction}

Usually we  are dealing with $d$--dimensional  Euclidean space $\BR^d$
with  the  group  of   transformations  $G$  containing  as  subgroups
rotations and translations.   One can then consider the  set of convex
bodies  embedded in this  space and,  as an  extension, the  so called
convex ring $\CR$ of all finite  unions of convex bodies.  In order to
characterize  a  body  $B$  from   the  convex  ring,  also  called  a
poly--convex body,  one looks for scalar functionals  $M$ that satisfy
the following requirements:
\begin{itemize}
\item{\em Motion Invariance\index{motion invariance}:} The functional 
should be independent of the body's position and orientation in space,
\begin{equation} 
M(gB) = M(B) \mbox{ for any } g\in G, \text{ and } B\in\CR .
\end{equation}
\item{\em Additivity\index{additivity}:} Uniting two bodies, one has 
to add their functionals and subtract the functional of the 
intersection,
\begin{equation} 
M(B_1\cup B_2) = M(B_1) + M(B_2) - M(B_1\cap B_2) \mbox{ for any }
B_1,\text{ and } B_2\in\CR.
\end{equation}
\item{\em  Conditional  (or  convex) continuity\index{conditional  (or
convex) continuity}:}  The functionals of  convex approximations
to   a   convex   body   converge   to  the   functionals   of   the 
body,
\begin{equation} 
M(K_i) \rightarrow M(K) \mbox{ as } K_i \rightarrow K \mbox{ for }
K,K_i\in\cal K.
\end{equation}
This applies  to convex bodies only, {\em  not} to the
whole convex ring.  The convergence  for bodies is with respect to
the Hausdorff--metric.
\end{itemize}
One might  think that these  fairly general requirements leave  a vast
choice  of  such functionals.   Surprisingly,  a  theorem by  Hadwiger
states that in fact there  are only $d+1$ independent such functionals
in $\BR^d$.   To be more precise:

\noindent{\em Hadwiger's theorem\index{Hadwiger's theorem}} 
{}\citep{hadwiger:vorlesung}: There exist  $d+1$ functionals $M_\mu$ on
the convex  ring $\CR$ such that  any functional $M$ on  $\CR$ that is
motion  invariant,  additive   and  conditionally  continuous  can  be
expressed as a linear combination of them:
\begin{equation} 
M = \sum_{\mu=0}^d c_\mu M_\mu, {\rm\ with\ numbers\ } c_\mu .
\end{equation}
In this sense  the $d+1$ Minkowski functionals give  a complete and up
to a  constant factor unique  characterization of a  poly--convex body
$B\in\CR$.  The four most  common normalizations are $M_\mu$, $V_\mu$,
$W_\mu$,  and  the  intrinsic  volumes $\overline  V_\mu$  defined  as
follows  ($\omega_\mu$ is  the volume  of the  $\mu$--dimensional unit
ball):
\begin{eqnarray*}
V_\mu = \frac{\omega_{d-\mu}}{\omega_d} M_\mu, \quad
\overline V_{d- \mu} = \frac{\omega_{d-\mu}}{\omega_d} 
\binko{d}{\mu} M_\mu, \\
W_\mu = \frac{\omega_\mu \omega_d}{\omega_{d-\mu}} M_\mu, \quad
\mbox{with} \quad \omega_\mu = \frac{\pi^{\mu/2}}{\Gamma (1+d/2)}.
\end{eqnarray*}
\begin{table}
\caption{The  most  common  notations  for  Minkowski  functionals  in
three--dimensional  space  expressed  in  terms of  the  corresponding
geometric quantities.}
\label{table:func-factors}
\begin{center}
\begin{tabular}{cl|c|c|c|c|c||c}
\\
& geometric quantity & $\mu$ & $M_\mu$ & $V_\mu$ & $W_\mu$ 
& $\overline V_{3 -\mu}$ & $\omega_\mu$ \\
\hline
$V$ & volume & 0 & $V$ & $V$ & $V$ & $V$ & 1\\
$A$ & surface & 1 & $A/8$ & $A/6$ & $A/3$ & $A/2$ & 2\\
$H$ & int.\ mean curvature & 2 & $H/2\pi^2$ & $H/3\pi$ & $H/3$ 
& $H/\pi$ & $\pi$ \\
$\chi$ & Euler characteristic\index{Euler characteristic} & 3 & 
$3\chi/4\pi$ & $\chi$ & $4\pi\chi/3$ 
& $\chi$ & $4\pi/3$
\end{tabular}
\end{center}
\end{table}
In three--dimensional Euclidean space, these functionals have a direct
geometric         interpretation          as         listed         in
Table~\ref{table:func-factors}.

%%%
\subsubsection{The germ-grain model\index{germ-grain model}}

Now the  Minkowski functionals are  used to describe the  geometry and
topology  of a  point set  $X=\{\bx_i\}_{i=1}^N$.   Direct application
gives  rather   boring  results,  $V_\mu(X)=0$   for  $\mu=0,1,2$  and
$V_3(X)=N$.   However, one  may think  of $X$  as a  skeleton  of more
complicated   spatial   structures   in   the  universe   (see   e.g.\
Fig~\ref{fig:cfa2}).   Decorating $X$  with balls  of radius  $r$ puts
``flesh'' on the skeleton in  a well defined way.  Also non--spherical
grains may be used.

The  Minkowski   functionals  for  the   union  set  of   these  balls
$A_r=\bigcup_{i=1}^N B_r(\bx_i)$  give non--trivial results, depending
on the point distribution considered.  We will use $r$ as a diagnostic
parameter  specifying   a  neighborhood  relations,   to  explore  the
connectivity and shape of $A_r$.

\begin{figure}
\begin{center}
\begin{minipage}{\mythirdpage}
\epsfig{file=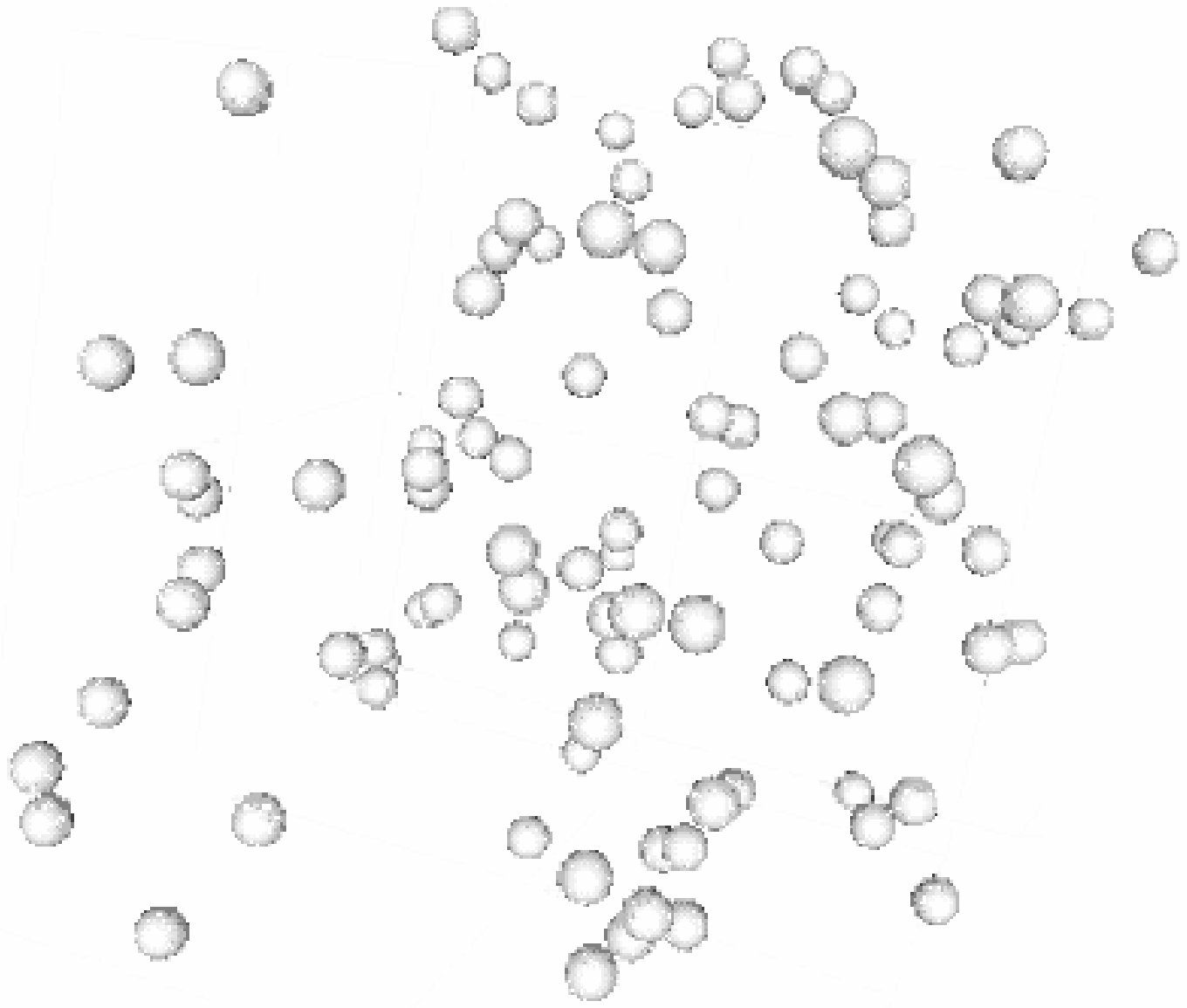,width=\mythirdpage}
\end{minipage} 
\hfill
\begin{minipage}{\mythirdpage}
\epsfig{file=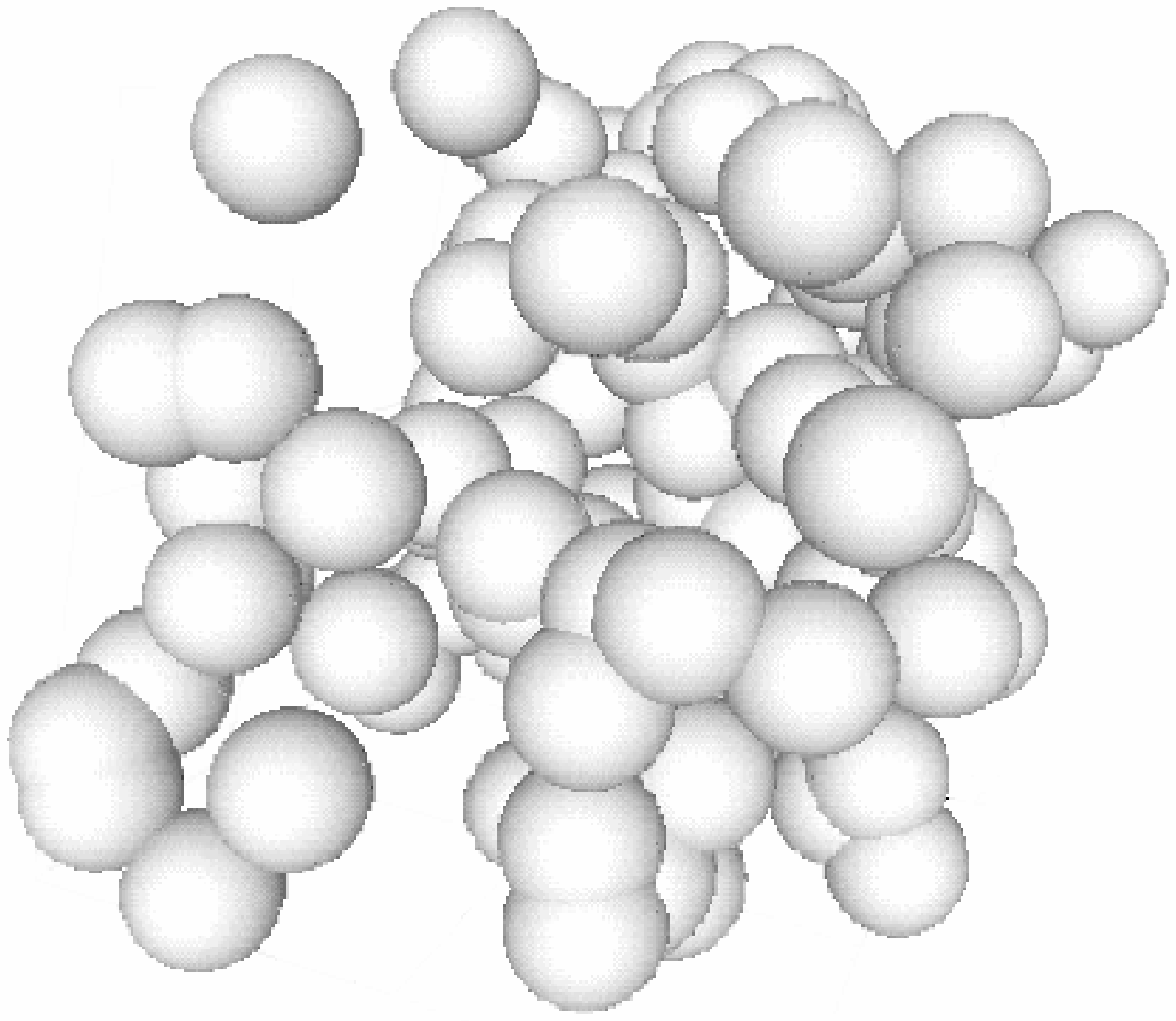,width=\mythirdpage}
\end{minipage} 
\hfill
\begin{minipage}{\mythirdpage}
\epsfig{file=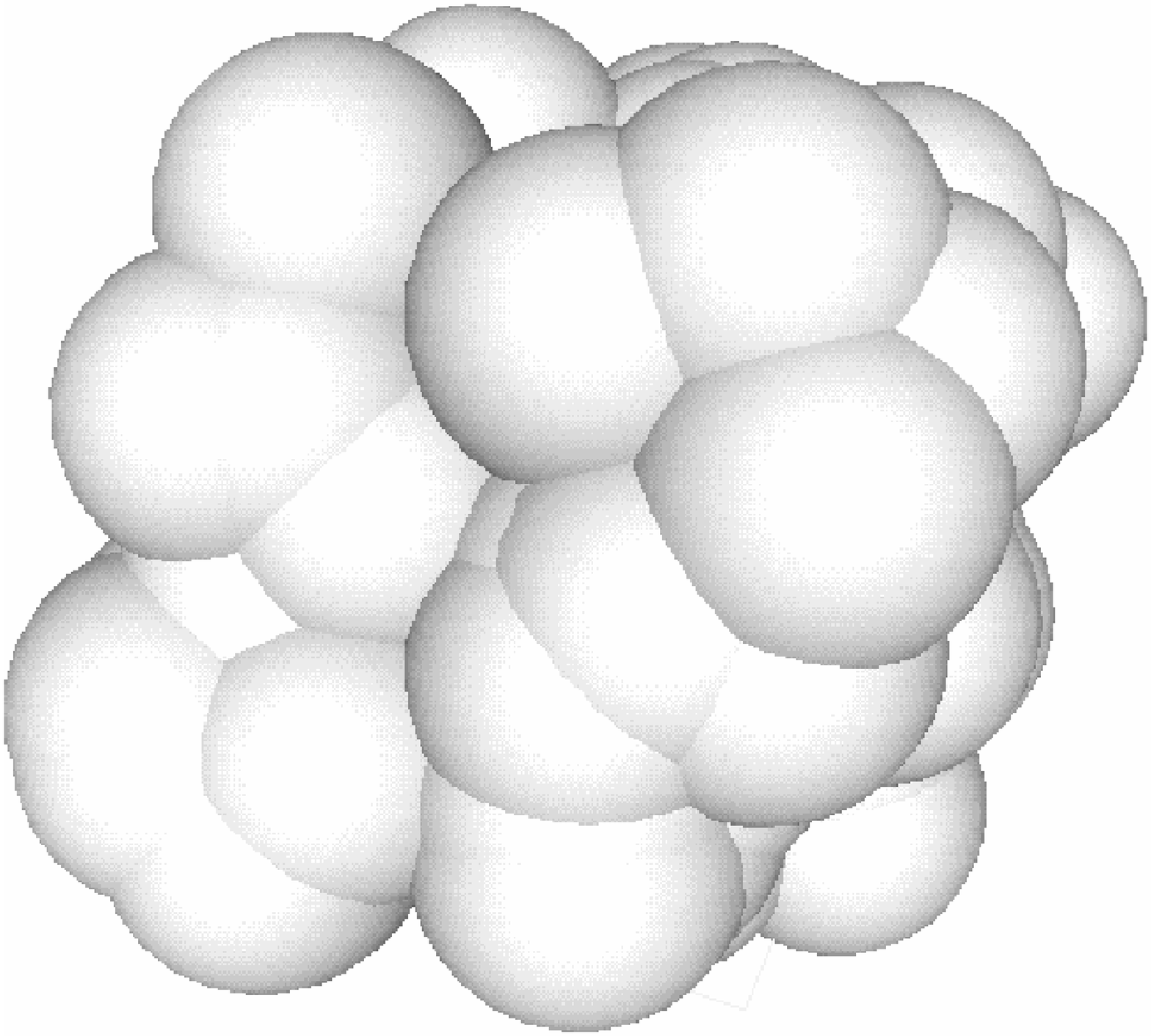,width=\mythirdpage}
\end{minipage} 
\end{center}
\caption{\label{fig:bobbel} Randomly distributed points decorated with
balls of  varying radius  $r$ -- a  realization of the  Booelean grain
model.}
\end{figure}

Let  $X$ be  a finite  subset of  a realization  of a  Poisson process
inside some finite domain $W$.  Then  $A_r$ is a part of a realization
of   the  {\em  Boolean   grain  model}\index{Boolean   grain  model},
illustrated in Figure~\ref{fig:bobbel}.
For these randomly  placed balls the mean volume  densities $m_\mu$ of
the   Minkowski   functionals\index{volume   densities  of   Minkowski
  functionals}     are     known     (e.g.\     {}\citealt{mecke:euler},
{}\citealt{schneider:integralgeometrie},  also   called  intensities  of
Minkowski functionals).
\begin{equation}
\label{eq:boolean-model}
\begin{array}{llll}
m_0(A_r) = & 1-\e^{-\rho  M_0}, & 
m_2(A_r) = & \e^{-\rho M_0 }\ (M_2 \rho - M_1^2 \rho^2 ) ,\\[2ex]
m_1(A_r) = & \e^{-\rho  M_0}\ M_1 \rho , & 
m_3(A_r) = & \e^{-\rho M_0}\ (M_3 \rho  -3 M_1 M_2 \rho^2  + M_1^3 \rho^3 ) ,
\end{array}
\end{equation}
with the number density  $\rho$ and 
\begin{equation}
M_0 = \frac{4\pi}{3} r^3,\quad
M_1 = \frac{\pi}{2} r^2,\quad
M_2 = \frac{4}{\pi} r,\quad
M_3 = \frac{3}{4\pi} .
\end{equation}

Starting from a general point  process, decorating it with spheres, we
arrive    at     the    {\em    germ--grain     model}    (see    also
{}\citealt{stoyan:stochgeom}).   The   Minkowski  functionals  or  their
volume densities calculated  for the set $A_r$ may be  use as tools to
describe  the underlying  point distribution,  directly  comparable to
standard  point  process statistics  like  the two--point  correlation
function   (Sect.~\ref{sect:twopoint})   or   the   nearest   neighbor
distribution   (Sect.~\ref{sect:J}).   Indeed,   the   volume  density
$m_0(A_r)$  equals the spherical  contact distribution\index{spherical
  contact  distribution} or  equivalently the  void  probability minus
one:         $m_0(A_r)=F(r)=H_s(r)=P_0(B_r)-1$        (see        also
Sect.~\ref{sect:higher-moments}).
Expressions relating  the Minkowski functionals  of such a  set $A_r$,
with   the  $n$--point   correlation  functions   of   the  underlying
point--process     may    be     found     in    {}\citet{mecke:diss},
{}\citet{mecke:robust},  and {}\citet{schmalzing:quantifying}  and the
contribution by K.~Mecke in this volume.

Already for moderate  radii $r$ nearly the whole  space is filled with
up by  $A_r$, leading to  $m_0(A_r)\approx1$ and $m_\mu(A_r)\approx0$,
with  $\mu>0$.  This  illustrates the  different role  the  radius $r$
plays for  the Minkowski functionals  compared to the distance  $r$ as
used  in the two--point  correlation function  $g(r)$.  Already  for a
fixed radius, the Minkowski functionals  of $A_r$ are sensitive to the
global geometry  and topology  of $A_r$ and,  hence, of  the decorated
point  set  (see  also Sect.~\ref{sect:minkowski-excursion}).   Indeed
point sets with an identical two--point correlation function, but with
clearly different large scale  morphology may be generated easily (see
e.g.\ {}\citealt{baddeley:cautionary}, and {}\citealt{szalay:walls}).

All galaxy  catalogues are spatially limited.  To  estimate the volume
densities  of Minkowski  functionals  for such  a  realization of  the
germ--grain  model  given  by  the  coordinates of  galaxies,  we  use
boundary   corrections\index{boundary    corrections   for   Minkowski
  functionals} based  on principal kinematical formula\index{principal
  kinematical     formula}    (see   {}\citealt{mecke:euler},
{}\citealt{stoyan:stochgeom}, {}\citealt{schmalzing:minkowski}):
\begin{equation}
m_\mu(A_r) = 
\frac{M_\mu(A_r \cap W)}{M_0(W)} - 
\sum_{\nu=0}^{\mu-1} 
\left(\!\!\!\begin{array}{c}{\mu}\\{\nu}\end{array}\!\!\!\right) 
m_\nu(A_r) \frac{M_{\mu-\nu}(W)}{M_0(W)},
\end{equation}
We  use  the convention  $\sum_{n=i}^jx_n=0$  for  $j<i$.  An  example
illustrating    these    boundary     corrections    is    given    in
{}\citet{kerscher:minkowski}.

In the  following an  application of these  methods to a  catalogue of
galaxy  clusters  {}\citep{kerscher:abell} (an  earlier  analysis of  a
smaller cluster catalogue was already given by {}\citealt{mecke:robust})
and  to  a  galaxy  catalogue  will  illustrate  the  qualitative  and
quantitative results obtainable with global Minkowski functionals.

%%%%
\subsubsection{Cluster catalogues}
\label{sect:cluster-minkowski}

The  spatial distribution of  centers of  galaxy clusters\index{galaxy
  cluster},     using    the     Abell/ACO    cluster     sample    of
{}\citet{plionis:evidence},     was     analyzed    with     Minkowski
functionals\index{Minkowski  functional}  applied  to the  germ--grain
model {}\citep{kerscher:abell}.   At first a  qualitative discussion of
the  observed features  is presented,  followed by  a  comparison with
models for the cluster distribution. 

\begin{figure}
\begin{center}
\begin{minipage}{\mymediumpage}
\epsfig{file=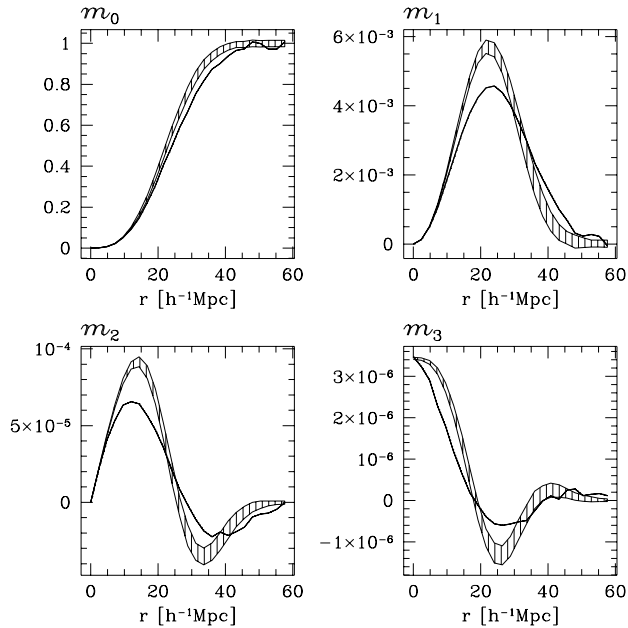,width=\mymediumpage}
\end{minipage} 
\caption{\label{fig:min-abell} Densities  of the Minkowski functionals
for the  Abell/ACO (solid  line) and a  Poisson process  (shaded area)
with the same  number density.  The shaded area  gives the statistical
variance  of  the  Poisson   process  calculated  from  100  different
realizations.}
\end{center}
\end{figure}
The  most  prominent feature  of  the  volume  densities of  all  four
Minkowski functionals  are the broader extrema for  the Abell/ACO data
as   compared  to   the   results  for   the   Poisson  process   (see
Fig.~\ref{fig:min-abell}).   This is a  first indication  for enhanced
clustering.  Let us now look at each functional in detail:\\
The density of the Minkowski  functional $m_0$ measures the density of
the covered volume.  On  scales between $25\hMpc$ and $40\hMpc$, $m_0$
as a function of $r$ lies  slightly below the Poisson data. The volume
density is lower because of the clumping of clusters on those
scales.\\
The  density of the  Minkowski functional  $m_1$ measures  the surface
density of the coverage.  It has a maximum at about $20\hMpc$ both for
the Poisson process and for the  cluster data.  This maximum is due to
the granular  structure of the union  set on the  relevant scales.  At
the   same  scales,   we   find  the   maximum   deviation  from   the
characteristics for  the Poisson process.   The lower values  of $m_1$
for  the  cluster  data with  respect  to  the  Poisson are  again  an
indication of a significant clumping of clusters at these scales.  The
functional $m_1$ shows  also a positive deviation from  the Poisson on
scales of $(35$\dots$50)\hMpc$ where  more coherent structures form in
the union
set than in the Poisson process, keeping the surface density larger.\\
The   densities  of   the  Minkowski   functionals  $m_2$   and  $m_3$
characterize in more  detail the kind of spatial  coverage provided by
the union set  of balls in the data sample.  The  density of the total
mean curvature $m_2$ of the  data reaches a maximum at about $10\hMpc$
produced by the dominance  of convex (positive $m_2$) structures.  The
density $m_2$  at the maximum is  reduced with respect  to the Poisson
process to about  70\% (or more than three  standard deviations).  The
integral  mean curvature  $m_2$ has  a zero  at a  scale  of $25\hMpc$
(almost  the   scale  of  maximum  of  $m_1$)   corresponding  to  the
turning--point  between  structures  with  mainly convex  and  concave
boundaries (negative $m_2$).   Significant deviations from the Poisson
process occur  between this  turning point and  $40 \hMpc$ due  to the
smaller mean curvature  of the union set of  the data, probably caused
by the interconnection of the void regions
in the cluster distribution.\\
The  density of  the Euler  characteristic\index{Euler characteristic}
$m_3$ describes  the global topology of the  cluster distribution.  On
small scales  all balls are  separated.  Therefore, each ball  gives a
contribution  of  unity  to  the  Euler characteristic  and  $m_3$  is
proportional to the cluster  number density.  As the radius increases,
more and more balls overlap and  $m_3$ decreases.  At a scale of about
$20\hMpc$ it drops  below zero due to the emergence  of tunnels in the
union set  (a double torus  has $\chi=-1$).  The positive  maximum for
the Poisson  process at scales  $\simeq 40\hMpc$ is the  signature for
the presence  of cavities.   The nearly linear  decrease of  the Euler
characteristic for the Abell/ACO sample indicates strong clustering on
scales $\le15\hMpc$.  The lack of a significant positive maximum after
the  minimum shows that  only a  few cavities  form.  This  suggests a
support dimension for the distribution of clusters of less than three.
The presence of  voids on scales of $30$ to $45\hMpc$  is shown by the
enhanced surface  area $m_1$ and  the reduced integral  mean curvature
$m_2$,  while  on  these  scales  the Euler  characteristic  $m_3$  is
approximately zero.

\begin{figure}
\begin{center}
\begin{minipage}{\mymediumpage}
\epsfig{file=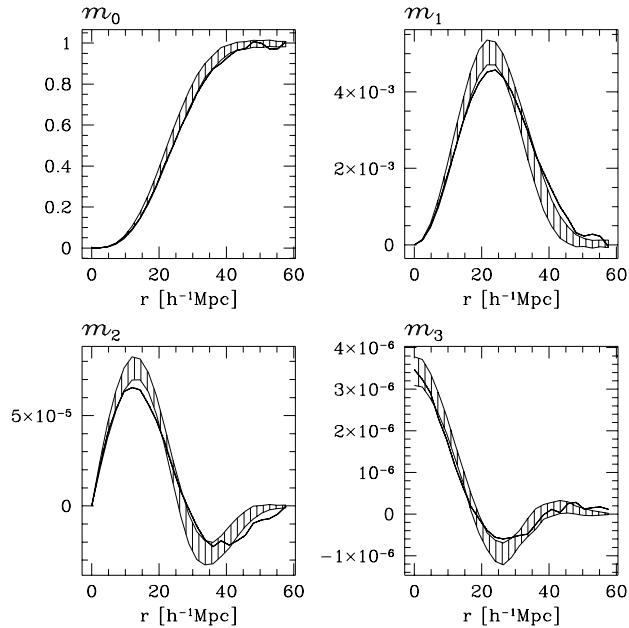,width=\mymediumpage}
\end{minipage} 
\caption{\label{fig:min-cdm-abell}   Densities    of   the   Minkowski
functionals for the Abell/ACO (solid  line in both panels) compared to
the  SCDM  (shaded  area  in   top  panel).   The  shaded  area  gives
1$\sigma$-error bars of the variance among different realizations.}
\end{center}
\end{figure}

The emphasis  of {}\citet{kerscher:abell}  was on the  comparison with
cosmological  model predictions. For  this purpose  artificial cluster
distributions  were constructed,  from  the density  field of  N--body
simulations. Such  simulations are  still quite costly,  and therefore
only four specific models were investigated.
In  Fig.~\ref{fig:min-cdm-abell} the  comparison  of the  observations
with  the Standard  Cold--Dark--Matter  (SCDM) model  is shown.   This
model shows  too little clustering on  small scales, as  it is clearly
seen by the enhanced maxima of the surface area $m_1$ and the integral
mean curvature $m_2$, as well as  in the flatter decrease of the Euler
characteristic $m_3$. Additionally,  the higher volume $m_0$ indicates
weak clumping and to few coherent structures also on large scales.
These deviations may be quantified  using some norm for the comparison
of the observational  data with the model prediction  (for details see
{}\citealt{kerscher:abell}).
A comparison  of the clusters distribution with  CDM--models using the
power   spectrum  {}\eqref{eq:power-spectrum}   lead   to  a   similar
conclusions {}\citep{retzlaff:constraining}.

\subsubsection{Large fluctuations}
\label{sect:fluctuations}

A physically interesting point is how well defined are the statistical
properties of the galaxy  or cluster distribution, determined from one
spatially limited realization only.  Or  in other words, how large are
the fluctuations\index{fluctuation} of the  morphology for a domain of
given  size?  {}\citet{kerscher:fluctuations}  investigate  this using
Minkowski       functionals\index{Minkowski      functional},      the
$J$--function\index{$J$--function}  (see section  {}\ref{sect:J}), and
the       two--point       statistic\index{$\sigma^2$}      $\sigma^2$
{}\eqref{eq:sigma2}.

By normalizing  with the functional  $M_\mu(B_r)$ of a single  ball we
can   introduce   normalized,   dimensionless  Minkowski   functionals
$\Phi_\mu(A_r)$,
\begin{equation}\label{eq:Phi--def}
\Phi_\mu(A_r) = \frac{m_\mu(A_r)}{\rho M_\mu(B_r)},
\end{equation}
where $\rho$ is the number density.   In the case of a Poisson process
the  exact  mean  values  are known  {}\eqref{eq:boolean-model}.   For
decorating spheres with radius $r$ one obtains:
\begin{equation}
\begin{array}{llll}
\Phi_0^P &=\left(1-\e^{-\eta}\right)~\eta^{-1}, &
\Phi_1^P &=\e^{-\eta}, \\
\Phi_2^P &=\e^{-\eta}~(1-\tfrac{3\pi^2}{32}\eta), &
\Phi_3^P &=\e^{-\eta}~(1-3\eta+\tfrac{3\pi^2}{32}\eta^2), 
\end{array}
\end{equation}
with  the  dimensionless   parameter  $\eta=\rho  M_0(B_r)=\rho\  4\pi
r^3/3$.   For  $\mu\ge1$ the  measures  $\Phi_{\mu}(A_r)$ contain  the
exponentially  decreasing   factor  $\e^{-\eta(r)}$.  We   employ  the
reduction
\begin{equation}
\phi_\mu(A_r) = \frac{\Phi_\mu(A_r)}{\Phi_1^P(A_r)}, \quad \mu \ge 1,
\end{equation}
and thereby remove the exponential decay and enhance the visibility of
differences in the displays shown below.

We  now apply  the  methods  introduced above  to  explore a  redshift
catalogue of  5313 IRAS selected  galaxies\index{galaxy} with limiting
flux of 1.2~Jy {}\citep{fisher:irasdata}.
A volume  limited sample of  100\hMpc\ depth contains 352  galaxies in
the northern part, and 358 galaxies in the southern part (with respect
to        galactic       coordinates),        as        shown       in
Fig.~\ref{fig:galaxy-volume-limited}.
As far as the number density\index{number density}, i.e.\ the first 
moment of the galaxy distribution is concerned, the sample does not 
reveal significant differences between north and south.  However, we 
want to assess the clustering properties of the data and, above all, 
tackle the question whether the southern and northern parts differ or 
not.
\begin{figure}
\begin{center}
\begin{minipage}{\mythirdpage}
\epsfig{file=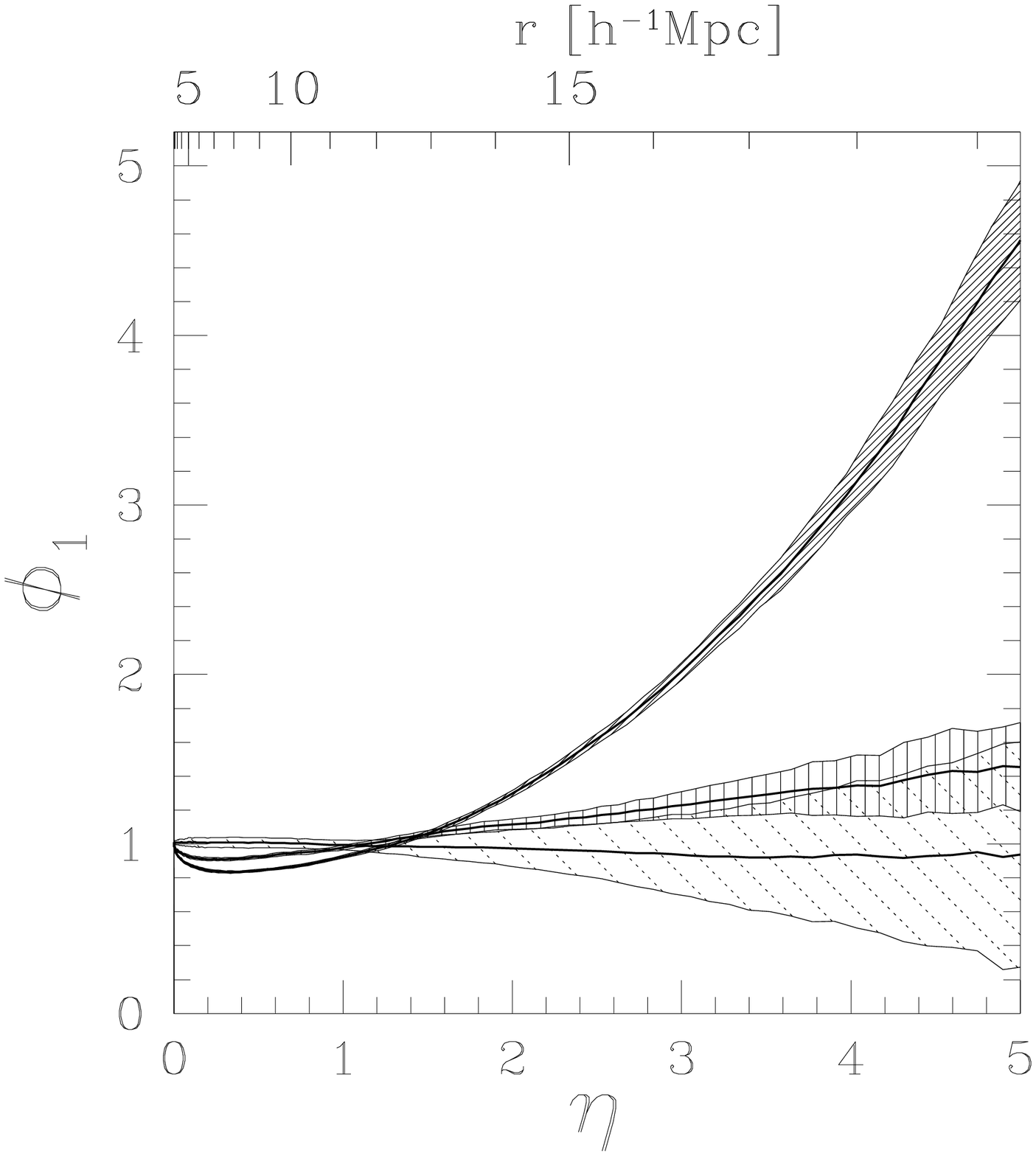,width=\mythirdpage}
\end{minipage} 
\hfill
\begin{minipage}{\mythirdpage}
\epsfig{file=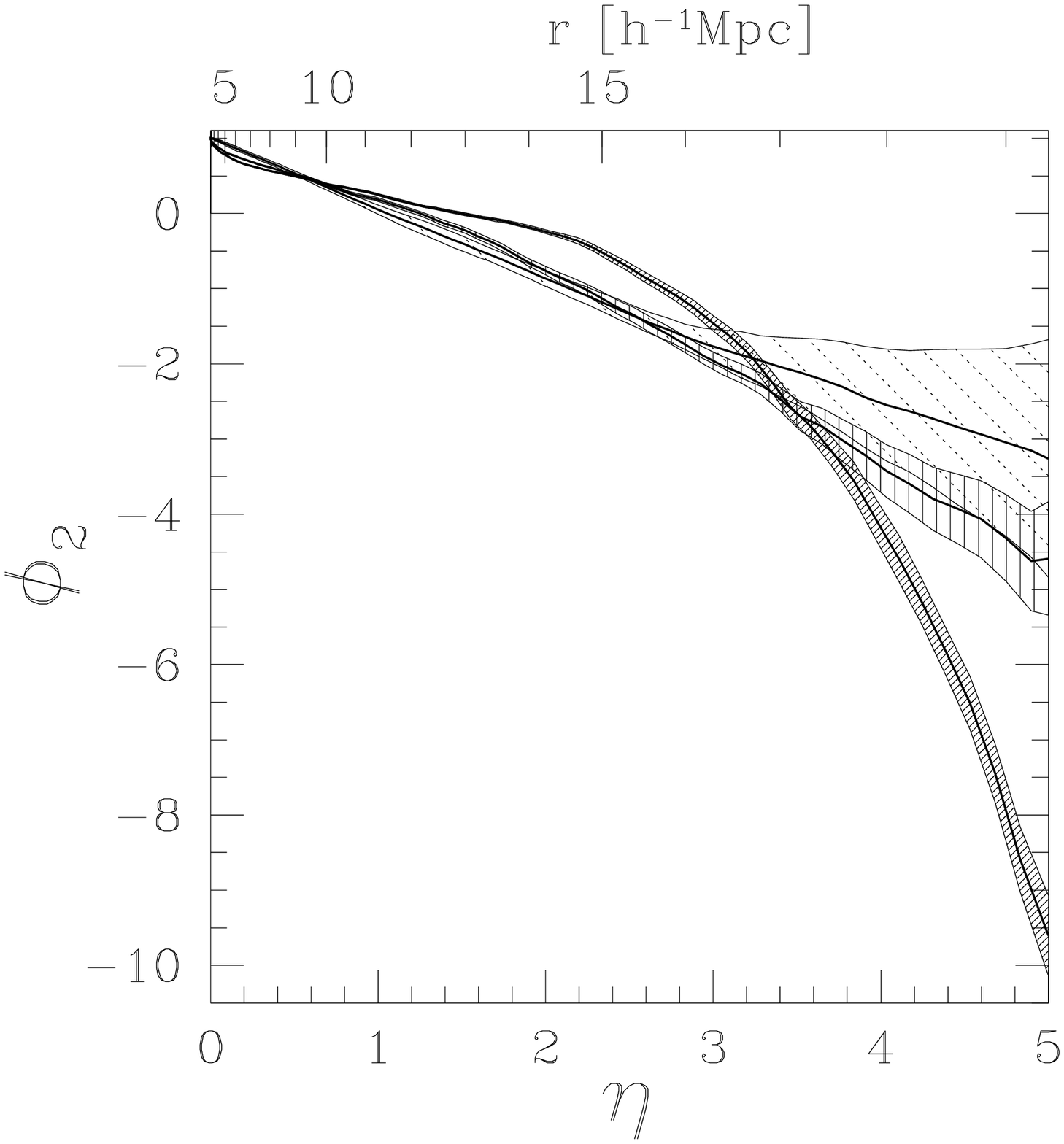,width=\mythirdpage}
\end{minipage} 
\hfill
\begin{minipage}{\mythirdpage}
\epsfig{file=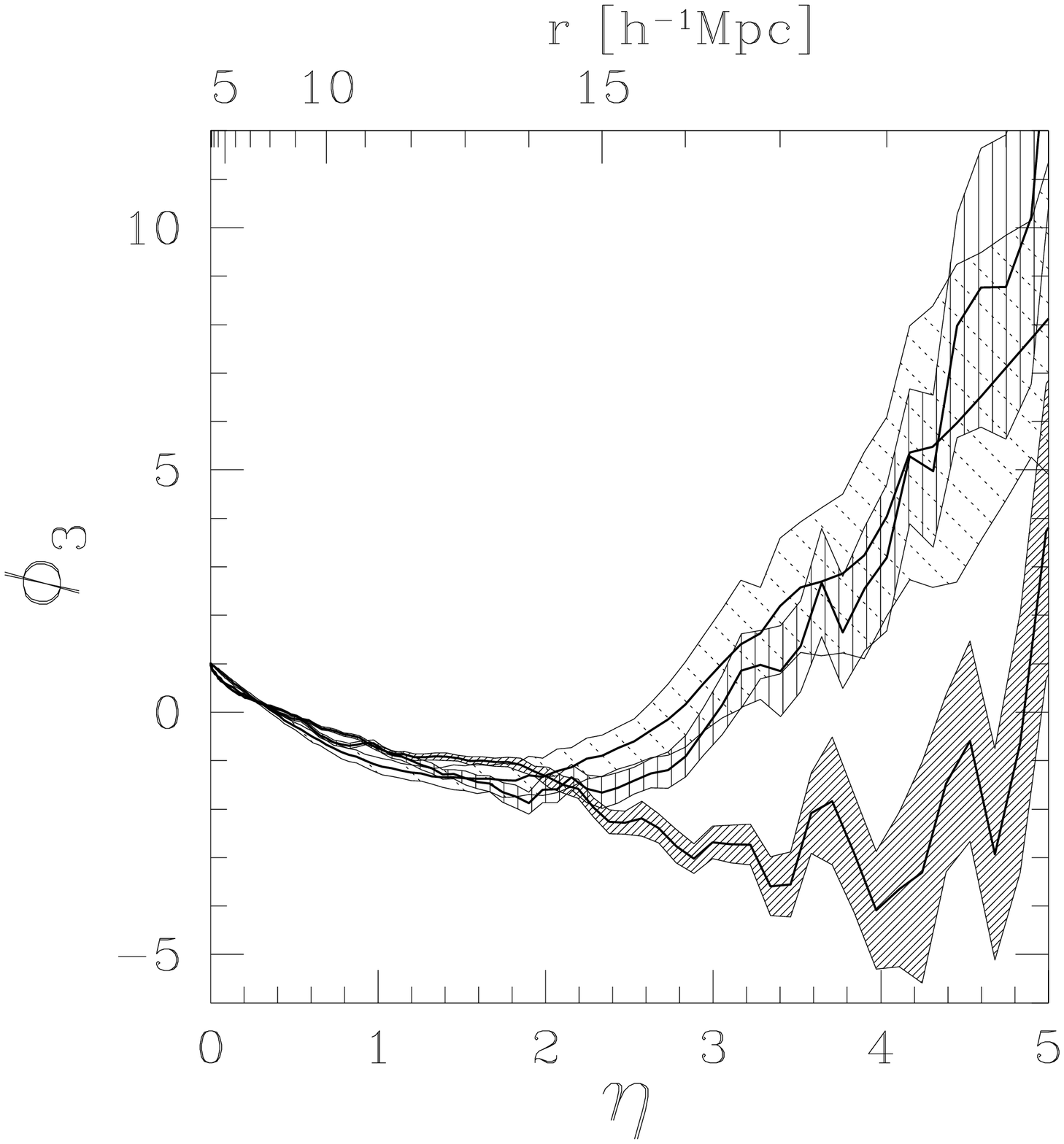,width=\mythirdpage}
\end{minipage} 
\end{center}
\caption{\label{fig:minjyv10}  Minkowski functionals  $\phi_\mu$  of a
volume  limited sample with  100\hMpc\ depth  extracted from  the IRAS
1.2~Jy catalogue;  the dark shaded areas represent  the southern part,
the medium shaded the northern  part, and the dotted a Poisson process
with  the same  number density.  The  shaded areas  are the  $1\sigma$
errors estimated from twenty  realizations for the Poisson process and
from twenty errors using a Jackknife procedure with 90\% sub--sampling,
for the data.}
\end{figure}
A  characterization  of  the  global morphology  using  the  Minkowski
functionals (Fig.~\ref{fig:minjyv10}) shows that  in both parts of the
1.2~Jy catalogue the  clustering of galaxies on scales  up to 10\hMpc\
is clearly stronger than in the case of a Poisson process, as inferred
from  the  lower values  of  the  functionals  for the  surface  area,
$\phi_1$,  the  integral  mean  curvature,  $\phi_2$,  and  the  Euler
characteristic, $\phi_3$.   Moreover, the northern  and southern parts
differ significantly,  with the northern  part being less  clumpy. The
most conspicuous  features are the  enhanced surface area  $\phi_1$ in
the southern  part on scales from 12  to 20\hMpc\ and the  kink in the
integral mean curvature $\phi_2$  at 14\hMpc.  This behavior indicates
that dense  substructures in the southern  part are filled  up at this
scale (i.e.~the  balls in these substructures  overlap without leaving
holes).

These strongly  fluctuating clustering properties are  also visible in
the $J$--function  (Sect.~\ref{sect:J}), and the  $\sigma^2(B_r)$ (see
{}\eqref{eq:sigma2}).   An analysis  of possible  contaminations and
systematic selection  effects showed that these  fluctuations are real
structural  differences  in  the  galaxy  distribution  on  scales  of
100\hMpc\     even     extending     to    200\hMpc\     (see     also
{}\citealt{kerscher:significance}).
It is interesting to note that an N--body simulation in a periodic box
with side--length of 250\hMpc\ {}\citep{kolatt:simulating} was not able
to reproduce these large--scale fluctuations.

%%%
\subsubsection{Minkowski functionals of excursion sets}
\label{sect:minkowski-excursion}

\begin{figure}
\begin{center}
\begin{minipage}{\mythirdpage}
\epsfig{file=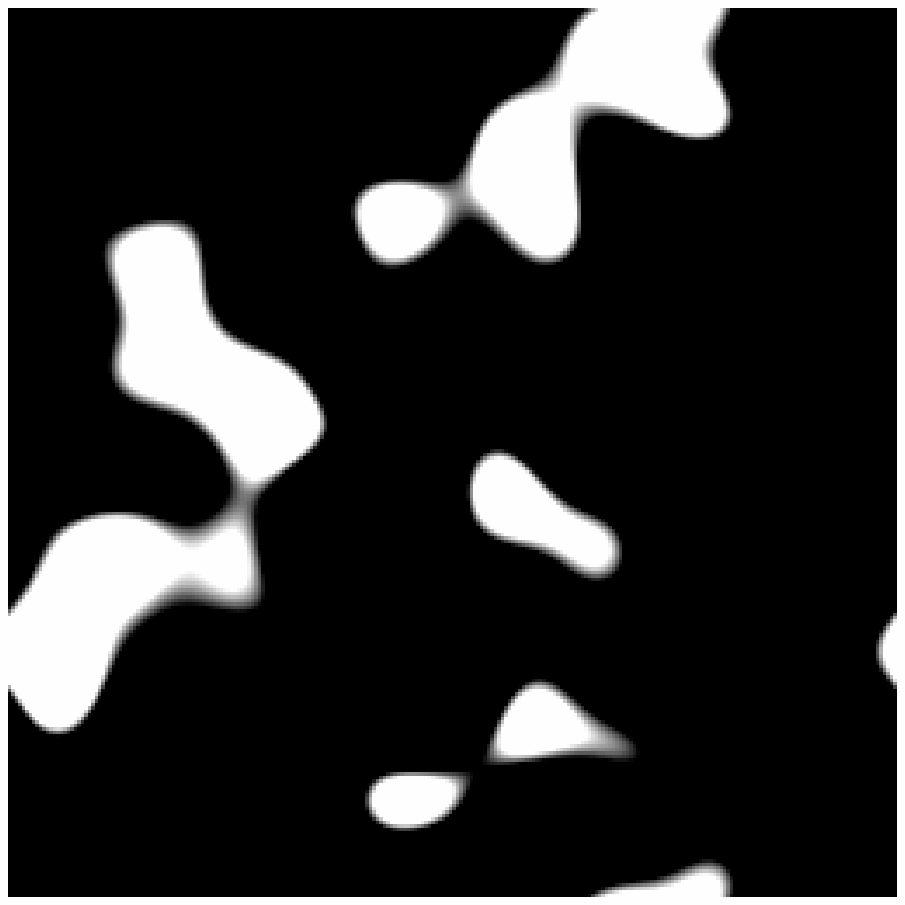,width=\mythirdpage}
\end{minipage} 
\hfill
\begin{minipage}{\mythirdpage}
\epsfig{file=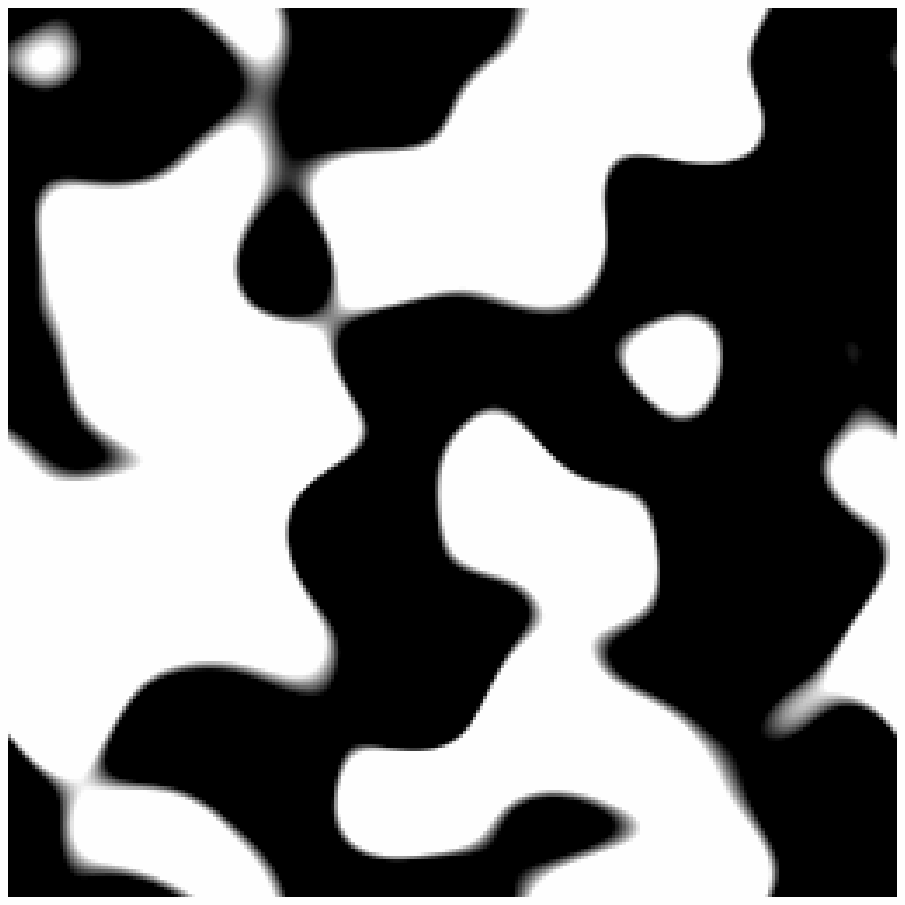,width=\mythirdpage}
\end{minipage} 
\hfill
\begin{minipage}{\mythirdpage}
\epsfig{file=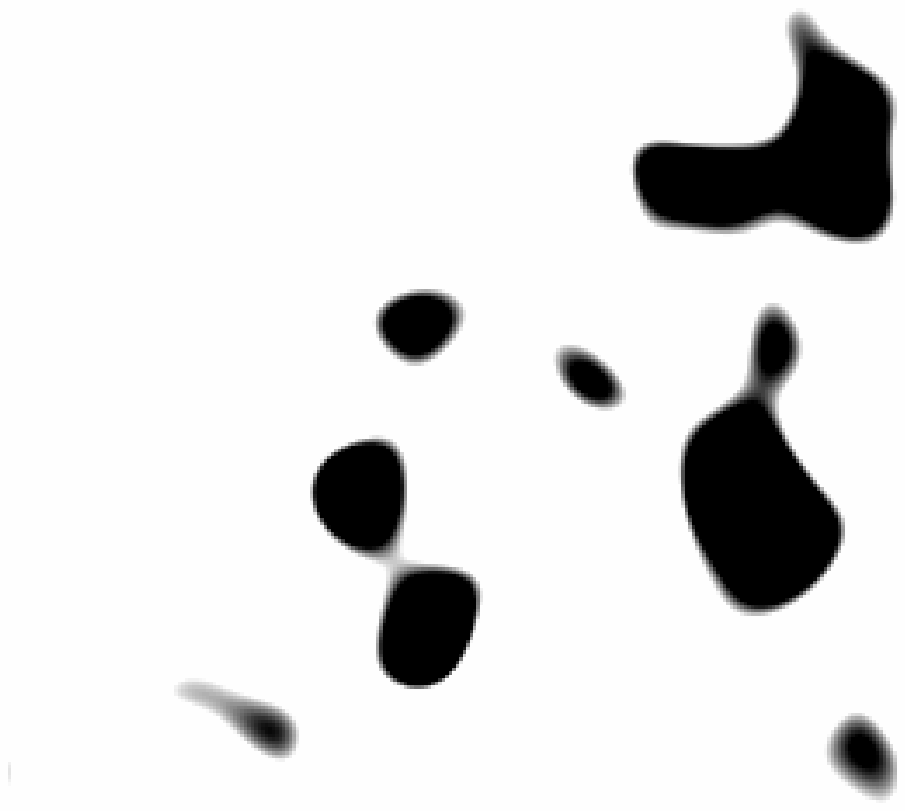,width=\mythirdpage}
\end{minipage} 
\end{center}
\caption{\label{fig:excursion} The  black set marks  the excursion set
$Q_\nu$ of a Gaussian density field with increasing $\nu$ from left to
right. Only the highest peaks remain for large $\nu$.}
\end{figure}

In  the  preceding section  the  Minkowski  functionals  were used  to
characterize the union  set of balls, the body  $A_r$.  Consider now a
smooth density  or temperature field  $u(\bx)$.  We wish  to calculate
the Minkowski functionals\index{Minkowski  functional} of an excursion
set\index{excursion  set} $Q_\nu$  over a  given threshold  $\nu$ (see
Fig.~\ref{fig:excursion}), defined by
\begin{equation}
Q_\nu = \{\bx~|~u(\bx)\ge\nu\} .
\end{equation}
This  threshold $\nu$  will be  used  as a  diagnostic parameter.
The  geometry  and  topology  of  random  fields  $u(\bx)$  and  their
excursion       sets       was       studied      extensively       by
{}\citet{adler:randomfields}.   Two complementary  calculation methods
for  the  Minkowski functionals  of  the  excursion  set $Q_\nu$  were
presented by {}\citet{schmalzing:beyond}.

Starting  with a  given  point  distribution a  density  field may  be
constructed with a folding  employing some kernel $k_\epsilon(\bz)$ of
width $\epsilon$
\begin{equation}
u(\by) = \sum_{i=1}^N\ k_\epsilon(\bx_i-\by).
\end{equation}
Often a  triangular or  a Gaussian kernel  sometimes with  an adaptive
smoothing scale  $\epsilon(\by)$ are used.  A  discussion of smoothing
techniques may be found in {}\citet{silverman:density}.

The  Euler characteristic\index{Euler  Characteristic}  $\chi$ of  the
excursion set is directly related to the genus $G$ of the iso--density
surface separating low from high density regions:
\begin{equation}
G(\partial Q_\nu) = 1-2\chi(Q_\nu) .
\end{equation}
The  analysis  of  cosmological  density  field  using  the  genus  of
iso--density  surfaces  is a  well  accepted  tool  in cosmology  (see
{}\citealt{weinberg:topologyI},                  {}\citealt{melott:review},
{}\citealt{coles:quantifying} and  refs.\ therein), now  incorporated in
the more general analysis using Minkowski functionals.  Especially the
Euler characteristic of excursion  sets has also applications in other
fields like medical image processing {}\citep{worsley:testing}.

%%%%%%
\subsubsection{Gaussianity of the cosmic microwave background}

As   already  mentioned   in  Sect.~\ref{sect:higher-moments}   it  is
physically very interesting, whether  the observed fluctuations in the
temperature    field    of    the    cosmic    microwave    background
radiation\index{cosmic microwave background radiation} (CMB), as shown
in  Fig.~\ref{fig:cobe},   are  compatible  with   a  Gaussian  random
field\index{Gaussian random field} model.  For a Gaussian random field
{}\citet{tomita:statistical}  obtained analytical expressions  for the
Minkowski   functionals\index{Minkowski  functional}  of   $Q_\nu$  in
arbitrary dimensions.
Since the temperature fluctuations  are given on the celestial sphere,
an adopted  integral geometry for spaces with  constant curvature must
be              used             ({}\citealt{santalo:integralgeometry}).
{}\citet{schmalzing:minkowski_cmb} took  this geometric constraint and
further complications due to boundary  and binning effects, as well as
noise contributions  into account. They find  no significant deviation
from a Gaussian random field for the resolution of the COBE data set.

Other methods to test for  Gaussianity are based on a wavelet analysis
{}\citep{hobson:wavelet}    on   high--order    correlation   functions
{}\citep{heavens:estimating} or on  the two--point correlation function
of peaks in the temperature fluctuations {}\citep{heavens:correlation}.

%%%
\subsubsection{Geometry of single objects -- shape--finders}

\begin{figure}
\begin{center}
\begin{minipage}{\myhalfpage}
\epsfig{figure=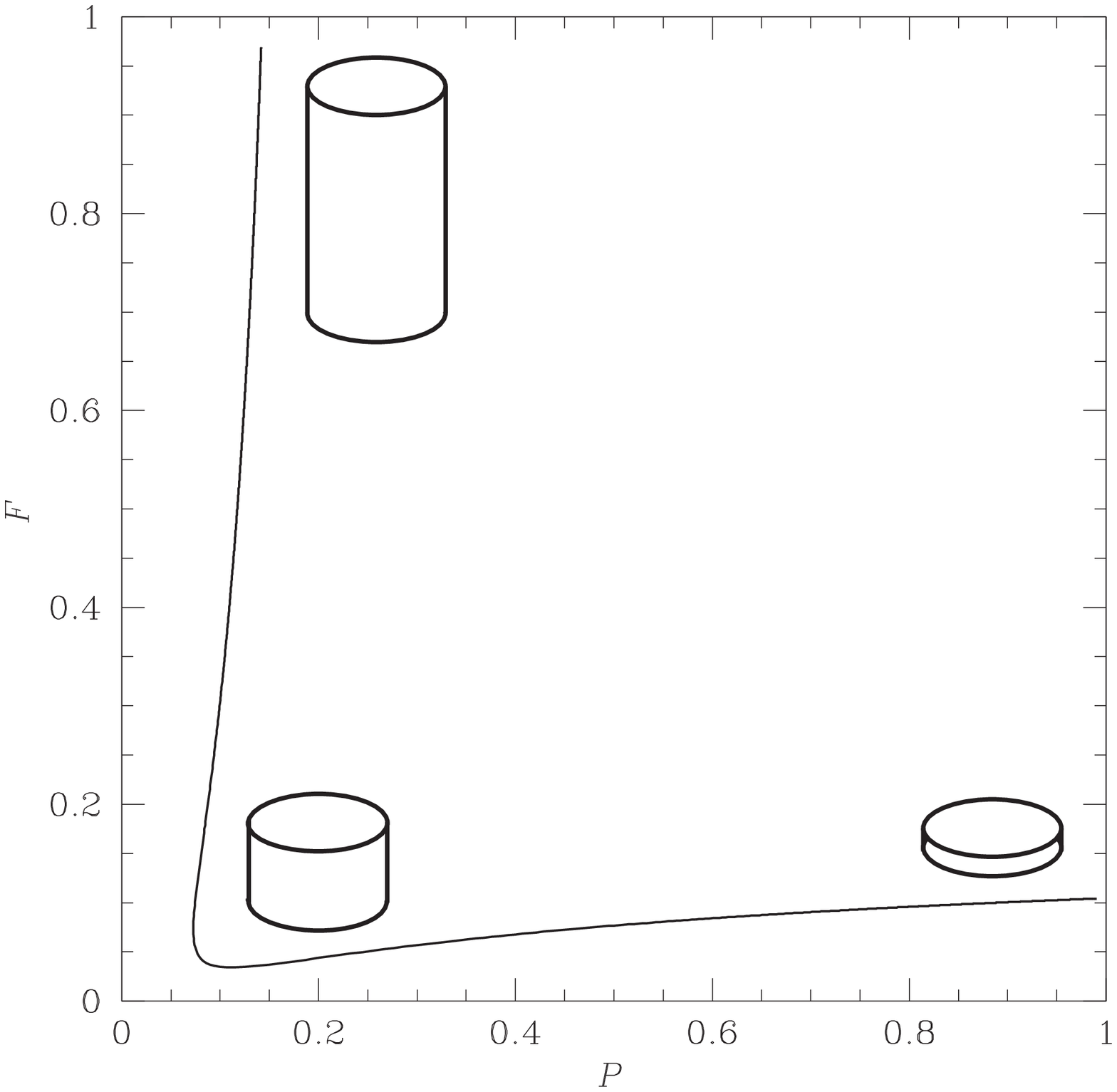,width=\myhalfpage}\end{minipage}
\hfill
\begin{minipage}{\myhalfpage}
\epsfig{figure=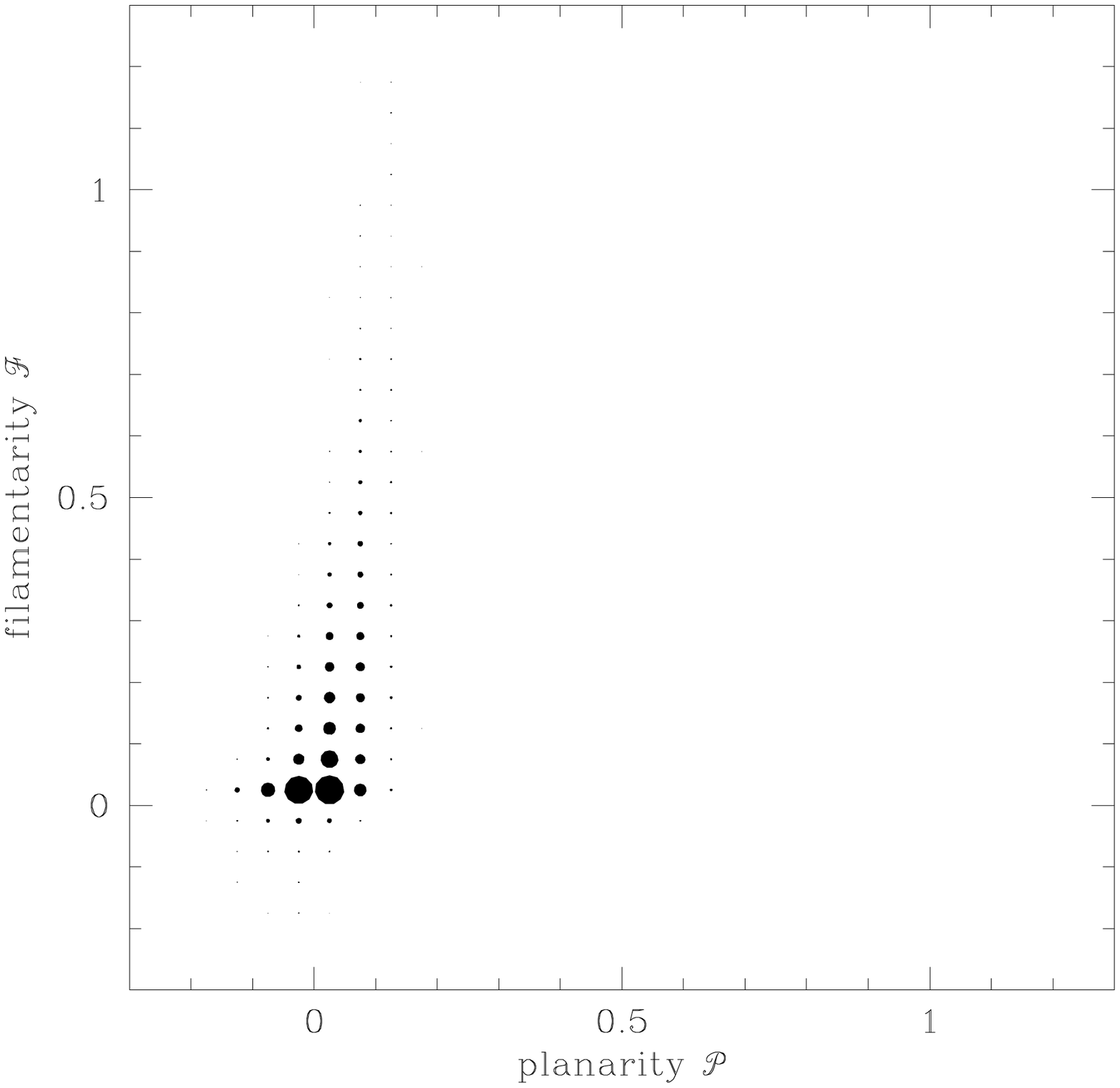,width=\myhalfpage}\end{minipage}
\end{center}
\caption{  On the  left  side a  plot  of the  shape--finders for  the
cylinder with  varying $\lambda$  is shown, illustrating  the turnover
from  $\lambda\approx0$,  a  plane  geometry  ($P\approx1$,  $F\ll1$),
through  a roughly  spherical ($P\approx0$,  $F\approx0$) to  a mainly
line like geometry ($P\ll1$, $F\approx1$) for $\lambda\gg1$.
On  the  right  side  a  frequency  histogram  of  the  shape--finders
determined from the excursion sets  of an N--body simulation is shown.
Larger circles correspond to more objects within the shape--finder bin
(from {}\protect\citealt{schmalzing:disentanglingI}).}
\label{fig:blaschke}
\end{figure}

Looking at high thresholds $\nu$, the excursion set is mainly composed
out   of  separated   regions  (see   Fig.~\ref{fig:excursion}).   The
morphology  of  these regions  may  be  characterized using  Minkowski
functionals   and   the  derived   shape--finders\index{shape--finder}
{}\citep{sahni:shapefinders}.
Employing     the     following     ratios    of     the     Minkowski
functionals\index{Minkowski        functional}       $H_1=V_0/(2V_1)$,
$H_2=2V_1/(\pi  V_2)$  and  $H_3=3V_2/(4V_3)$  one may  construct  the
dimensionless   shape--finders   {\em    planarity}   $P$   and   {\em
  filamentarity} $F$
\begin{equation}
P = \frac{H_2-H_1}{H_2+H_1} \quad \text{ and } F = \frac{H_3-H_2}{H_3+H_2} .
\end{equation}
A simple  example {}\citep{schmalzing:disentanglingI} is  provided by a
cylinder  of radius  $r$ and  height  $\lambda r$  with the  Minkowski
functionals
\begin{equation}
V_0=\pi r^3\lambda, \quad V_1=\tfrac{\pi}{3}r^2(1+\lambda), \quad
V_2=\tfrac{1}{3}r(\pi+\lambda), \quad V_3=1.
\end{equation}
The  shape--finders  planarity  $P$  and filamentarity  $F$  for  this
specific    example    are    plotted    against   each    other    in
Fig.~\ref{fig:blaschke}.  Indeed this is  nothing else but an inverted
Blaschke   diagram\index{Blaschke  diagram}   for  the   form  factors
({}\citealt{Hadwiger:altes}, {}\citealt{schneider:brunn}).
Following  {}\citet{schmalzing:disentanglingI} the  shape--finders may
be written in terms of the form factors. With
\begin{equation}
x=\frac{\pi V_0 V_2}{4 V_1^2}, \quad y=\frac{8V_1 V_3}{3\pi V_2^2}
\end{equation}
one obtains
\begin{equation}
P = \frac{1-x}{1+x} \quad \text{ and } F = \frac{1-y}{1+y} .
\end{equation}
The isoperimetric inequalities {}\citep{schneider:brunn} assure that 
$0\le P,F\le1$ for convex bodies.  For a sphere one gets $P=0=F$.

One of the results obtained  with the shape--finders applied to single
objects    in   the    excursion   sets    of    N--body   simulations
{}\citep{schmalzing:disentanglingI}         is         given         in
Fig.~\ref{fig:blaschke}.   This histogram shows  that the  majority of
the regions inside  the excursion set has $P\approx0\approx  F$, and a
smaller fraction  has $P\approx0$,  $F>0$, whereas only  a few  of the
regions have $F\approx0$ $P>0$.
Interpreting regions  with e.g.\ $P\approx0$, $F>0$  as filamentary or
line--like   structures  is   tempting  but   dangerous,   since  also
non--convex   regions  are  considered.    Also,  the   histogram  was
constructed  from   the  excursion   sets  of  all   thresholds  under
consideration.

It does not  seem to be possible to  construct shape--finders based on
the  global   scalar  Minkowski  functionals   facilitating  a  unique
interpretation  for non--convex  sets.  Abandoning  the  density field
approach, and going  back to the germ--grain model,  and the Minkowski
functional of  a union set of  balls $A_r=\bigcup_{i=1}^N B_r(\bx_i)$,
one may assign  a partial Minkowski functional\index{partial Minkowski
  functional} to each ball.  These {\em partial} Minkowski functionals
may be used  to extract information on the  spatial structure elements
-- whether the ball  around $\bx_i$ is inside a cluster,  a sheet or a
filament   (see  {}\citealt{mecke:diss},  {}\citealt{platzoeder:ringberg},
{}\citealt{schmalzing:cfa2}).
Another  promising  {\em  global}  method  for  extracting  shape  and
symmetry information  from {\em non-convex} bodies is  provided by the
global       Querma\ss{}       vectors\index{Querma\ss{}       vector}
{}\citep{beisbart:morphological}.

%%%
\subsubsection{Other applications of Minkowski functionals}

In the preceding applications we analyzed the union set of balls $A_r$
or  the excursion  set  $Q_\nu$ with  Minkowski functionals.   Another
possibility is to consider  Minkowski functionals of the Delauney-- or
Voronoi--cells,     as    determined     from     the    corresponding
tesselation\index{tesselation} defined by the given point distribution
({}\citealt{muche:distributional},           {}\citealt{muche:fragmenting},
{}\citealt{kerscher:diss}).

Going beyond motion invariance, instead demanding motion equivariance,
one   can  construct  vector--valued   extensions  of   the  Minkowski
functionals,  the {\em Querma\ss{}  vectors}\index{Querma\ss{} vector}
({}\citealt{hadwiger:vektorielle},     {}\citealt{beisbart:quermass}).     
{}\citet{beisbart:morphological}  investigate the  dynamical evolution
of the substructure in  galaxy clusters using Querma\ss{} vectors (see
also {}\citealt{beisbart:characterizing}).

%%%
\subsection{The $J$ function}
\label{sect:J}

Other methods to characterize the spatial distribution of points, well
known   in    spatial   statistics,   are    the   spherical   contact
distribution\index{spherical   contact   distribution}  $F(r)$   (also
denoted by $H_s(r)$), i.e.\ the distribution function of the distances
$r$ between an arbitrary point and the nearest object in the point set
$X$,  and  the  nearest neighbor  distance  distribution\index{nearest
  neighbor  distance  distribution} $G(r)$,  that  is  defined as  the
distribution  function of  distance $r$  of an  object in  $X$  to the
nearest other object in $X$.
$F(r)$   is   related   to   the  void   probability   $P_0(B_r)$   by
$F(r)=1-P_0(B_r)$.
For a   Poisson distribution\index{Poisson process} it is simply
\begin{equation} \label{eq:FG_poi}
G(r) = F(r) = 1 - \exp\left(- \rho \frac{4 \pi}{3} r^3\right) .
\end{equation}
Recently, {}\citet{vanlieshout:j} suggested to use the ratio
\begin{equation}
\label{eq:def-J}
J(r) = \frac{1-G(r)}{1-F(r)}
\end{equation}
as a further distributional characteristic\index{$J$--function}.  
For  a Poisson  distribution\index{Poisson  process} $J(r)=1$  follows
directly      from     {}\eqref{eq:FG_poi}.       As      shown     by
{}\citet{vanlieshout:j},   a  clustered  point   distribution  implies
$J(r)\le1$, whereas  regular structures  are indicated by  $J(r)\ge1$. 
However, {}\citet{bedford:remark}  showed that $J=1$ does  not imply a
Poisson process.  For several point process models $J(r)$, or at least
limiting values for $J(r)$, are known {}\citep{vanlieshout:j}.  The $J$
function  was considered by  {}\citet{white:hierarchy} as  the ``first
conditional correlation  function'' and used  by {}\citet{sharp:holes}
to  test hierarchical  models.  The  relation between  $J(r)$  and the
cumulants  $\xi_n(r)$  was  used  by  {}\citet{kerscher:regular}.   An
empirical study  of the performance  of the $J$--function  for several
point  process models is  given by  {}\citet{thoennes:comparative}.  A
refined definition of the $J$-function ``without edge correction'' may
be   especially   useful   for    a   test   on   spatial   randomness
{}\citep{baddeley:estimating}.

\subsubsection{Clustering of galaxies}
\label{sect:J-galaxies}

\begin{figure}
\begin{center}
\begin{minipage}{\myhalfpage}
\epsfig{figure=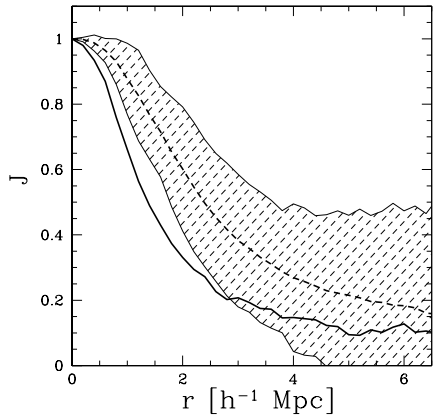,width=\myhalfpage}\end{minipage}
\end{center}
\caption{
\label{fig:redshift-data} 
$J(r)$  for the  volume limited  sample from  Perseus--Pisces redshift
survey  (solid line) and  the 1$\sigma$  range determined  from galaxy
samples generated by a mixed dark matter simulation.}
\end{figure}

The $J$--function\index{$J$--function} may be used to characterize the
distribution of galaxies\index{galaxy} or  galaxy clusters and for the
comparison  with   the  results  from  simulations,   similar  to  the
application       of      the      Minkowski       functionals      in
sect.~\ref{sect:cluster-minkowski}.   This  approach  was  pursued  by
{}\citet{kerscher:global}.    The   Perseus--Pisces  redshift   survey
({}\citealt{wegner:survey} and refs.   therein) was compared with galaxy
samples constructed from a mixed dark matter simulation.
The  observed $J(r)$  determined  from a  volume  limited sample  with
79\hMpc\  depth   differs  significantly  from  the   results  of  the
simulations (Fig.~\ref{fig:redshift-data}). Especially on small scales
the  galaxy  distribution shows  a  stronger  clustering,  as seen  by
steeper decreasing $J(r)$.
We also could show that modeling the galaxy distribution with a simple 
Poisson cluster process is not appropriate.

\subsubsection{Regularity in the distribution of super--clusters?}

{}\citet{einasto:120mpc} report a peak  in the 3D--power spectrum (the
Fourier transform of $\xi_2$) of a catalogue of clusters on a scale of
120\hMpc.   {}\citet{broadhurst:large-scale}  observed periodicity  on
approximately  the  same scale  in  an  analysis  of 1D--data  from  a
pencil--beam redshift  survey.  As  is well known  from the  theory of
fluids, the  regular distribution (e.g.\ of molecules  in a hard--core
fluid)  reveals  itself   in  an  oscillating  two--point  correlation
function  and  a peak  in  the  structure  function respectively  (see
e.g.~\citealt{hansen:theory},  and the  contribution of  H.~L{\"o}wen in
this  volume).   In accordance  with  this  an oscillating  two--point
correlation function $\xi_2(r)$ or at  least a first peak was reported
on approximately the same scale (e.g.\ {}\citealt{mo:typical_scales} and
{}\citealt{einasto:supercluster_II}).
The existence  of regularity\index{regular} on large  scales implies a
preferred scale  in the  initial conditions, which  would be  of major
physical interest.

\begin{figure}
\begin{center}
\begin{minipage}{\myhalfpage}
\epsfig{figure=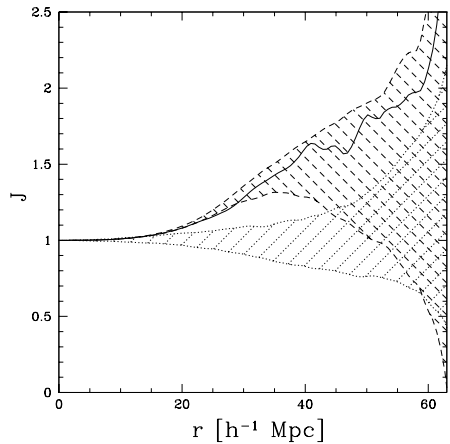,width=\myhalfpage}\end{minipage}
\end{center}
\caption{\label{fig:J-sc}    $J(r)$    determined   from    the
  super--cluster  sample  (solid  line)  is shown  together  with  the
  1--$\sigma$  range determined  from a  pure Poisson  process (dotted
  area)   and    a   Poisson    process   followed   by    a   similar
  friend--of--friends procedure (dashed area) as used to construct the
  super--cluster catalogue.}
\end{figure}

Using             the            $J(r)$--function\index{$J$--function}
{}\citet{kerscher:regular}               investigates              the
super--cluster\index{super--cluster}                          catalogue
{}\citep{einasto:supercluster_data} constructed from an earlier version
of  the  cluster  catalogue  by  {}\citet{andernach:current}  using  a
friend--of--friends  procedure. (The friend--of--friends  procedure is
called  single linkage  clustering  in the  mathematical literature).  
Comparing with Poisson distributed  points one clearly recognizes that
the   super--cluster  catalogue  is   a  regular   point  distribution
(Fig.~\ref{fig:J-sc}).
However, a similar signal for  $J(r)$ may be obtained by starting with
a Poisson process followed by a friend--of--friends procedure with the
same linking length as used  in the construction of the super--cluster
catalogue.  Only  some indication for a regular  distribution on large
scales  remains,   showing  that  this   super--cluster  catalogue  is
seriously affected by the construction method.

\subsubsection{$G_n$ and $F_n$}
\label{sect:GnFn}

\begin{figure}
\begin{center}
\begin{minipage}{\myhalfpage}
\epsfig{figure=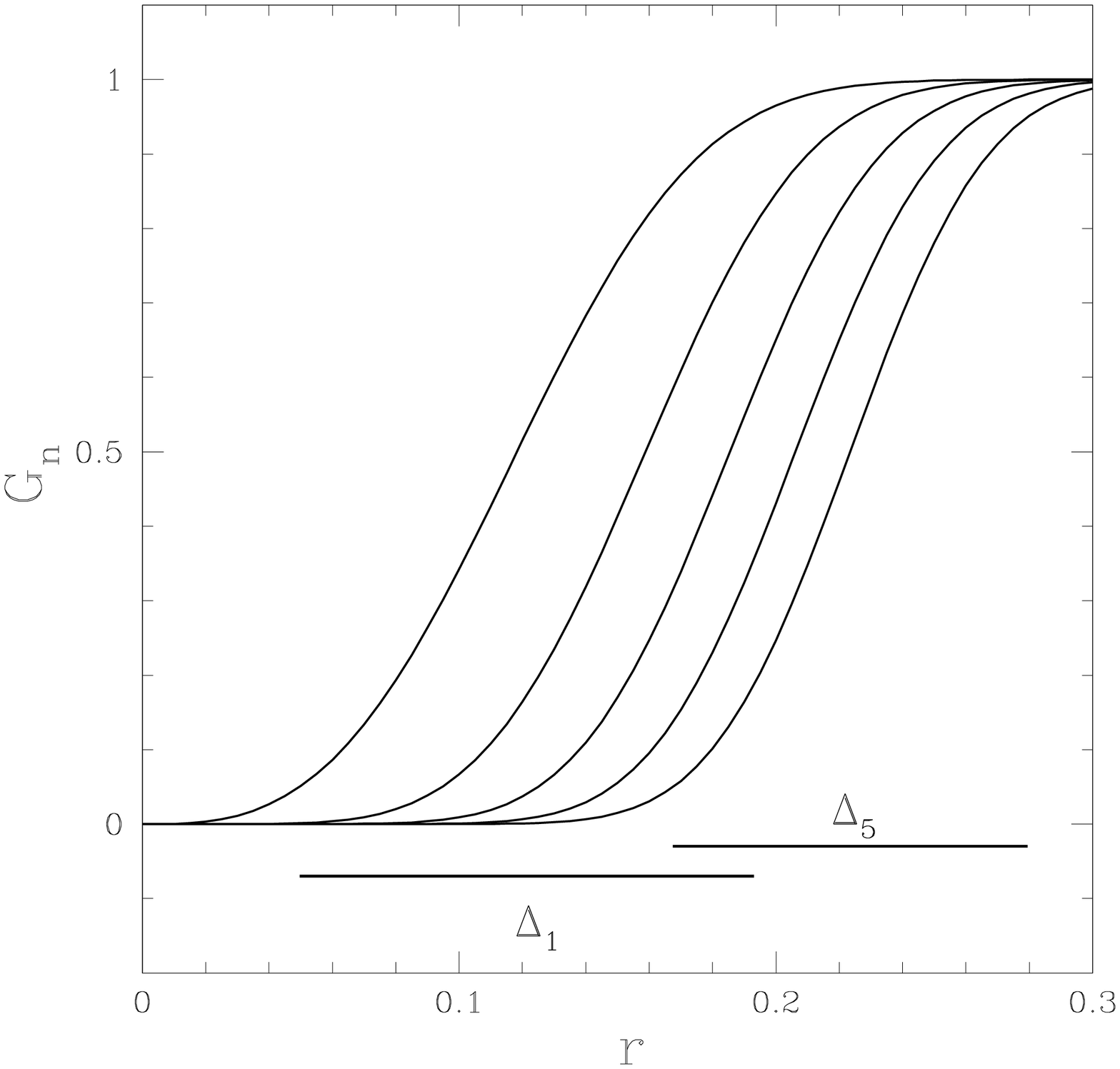,width=\myhalfpage}\end{minipage}
\hfill
\begin{minipage}{\myhalfpage}
\epsfig{figure=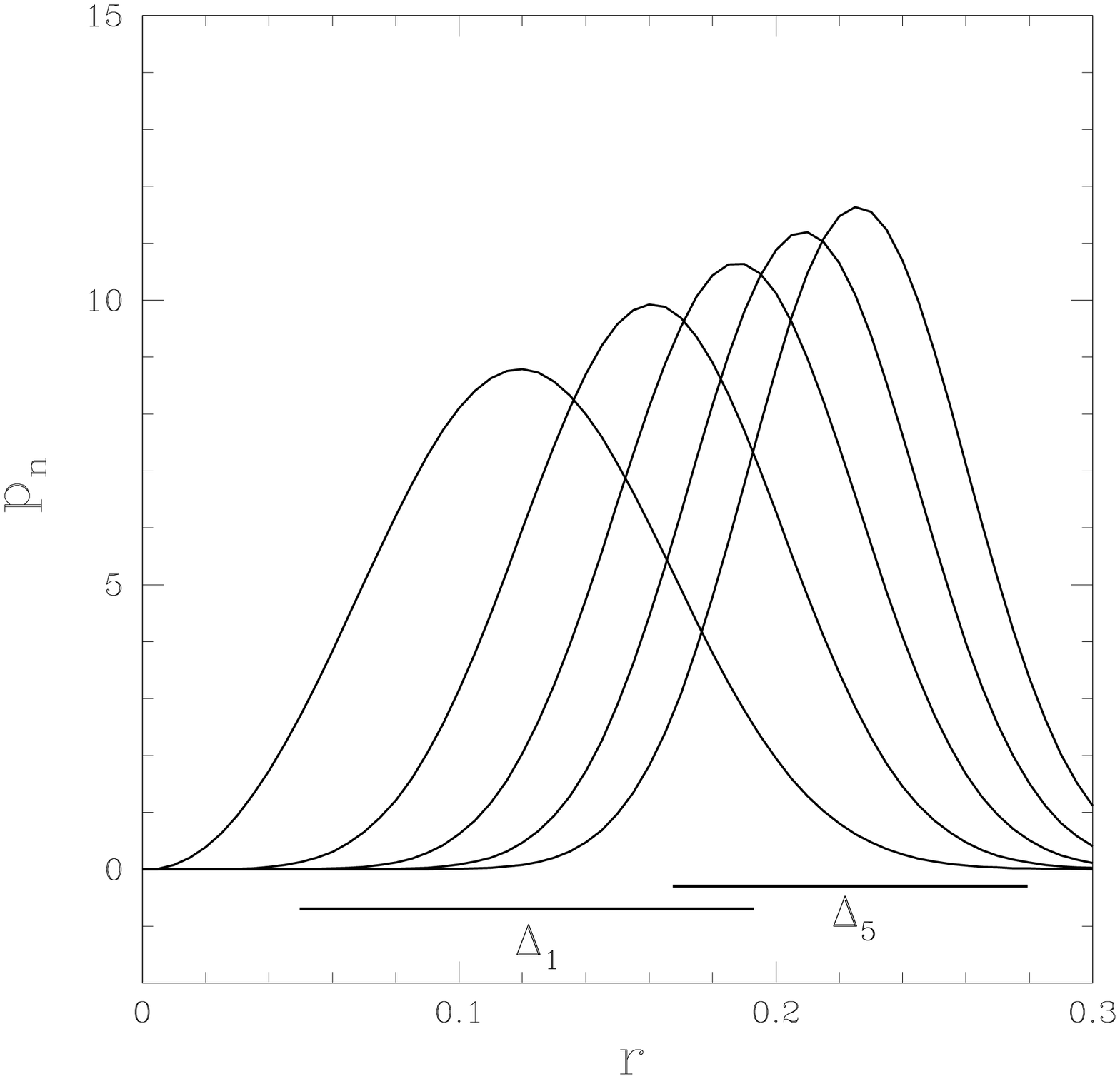,width=\myhalfpage}\end{minipage}
\end{center}
\caption{\label{fig:Gn-poisson} In the left  plot you see the $G_n(r)$
with $n=1,\dots,5$ for a  Poisson process with $\rho=100$.
In the right plot the corresponding densities $p_{n}(r)$ are shown.}
\end{figure}

As  a   direct  generalization   of  the  nearest   neighbor  distance
distribution   one  may   consider  the   $n$--th   neighbor  distance
distributions\index{$n$--th     neighbor    distance    distributions}
$G_{n}(r)$,  the  distribution of  the  distance  $r$  to the  $n$--th
nearest  point  (e.g.\   {}\citealt{stoyan:fractals}).   For  a  Poisson
process\index{Poisson     process}    in    three     dimensions    we
have
\begin{equation} 
\label{eq:Gn-poisson}
G_n(r) = 
1 - \frac{\Gamma\left(n,\rho\frac{4\pi}{3} r^3\right)}
{\Gamma(n)},
\end{equation}
shown  in Fig.~\ref{fig:Gn-poisson}.  $\Gamma(n,x)=\int_x^\infty\d{s}\ 
s^{n-1}e^{-s}$      is       the      incomplete      Gamma--function,
$\Gamma(n)=\Gamma(n,0)$  the   complete.  Clearly  $G_1(r)=G(r)$.   In
Fig.~\ref{fig:Gn-poisson} the curves for the first five $G_n(r)$ for a
Poisson  process are  shown,  together with  their densities  $p_n(r)$
defined by
\begin{equation}
G_n(r) = \int_{0}^r\d s\ p_{n}(s) .
\end{equation}
The sum of these densities is directly related to the two--point 
correlation function\index{two--point correlation function} 
{}\citep{mazur:neighborship}
\begin{equation}
\label{eq:g-pn}
g(r)\ \rho\ 4\pi r^2 = \sum_{n=1}^\infty\ p_n(r) .
\end{equation}

The  $n$--th  spherical  contact distribution\index{$n$--th  spherical
  contact distribution}  $F_n(r)$ is the distribution  function of the
distances  $r$ between  an  arbitrary point  and  the $n$--th  closest
object in the point set $X$  (we assume that the $n$--th closest point
is unique).  Clearly $F_1(r)=F(r)$.
For  stationary   and  isotropic  point  processes   $F_n(r)$  is  the
probability to find  at least $n$ points inside  a sphere $B_{r}$ with
radius $r$, and therefore
\begin{equation}
\label{eq:fgj:FFn:Fn-counts}
F_n(r) = \sum_{i=n}^\infty P_i(B_r)
= 1-\sum_{i=0}^{n-1} P_i(B_r) ,
\end{equation}
where   $P_i(B_r)$   are  the   counts--in--cells   as  discussed   in
Sect.~\ref{sect:higher-moments}.

For a Poisson process\index{Poisson process} with number density 
$\rho$ we obtain directly from {}\eqref{eq:PN-poisson}
\begin{equation}
F_n(r) = 1- \exp(-\rho|B_r|) 
\sum_{i=0}^{n-1} \frac{(\rho|B_r|)^i}{i!} ,
\end{equation}
which  is essentially  the series  expansion of  the  incomplete gamma
function (see e.g.\ {}\citealt{abramowitz:pocketbook}). Therefore,
\begin{equation} 
F_n(r) = 
1 - \frac{\Gamma\left(n,\rho\frac{4\pi}{3} r^3\right)}{\Gamma(n)},
\end{equation}
and we explicitely see that for a Poisson process
\begin{equation}
\label{eq:FnGn}
F_n(r) = G_n(r) 
\end{equation}
This is a special case of the ``Slivnyak's theorem'' 
{}\citep{stoyan:stochgeom}.

A  very interesting  feature of  the  $G_n(r)$ and  $F_n(r)$ is  their
sensitivity to structures on large  scales increasing with $n$.  As an
illustration consider the interval $\Delta_n\subset\BR^+$ specified by
$\int_{\Delta_n}\d s\ p_{n}(s)=0.9$.   Then $\Delta_n$ is the interval
in  which 90\%  of the  distances to  the $n$--th  neighbor  lie. (The
choice of 0.9 is arbitrary and may certainly be adopted to the problem
considered. Also  the interval $\Delta_n$ is ``centered''  as shown in
Fig.~\ref{fig:Gn-poisson}.)  The  empirical $G_n(r)$  may  be used  to
probe structures  within this specific radial range  as illustrated in
Fig.~\ref{fig:Gn-poisson}.  Going to larger $n$ one considers distance
intervals for larger radii.

\subsubsection{The $J_n$ function}

A drawback of the $J(r)$--function in empirical investigations is that
it becomes  ill defined  for large radii,  since the  empirical $F(r)$
reaches unity and the quotient in {}\eqref{eq:def-J} diverges.
In the following we will discuss the straightforward generalization of
the    $J$--function~\eqref{eq:def-J},   introducing    the   $J_n(r)$
functions\index{$J_n$--function}:
\begin{equation}
J_n(r) = \frac{1-G_n(r)}{1-F_n(r)}.
\end{equation}
From {}\eqref{eq:FnGn} we obtain directly for a Poisson 
process\index{Poisson process}
\begin{equation}
J_n(r) = 1 \quad \text{ for all $n$}.
\end{equation}
Qualitatively we  expect the same behavior  of the $J_n(r)$--functions
as for the $J(r)$--function, but now  for a radius $r$ in the interval
$\Delta_n$ (defined at the end of Sect.~\ref{sect:GnFn}).
\begin{itemize}
\item
If a point distribution shows  clustering on scales $r$ in $\Delta_n$,
the $G_n(r)$  increases faster  than for a  Poisson process  since the
$n$--th nearest neighbor is typically closer.  $F_n(r)$ increases more
slowly  than for  a  random  distribution. Both  effects  result in  a
$J_n(r)\le1$.
\item    On   the    other    hand,   for    a   point    distribution
  regular\index{regular}  on  the scale  $r$  in $\Delta_n$,  $G_n(r)$
  increases more slowly than for  a Poisson process, since the $n$--th
  neighbor  is found  at a  finite characteristic  distance.  $F_n(r)$
  increases stronger  since the  distance from a  random point  to the
  $n$--th closest point on the regular structure is typically smaller.
  This results in $J_n(r)\ge1$.
\item
$J_n(r)  =  1$ indicates  the  transition  from  regular to  clustered
structures on scales $r$ in $\Delta_n$.
\end{itemize}

\begin{figure}
\begin{center}
\begin{minipage}{\myhalfpage}
\epsfig{figure=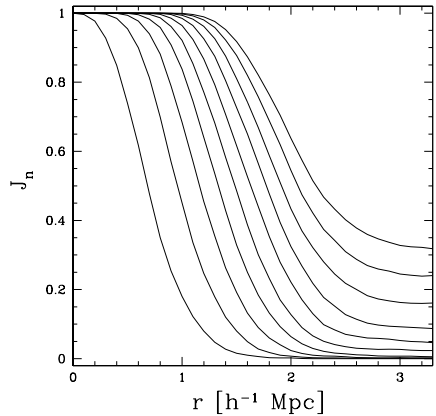,width=\myhalfpage}\end{minipage}
\end{center}
\caption{\label{fig:Jn-matern}   The   $J_n(r)$  with   $n=1,\ldots10$
(bending up successively) for a Mat\'ern cluster process with $\mu=10$
and $R=1.5\hMpc$ calculated using the reduced sample estimators.}
\end{figure}

With a simple point process  model we illustrate these properties.  In
a  Mat\'ern cluster process\index{Mat\'ern  cluster process}  a single
cluster consists out of $\mu$ points in the mean, randomly distributed
inside a  sphere of radius $R$,  where the number of  points follows a
Poisson  distribution.  The  clusters  centers (not  belonging to  the
point process)  form a  Poisson process with  a density  of $\rho/\mu$
{}\citep{stoyan:stochgeom}.
In  Fig.~\ref{fig:Jn-matern}  the strong  clustering  in the  Mat\'ern
cluster  process is  visible  from  a decline  of  the $J_n(r)$.  This
decline becomes  weaker with increasing  $n$. For large radii  $r$ the
$J_n$ acquire  a constant  value.  Investigating larger  scales, i.e.\ 
for  large $n$,  the constant  value of  $J_n$ shows  a  trend towards
unity,  i.e.\ we  start to  ``see''  the Poisson  distribution of  the
clusters centers.

\subsubsection{On our way to large scales}

\begin{figure}
\begin{center}
\begin{minipage}{\myhalfpage}
\epsfig{figure=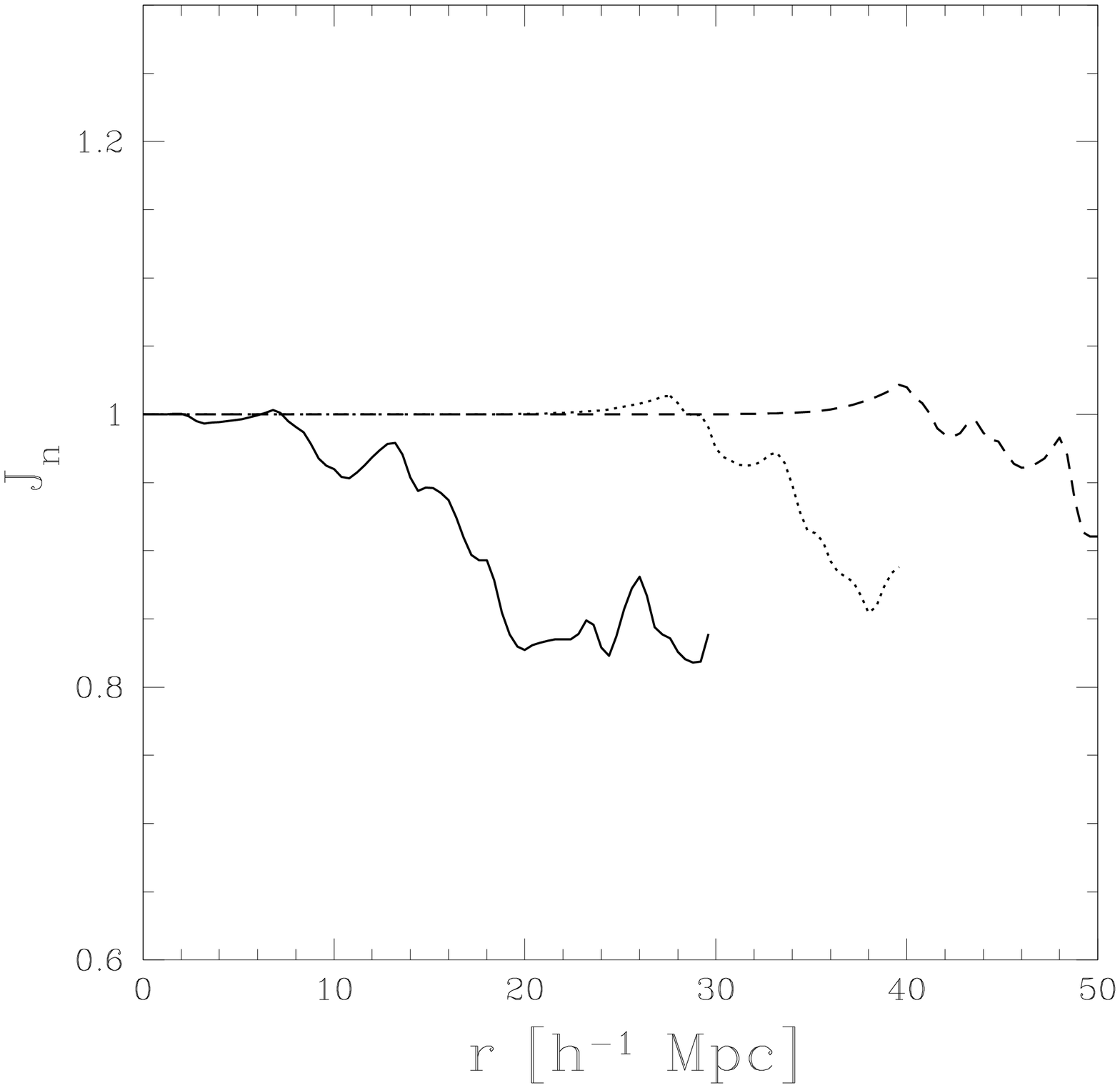,width=\myhalfpage}\end{minipage}
\hfill
\begin{minipage}{\myhalfpage}
\epsfig{figure=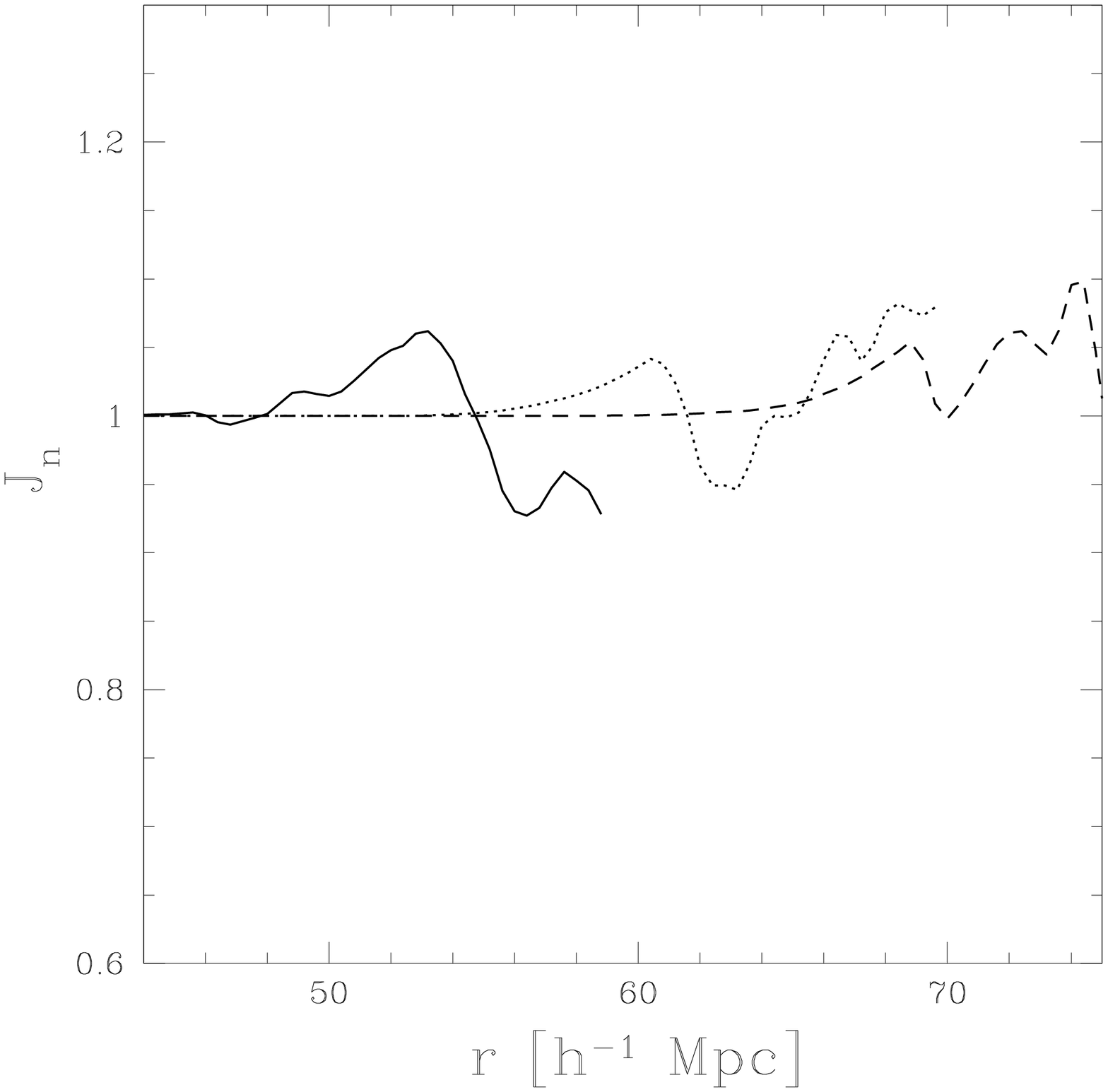,width=\myhalfpage}\end{minipage}
\end{center}
\caption{\label{fig:Jn-jy12} In  the left plot  the $J_n$ of  the IRAS
galaxies  are shown  with $n=1,4,7$  (solid, dotted,  dashed),  in the
right plot the $J_n$ with $n=10,15,20$ (solid, dotted, dashed).}
\end{figure}

A similar behavior  may be identified in the  galaxy distribution.  We
calculate  the  $J_n$--functions\index{$J_n$--function}  for a  volume
limited  sample  of  galaxies\index{galaxy}  extracted from  the  IRAS
1.2~Jy  catalogue  with  200\hMpc\  depth  using  the  reduced  sample
estimator  for both  $F_n$  and  $G_n$.  For  small  $n$, i.e.\  small
scales, the $J_n(r)$ are all smaller than unity, indicating clustering
out to scales of  40\hMpc (see Fig.~\ref{fig:Jn-jy12}).  For large $n$
the $J_n$ are consistent with no clustering, i.e.\ $J_n=1$.  However a
trend    towards    a   $J_n$    larger    than   unity,    indicating
regularity\index{regular} is observed.   Clearly, the results obtained
from this sparse sample with 280  galaxies only may serve mainly as an
illustration of the method --  to obtain decisive results we will have
to wait for deeper surveys.

%%%%%%%
\section{Summary and Outlook}
\label{sect:summary}

In  Sections~\ref{sect:cluster-minkowski}  and {}\ref{sect:J-galaxies}
we  discussed that  advanced  geometrical methods  like the  Minkowski
functionals and the $J$--function  are able to constrain parameters of
cosmological models.   However, these  geometric methods are  not only
limited to the parameter  estimation in cosmological simulations, they
are also  valuable tools as  point process statistics in  general. The
direct probe  of galaxy surveys  with geometrical methods  showed that
the large--scale structure  exhibits strong morphological fluctuations
(Sect.~\ref{sect:fluctuations}).    Such    fluctuations   are   often
attributed to  ``cosmic variance'' in an Universe  homogeneous on very
large scales.   However the fluctuations are  astonishingly large even
on  scales  of 200\hMpc.   A  preferred scale,  may  be  viewed as  an
indication  for a  homogeneous galaxy  distribution on  large  scales. 
Especially geometric  methods like the $J$--  and $J_n$--functions may
be helpful to identify a preferred scale in the galaxy distribution.

Perspectives for future research might be as follows:\\
Starting  with  the  Minkowski   functionals  or  other  well  founded
geometrical  tools, more  specialized  methods may  be constructed  to
understand certain  features in the galaxy distribution  in detail. An
example are the vector  valued extension of the Minkowski functionals,
the Querma{\ss}--vector, used in the investigation of the substructure
in galaxy clusters.\\
In  empirical work, one  has to  determine these  geometrical measures
from  a given  point set.   The construction  of estimators  with well
understood  distributional properties is  crucial to  be able  to draw
decisive conclusions from the data.\\
Using  these  geometrical  methods   as  tools  for  constraining  the
cosmological  parameters will  be one  way to  go.  Currently  this is
mainly performed  by comparisons with N--body  simulations.  Clearly a
more direct  link between the geometry  and the dynamics  of matter in
the  Universe  promoting  our  understanding how  structures  form  is
desirable.  Carefully constructed approximations may be the key
ingredient.\\
Another way in trying to understand structure formation is to directly
investigate   the  appearance  of   geometric  features   like  walls,
filaments, and clusters -- or to identify a preferred scale showing up
in a regular distribution  on large scales.  Such  findings will guide
us in the construction of  approximations, which are able to reproduce
such geometric features.

%%%
\section*{Acknowledgements}

I would like to thank Claus Beisbart, J\"org Retzlaff, Dietrich Stoyan
for comments on the manuscript and Jens Schmalzing who kindly provided
the           Figures~\ref{fig:cobe},           {}\ref{fig:excursion},
{}\ref{fig:blaschke}.  
%Especially  I would  like to thank  Klaus Mecke
%and  Dietrich  Stoyan  for   organizing  this  interesting  an  highly
%motivating  conference.  
%
Special thanks  to Thomas Buchert and Herbert  Wagner.  Their constant
support,  the   inspiring  discussions,  and   the  helpful  criticism
significantly influenced  my understanding of  physics, cosmology, and
statistics; the emphasis of  morphological measures was always a major
concern.

%%%%%%%%%%%%%%%%%%%%%%%%%%%%
%\bibliographystyle{wuppertal}
%\bibliography{my}

\begin{thebibliography}{130}
\expandafter\ifx\csname natexlab\endcsname\relax\def\natexlab#1{#1}\fi

\bibitem[{Abramowitz and Stegun(1984)}]{abramowitz:pocketbook}
Abramowitz, M.~X. and Stegun, I.~A. (1984): {\em Pocketbook of Mathematical
  Functions\/}.
\newblock Harri Deutsch, Thun, Frankfurt/Main.

\bibitem[{Adler(1981)}]{adler:randomfields}
Adler, R.~J. (1981): {\em The geometry of random fields\/}.
\newblock John Wiley \& Sons, Chichester.

\bibitem[{Andernach and Tago(1998)}]{andernach:current}
Andernach, H. and Tago, E. (1998): Current status of the {ACO} cluster redshift
  compilation.
\newblock In V.~M\"uller, S.~Gottl\"ober, J.~P. M\"ucket, and J.~Wambsgans,
  eds., {\em Proc.\ of the 12th Potsdam Cosmology Workshop 1997, Large Scale
  Structure: Tracks and Traces\/}. World Scientific, Singapore.
\newblock Astro-ph/9710265.

\bibitem[{Baddeley et~al.(1999)Baddeley, Kerscher, Schladitz, and
  Scott}]{baddeley:estimating}
Baddeley, A.~J., Kerscher, M., Schladitz, K., and Scott, B. (1999): Estimating
  the {$J$} function without edge correction.
\newblock { }Statist.\ Neerlandica, in press{ }.

\bibitem[{Baddeley and Silverman(1984)}]{baddeley:cautionary}
Baddeley, A.~J. and Silverman, B.~W. (1984): A cautionary example on the use of
  second--order methods for analyzing point patterns.
\newblock {\em Biometrics\/}, {\bf 40}, 1089--1093.

\bibitem[{Balian and Schaeffer(1989)}]{balian:I}
Balian, R. and Schaeffer, R. (1989): Scale--invariant matter distribution in
  the {U}niverse {I}.~counts in cells.
\newblock {\em \providecommand{\aanda}{Astron.\ Astrophys.}{\aanda}\/}, {\bf
  220}, 1--29.

\bibitem[{Bardeen et~al.(1986)Bardeen, Bond, Kaiser, and
  Szalay}]{bardeen:gauss}
Bardeen, J.~M., Bond, J.~R., Kaiser, N., and Szalay, A.~S. (1986): The
  statistics of peaks of {G}aussian random fields.
\newblock {\em \providecommand{\apj}{Ap.\ J.}{\apj}\/}, {\bf 304}, 15--61.

\bibitem[{Barrow et~al.(1985)Barrow, Sonoda, and
  Bhavsar}]{barrow:minimalspanning}
Barrow, J.~D., Sonoda, D.~H., and Bhavsar, S.~P. (1985): Minimal spanning
  trees, filaments and galaxy clustering.
\newblock {\em \providecommand{\mnras}{Mon.\ Not.\ Roy.\ Astron.\
  Soc.}{\mnras}\/}, {\bf 216}, 17--35.

\bibitem[{Bedford and {van den Berg}(1997)}]{bedford:remark}
Bedford, T. and {van den Berg}, J. (1997): A remark on the van {L}ieshout and
  {B}addeley {J}--function for point processes.
\newblock {\em Adv.\ Appl.\ Prob.\/}, {\bf 29}, 19--25.

\bibitem[{Beisbart and Buchert(1998)}]{beisbart:characterizing}
Beisbart, C. and Buchert, T. (1998): Characterizing cluster morphology using
  vector--valued minkowski functionals.
\newblock In V.~M\"uller, S.~Gottl\"ober, J.~P. M\"ucket, and J.~Wambsgans,
  eds., {\em Proc.\ of the 12th Potsdam Cosmology Workshop 1997, Large Scale
  Structure: Tracks and Traces\/}. World Scientific, Singapore.
\newblock { }astro-ph/9711034{ }.

\bibitem[{Beisbart et~al.(2000)Beisbart, Buchert, Bartelmann, Colberg, and
  Wagner}]{beisbart:morphological}
Beisbart, C., Buchert, T., Bartelmann, M., Colberg, J.~G., and Wagner, H.
  (2000): Morphological evolution of galaxy clusters: First application of the
  querma\ss{} vectors.
\newblock { }in preparation{ }.

\bibitem[{Beisbart et~al.(1999)Beisbart, Buchert, and
  Wagner}]{beisbart:quermass}
Beisbart, C., Buchert, T., and Wagner, H. (1999): Morphometry of galaxy
  clusters.
\newblock { }submitted{ }.

\bibitem[{Bernardeau and Kofman(1995)}]{bernardeau:properties}
Bernardeau, F. and Kofman, L. (1995): Properties of the cosmological density
  distribution function.
\newblock {\em \providecommand{\apj}{Ap.\ J.}{\apj}\/}, {\bf 443}, 479--498.

\bibitem[{Bertschinger(1992)}]{bertschinger:largescale}
Bertschinger, E. (1992): Large-scale structures and motions: Linear theory and
  statistics.
\newblock In V.~Martinez, M.~Portilla, and D.~Saez, eds., {\em New Insights
  into the {U}niverse\/}, number 408 in { }Lecture Notes in Physics{ }, pages
  65--126. Springer Verlag, Berlin.

\bibitem[{Borgani(1995)}]{borgani:scaling}
Borgani, S. (1995): Scaling in the universe.
\newblock {\em Physics Rep.\/}, {\bf 251}, 1--152.

\bibitem[{B\"orner(1993)}]{boerner:early}
B\"orner, G. (1993): {\em The Early {U}niverse, Facts and Fiction\/}.
\newblock Springer Verlag, Berlin, 3rd edition.

\bibitem[{Bouchet et~al.(1992)Bouchet, Juszkiewicz, Colombi, and
  Pellat}]{bouchet:weakly}
Bouchet, F.~R., Juszkiewicz, R., Colombi, S., and Pellat, R. (1992): Weakly
  nonlinear gravitational instability for arbitrary {O}mega.
\newblock {\em \providecommand{\apjl}{Ap.\ J.\ Lett.}{\apjl}\/}, {\bf 394},
  L5--L8.

\bibitem[{Broadhurst et~al.(1990)Broadhurst, Ellis, Koo, and
  Szalay}]{broadhurst:large-scale}
Broadhurst, T.~J., Ellis, R.~S., Koo, D.~C., and Szalay, A.~S. (1990):
  Large--scale distribution of galaxies at the galactic poles.
\newblock {\em \providecommand{\nat}{Nature}{\nat}\/}, {\bf 343}, 726.

\bibitem[{Buchert(1996)}]{buchert:lagrangian}
Buchert, T. (1996): {L}agrangian perturbation approach to the formation of
  large--scale structure.
\newblock In S.~Bonometto, J.~Primack, and A.~Provenzale, eds., {\em
  Proceedings of the international school of physics Enrico Fermi. Course
  CXXXII: Dark matter in the {U}niverse\/}. Societ{\`a} Italiana di Fisica,
  Varenna sul Lago di Como.

\bibitem[{Buchert(1999)}]{buchert:onaverage}
Buchert, T. (1999): On average properties of inhomogeneous fluids in general
  relativity: I.\ dust cosmologies.
\newblock G.R.G. in press, gr-qc/9906015.

\bibitem[{Buchert and Ehlers(1997)}]{buchert:averaging}
Buchert, T. and Ehlers, J. (1997): Averaging inhomogeneous {N}ewtonian
  cosmologies.
\newblock {\em \providecommand{\aanda}{Astron.\ Astrophys.}{\aanda}\/}, {\bf
  320}, 1--7.

\bibitem[{Cappi et~al.(1998)Cappi, Benoist, {Da Costa}, and
  Maurogordato}]{cappi:fractal}
Cappi, A., Benoist, C., {Da Costa}, L., and Maurogordato, S. (1998): Is the
  universe a fractal? results from the {SSRS2}.
\newblock {\em \providecommand{\aanda}{Astron.\ Astrophys.}{\aanda}\/}, {\bf
  335}, 779--788.

\bibitem[{Coles et~al.(1996)Coles, Davies, and Pearson}]{coles:quantifying}
Coles, P., Davies, A., and Pearson, R.~C. (1996): Quantifying the topology of
  large--scale structure.
\newblock {\em \providecommand{\mnras}{Mon.\ Not.\ Roy.\ Astron.\
  Soc.}{\mnras}\/}, {\bf 281}, 1375--1384.

\bibitem[{Coles and Lucchin(1994)}]{coles:cosmology}
Coles, P. and Lucchin, F. (1994): {\em Cosmology: The origin and evolution of
  cosmic structure\/}.
\newblock John Wiley \& Sons, Chichester.

\bibitem[{Colombi et~al.(1998)Colombi, Szapudi, and Szalay}]{colombi:effects}
Colombi, S., Szapudi, I., and Szalay, A.~S. (1998): Effects of sampling on
  statistics of large scale structure.
\newblock {\em \providecommand{\mnras}{Mon.\ Not.\ Roy.\ Astron.\
  Soc.}{\mnras}\/}, {\bf 296}, 253.

\bibitem[{{da Costa} et~al.(1998){da Costa}, Willmer, Pellegrini, Chaves, Rite,
  Maia, Geller, Latham, Kurtz, Huchra, Ramella, Fairall, Smith, and
  Lipari}]{dacosta:southern}
{da Costa}, L.~N., Willmer, C. N.~A., Pellegrini, P., Chaves, O.~L., Rite, C.,
  Maia, M. A.~G., Geller, M.~J., Latham, D.~W., Kurtz, M.~J., Huchra, J.~P.,
  Ramella, M., Fairall, A.~P., Smith, C., and Lipari, S. (1998): The {S}outhern
  {S}ky {R}edshift {S}urvey.
\newblock {\em \providecommand{\aj}{A.\ J.}{\aj}\/}, {\bf 116}, 1--7.

\bibitem[{Daley and Vere-Jones(1988)}]{daley:introduction}
Daley, D.~J. and Vere-Jones, D. (1988): {\em An Introduction to the Theory of
  Point Processes\/}.
\newblock Springer Verlag, Berlin.

\bibitem[{Dolgov et~al.(1999)Dolgov, Doroshkevich, Novikov, and
  Novikov}]{dolgov:geometry}
Dolgov, A., Doroshkevich, A., Novikov, D., and Novikov, I. (1999): Geometry and
  statistics of cosmic microwave polarization.
\newblock Astro-ph/9901399.

\bibitem[{Doroshkevich et~al.(1999)Doroshkevich, , M{\"u}ller, Retzlaff, and
  Turchaninov}]{doroshkevich:superlargenbody}
Doroshkevich, A.~G., , M{\"u}ller, V., Retzlaff, J., and Turchaninov, V.
  (1999): Superlarge--scale structure in n--body simulations.
\newblock {\em \providecommand{\mnras}{Mon.\ Not.\ Roy.\ Astron.\
  Soc.}{\mnras}\/}, {\bf 306}, 575--591.

\bibitem[{Efstathiou(1996)}]{efstathiou:observations}
Efstathiou, G. (1996): Observations of large--scale structure in the
  {U}niverse.
\newblock In R.~Schaeffer, J.~Silk, M.~Spiro, and J.~Zinn-Justin, eds., {\em
  LesHouches Session {LX}: cosmology and large scale structure, august 1993\/},
  pages 133--252. Elsevier, Amsterdam.

\bibitem[{Einasto et~al.(1997{\natexlab{a}})Einasto, Einasto, Frisch,
  Gottl\"ober, M\"uller, Saar, Starobinsky, Tago, and
  Andernach}]{einasto:supercluster_II}
Einasto, J., Einasto, M., Frisch, P., Gottl\"ober, S., M\"uller, V., Saar, V.,
  Starobinsky, A.~A., Tago, E., and Andernach, D. T.~H. (1997{\natexlab{a}}):
  The supercluster--void network. {II}. an oscillating cluster correlation
  function.
\newblock {\em \providecommand{\mnras}{Mon.\ Not.\ Roy.\ Astron.\
  Soc.}{\mnras}\/}, {\bf 289}, 801--812.

\bibitem[{Einasto et~al.(1997{\natexlab{b}})Einasto, Einasto, Gottl{\"o}ber,
  M{\"u}ller, Saar, Starobinsky, Tago, Tucker, Andernach, and
  Frisch}]{einasto:120mpc}
Einasto, J., Einasto, M., Gottl{\"o}ber, S., M{\"u}ller, V., Saar, V.,
  Starobinsky, A.~A., Tago, E., Tucker, D., Andernach, H., and Frisch, P.
  (1997{\natexlab{b}}): A 120-{Mpc} periodicity in the three--dimensional
  distribution of galaxy superclusters.
\newblock {\em \providecommand{\nat}{Nature}{\nat}\/}, {\bf 385}, 139--141.

\bibitem[{Einasto et~al.(1997{\natexlab{c}})Einasto, Tago, Jaaniste, Einasto,
  and Andernach}]{einasto:supercluster_data}
Einasto, M., Tago, E., Jaaniste, J., Einasto, J., and Andernach, H.
  (1997{\natexlab{c}}): The supercluster--void network. {I}. the supercluster
  catalogue and large--scale distribution.
\newblock {\em \providecommand{\aas}{Astron.\ Astrophys.\ Suppl.}{\aas}\/},
  {\bf 123}, 119.

\bibitem[{Fiksel(1988)}]{fiksel:edge}
Fiksel, T. (1988): Edge--corrected density estimators for point processes.
\newblock {\em Statistics\/}, {\bf 19}, 67--75.

\bibitem[{Fisher et~al.(1995)Fisher, Huchra, Strauss, Davis, Yahil, and
  Schlegel}]{fisher:irasdata}
Fisher, K.~B., Huchra, J.~P., Strauss, M.~A., Davis, M., Yahil, A., and
  Schlegel, D. (1995): The {IRAS} 1.2 {J}y survey: Redshift data.
\newblock {\em \providecommand{\apjs}{Ap.\ J.\ Suppl.}{\apjs}\/}, {\bf 100},
  69.

\bibitem[{Gaite et~al.(1999)Gaite, Dominguez, and
  Perez-Mercader}]{gaite:fractal}
Gaite, J., Dominguez, A., and Perez-Mercader, J. (1999): The fractal
  distribution of galaxies and the transition to homogeneity.
\newblock {\em \providecommand{\apj}{Ap.\ J.}{\apj}\/}, {\bf 522}, L5--L8.

\bibitem[{Goldenfeld(1992)}]{goldenfeld:lectures}
Goldenfeld, N. (1992): {\em Lectures on Phase Transitions and the
  Renormalization Group\/}.
\newblock Addison--Wesley, Reading, MA.

\bibitem[{Grassberger and Procaccia(1984)}]{grassberger:dimensions}
Grassberger, P. and Procaccia, I. (1984): Dimensions and entropies of strange
  attractors from fluctuating dynamics approach.
\newblock {\em Physica D\/}, {\bf 13}, 34--54.

\bibitem[{Gunn(1995)}]{gunn:sdss}
Gunn, J.~E. (1995): The {S}loan {D}igital {S}ky {S}urvey.
\newblock {\em \providecommand{\baas}{Bull.\ American Astron.\ Soc.}{\baas}\/},
  {\bf 186}, 875.

\bibitem[{Hadwiger(1955)}]{Hadwiger:altes}
Hadwiger, H. (1955): {\em Altes und {N}eues \"uber konvexe {K}\"orper\/}.
\newblock Birkh{\"a}user, Basel.

\bibitem[{Hadwiger(1957)}]{hadwiger:vorlesung}
Hadwiger, H. (1957): {\em Vorlesungen {\"u}ber {I}nhalt, {O}berfl{\"a}che und
  {I}soperimetrie\/}.
\newblock Springer Verlag, Berlin.

\bibitem[{Hadwiger and Schneider(1971)}]{hadwiger:vektorielle}
Hadwiger, H. and Schneider, R. (1971): Vektorielle {I}ntegralgeometrie.
\newblock {\em Elemente der Mathematik\/}, {\bf 26}, 49--72.

\bibitem[{Hamilton(1993)}]{hamilton:towards}
Hamilton, A. J.~S. (1993): Toward better ways to measure the galaxy correlation
  function.
\newblock {\em \providecommand{\apj}{Ap.\ J.}{\apj}\/}, {\bf 417}, 19--35.

\bibitem[{Hansen and McDonnald(1986)}]{hansen:theory}
Hansen, J.~P. and McDonnald, I.~R. (1986): {\em Theory of Simple Liquids\/}.
\newblock Academic Press, New York and London.

\bibitem[{Heavens(1999)}]{heavens:estimating}
Heavens, A.~F. (1999): Estimating non--{G}aussianity in the microwave
  background.
\newblock {\em \providecommand{\mnras}{Mon.\ Not.\ Roy.\ Astron.\
  Soc.}{\mnras}\/}, {\bf 299}, 805--808.

\bibitem[{Heavens and Sheth(1999)}]{heavens:correlation}
Heavens, A.~F. and Sheth, R.~K. (1999): The correlation of peaks in the
  microwave background.
\newblock { }submitted, astro-ph/9906301{ }.

\bibitem[{Hobson et~al.(1999)Hobson, Jones, and Lasenby}]{hobson:wavelet}
Hobson, M.~P., Jones, A.~W., and Lasenby, A.~N. (1999): Wavelet analysis and
  the detection of non--{G}aussianity in the {CMB}.
\newblock {\em \providecommand{\mnras}{Mon.\ Not.\ Roy.\ Astron.\
  Soc.}{\mnras}\/}, {\bf 309}, 125--140.

\bibitem[{Huchra et~al.(1995)Huchra, Geller, and {Corwin~Jr.}}]{huchra:cfa2s2}
Huchra, J.~P., Geller, M.~J., and {Corwin~Jr.}, H.~G. (1995): The {CfA}
  redshift survey: Data for the {NGP} + 36 zone.
\newblock {\em \providecommand{\apjs}{Ap.\ J.\ Suppl.}{\apjs}\/}, {\bf 99},
  391.

\bibitem[{Huchra et~al.(1990)Huchra, Geller, {De~Lapparent}, and
  {Corwin~Jr.}}]{huchra:cfa2s1}
Huchra, J.~P., Geller, M.~J., {De~Lapparent}, V., and {Corwin~Jr.}, H.~G.
  (1990): The {CfA} redshift survey -- data for the {NGP} + 30 zone.
\newblock {\em \providecommand{\apjs}{Ap.\ J.\ Suppl.}{\apjs}\/}, {\bf 72},
  433--470.

\bibitem[{Jing and B{\"o}rner(1998)}]{jing:threepointlcrs}
Jing, Y.~P. and B{\"o}rner, G. (1998): The three--point correlation function of
  galaxies.
\newblock {\em \providecommand{\apj}{Ap.\ J.}{\apj}\/}, {\bf 503}, 37.

\bibitem[{Juszkiewicz et~al.(1995)Juszkiewicz, Weinberg, Amsterfamski,
  Chodorowski, and Bouchet}]{juszkiewicz:weakly}
Juszkiewicz, R., Weinberg, D.~H., Amsterfamski, P., Chodorowski, M., and
  Bouchet, F.~R. (1995): Weakly nonlinear {G}aussian fluctuations and the
  {E}dgeworth expansion.
\newblock {\em \providecommand{\apj}{Ap.\ J.}{\apj}\/}, {\bf 442}, 39--56.

\bibitem[{Kates et~al.(1991)Kates, Kotok, and Klypin}]{kates:highres}
Kates, R., Kotok, E., and Klypin, A. (1991): High--resolution simulations of
  galaxy formation in a cold dark matter scenario.
\newblock {\em \providecommand{\aanda}{Astron.\ Astrophys.}{\aanda}\/}, {\bf
  243}, 295--308.

\bibitem[{Kauffmann et~al.(1997)Kauffmann, Nusser, and
  Steinmetz}]{kauffmann:galaxy}
Kauffmann, G., Nusser, A., and Steinmetz, M. (1997): Galaxy formation and
  large-scale bias.
\newblock {\em \providecommand{\mnras}{Mon.\ Not.\ Roy.\ Astron.\
  Soc.}{\mnras}\/}, {\bf 286}, 795--811.

\bibitem[{Kerscher(1998{\natexlab{a}})}]{kerscher:diss}
Kerscher, M. (1998{\natexlab{a}}): {\em Morphologie gro\ss r\"aumiger
  {S}trukturen im {U}niversum\/}.
\newblock GCA--Verlag, Herdecke.

\bibitem[{Kerscher(1998{\natexlab{b}})}]{kerscher:regular}
Kerscher, M. (1998{\natexlab{b}}): Regularity in the distribution of
  superclusters?
\newblock {\em \providecommand{\aanda}{Astron.\ Astrophys.}{\aanda}\/}, {\bf
  336}, 29--34.

\bibitem[{Kerscher(1999)}]{kerscher:twopoint}
Kerscher, M. (1999): The geometry of second--order statistics -- biases in
  common estimators.
\newblock {\em \providecommand{\aanda}{Astron.\ Astrophys.}{\aanda}\/}, {\bf
  343}, 333--347.

\bibitem[{Kerscher et~al.(1999{\natexlab{a}})Kerscher, Pons-Border\'{\i}a,
  Schmalzing, Trasarti-Battistoni, Mart\'{\i}nez, Buchert, and
  Valdarnini}]{kerscher:global}
Kerscher, M., Pons-Border\'{\i}a, M.~J., Schmalzing, J., Trasarti-Battistoni,
  R., Mart\'{\i}nez, V.~J., Buchert, T., and Valdarnini, R.
  (1999{\natexlab{a}}): A global descriptor of spatial pattern interaction in
  the galaxy distribution.
\newblock {\em \providecommand{\apj}{Ap.\ J.}{\apj}\/}, {\bf 513}, 543--548.

\bibitem[{Kerscher et~al.(1996{\natexlab{a}})Kerscher, Schmalzing, and
  Buchert}]{kerscher:minkowski}
Kerscher, M., Schmalzing, J., and Buchert, T. (1996{\natexlab{a}}): Analyzing
  galaxy catalogues with {M}inkowski functionals.
\newblock In P.~Coles, V.~{Mart{\'\i}nez}, and M.~J. {Pons Border{\'\i}a},
  eds., {\em Mapping, measuring and modelling the {U}niverse\/}, pages
  247--252. Astronomical Society of the Pacific, Valencia.

\bibitem[{Kerscher et~al.(1996{\natexlab{b}})Kerscher, Schmalzing, Buchert, and
  Wagner}]{kerscher:significance}
Kerscher, M., Schmalzing, J., Buchert, T., and Wagner, H. (1996{\natexlab{b}}):
  The significance of the fluctuations in the {IRAS} 1.2 {J}y galaxy catalogue.
\newblock In R.~Bender, T.~Buchert, and P.~Schneider, eds., {\em Proc.\ $2^{\rm
  nd}$ SFB workshop on {\em Astro--particle physics} Ringberg 1996, Report
  SFB375/P002\/}, pages 83--98. Ringberg, Tegernsee.

\bibitem[{Kerscher et~al.(1998)Kerscher, Schmalzing, Buchert, and
  Wagner}]{kerscher:fluctuations}
Kerscher, M., Schmalzing, J., Buchert, T., and Wagner, H. (1998): Fluctuations
  in the 1.2 {J}y galaxy catalogue.
\newblock {\em \providecommand{\aanda}{Astron.\ Astrophys.}{\aanda}\/}, {\bf
  333}, 1--12.

\bibitem[{Kerscher et~al.(1997)Kerscher, Schmalzing, Retzlaff, Borgani,
  Buchert, Gottl{\"o}ber, M{\"u}ller, Plionis, and Wagner}]{kerscher:abell}
Kerscher, M., Schmalzing, J., Retzlaff, J., Borgani, S., Buchert, T.,
  Gottl{\"o}ber, S., M{\"u}ller, V., Plionis, M., and Wagner, H. (1997):
  {M}inkowski functionals of {A}bell/{ACO} clusters.
\newblock {\em \providecommand{\mnras}{Mon.\ Not.\ Roy.\ Astron.\
  Soc.}{\mnras}\/}, {\bf 284}, 73--84.

\bibitem[{Kerscher et~al.(1999{\natexlab{b}})Kerscher, Szapudi, and
  Szalay}]{kerscher:comparison}
Kerscher, M., Szapudi, I., and Szalay, A. (1999{\natexlab{b}}): A comparison of
  estimators for the two-point correlation function: dispelling the myths.
\newblock { }submitted{ }.

\bibitem[{Klain and Rota(1997)}]{klain:introduction}
Klain, D.~A. and Rota, C.-C. (1997): {\em Introduction to Geometric
  Probability\/}.
\newblock Cambridge University Press, Cambridge.

\bibitem[{Kolatt et~al.(1996)Kolatt, Dekel, Ganon, and
  Willick}]{kolatt:simulating}
Kolatt, T., Dekel, A., Ganon, G., and Willick, J.~A. (1996): Simulating our
  cosmological neighborhood: Mock catalogs for velocity analysis.
\newblock {\em \providecommand{\apj}{Ap.\ J.}{\apj}\/}, {\bf 458}, 419--434.

\bibitem[{Landy and Szalay(1993)}]{landy:bias}
Landy, S.~D. and Szalay, A.~S. (1993): Bias and variance of angular correlation
  functions.
\newblock {\em \providecommand{\apj}{Ap.\ J.}{\apj}\/}, {\bf 412}, 64--71.

\bibitem[{Maddox(1998)}]{maddox:2df}
Maddox, S. (1998): The 2df galaxy redshift survey: Preliminary results.
\newblock In V.~M\"uller, S.~Gottl\"ober, J.~P. M\"ucket, and J.~Wambsgans,
  eds., {\em Proc.\ of the 12th Potsdam Cosmology Workshop 1997, Large Scale
  Structure: Tracks and Traces\/}. World Scientific, Singapore.
\newblock Astro-ph/9711015.

\bibitem[{Mandelbrot(1982)}]{mandelbrot:fractal}
Mandelbrot, B. (1982): {\em The Fractal Geometry of Nature\/}.
\newblock Freeman, San Francisco.

\bibitem[{{Mart{\'\i}nez}(1996)}]{martinez:measures}
{Mart{\'\i}nez}, V.~J. (1996): Measures of galaxy clustering.
\newblock In S.~Bonometto, J.~Primack, and A.~Provenzale, eds., {\em
  Proceedings of the international school of physics Enrico Fermi. Course
  CXXXII: Dark matter in the {U}niverse\/}. Societ{\`a} Italiana di Fisica,
  Varenna sul Lago di Como.

\bibitem[{Matheron(1989)}]{matheron:estimating}
Matheron, G. (1989): {\em {E}stimating and {C}hoosing: {A}n {E}ssay on
  {P}robability in {P}ractice\/}.
\newblock Springer Verlag, Berlin.

\bibitem[{Mazur(1992)}]{mazur:neighborship}
Mazur, S. (1992): Neighborship partition of the radial distribution function
  for simple liquids.
\newblock {\em J.\ Chem.\ Phys.\/}, {\bf 97}, 9276--9282.

\bibitem[{McCauley(1997)}]{mccauley:galaxy}
McCauley, J.~L. (1997): Are galaxy distributions scale invariant? a perspective
  from dynamical systems theory.
\newblock { }submitted, astro-ph/9703046{ }.

\bibitem[{McCauley(1998)}]{mccauley:thegalaxy}
McCauley, J.~L. (1998): The galaxy distributions: Homogeneous, fractal, or
  neither?
\newblock {\em Fractals\/}, {\bf 6}, 109--119.

\bibitem[{Mecke(1994)}]{mecke:diss}
Mecke, K.~R. (1994): {\em {I}ntegralgeometrie in der {S}tatistischen {P}hysik:
  {P}erkolation, komplexe {F}l{\"u}ssigkeiten und die {S}truktur des
  {U}niversums\/}.
\newblock Harri Deutsch, Thun, Frankfurt/Main.

\bibitem[{Mecke et~al.(1994)Mecke, Buchert, and Wagner}]{mecke:robust}
Mecke, K.~R., Buchert, T., and Wagner, H. (1994): Robust morphological measures
  for large--scale structure in the {U}niverse.
\newblock {\em \providecommand{\aanda}{Astron.\ Astrophys.}{\aanda}\/}, {\bf
  288}, 697--704.

\bibitem[{Mecke and Wagner(1991)}]{mecke:euler}
Mecke, K.~R. and Wagner, H. (1991): {E}uler characteristic and related measures
  for random geometric sets.
\newblock {\em J.\ Stat.\ Phys.\/}, {\bf 64}, 843.

\bibitem[{Melott(1990)}]{melott:review}
Melott, A.~L. (1990): The topology of large--scale structure in the {U}niverse.
\newblock {\em Physics Rep.\/}, {\bf 193}, 1--39.

\bibitem[{Milne and Westcott(1972)}]{milne:further}
Milne, R.~K. and Westcott, M. (1972): Further results for {G}auss--poisson
  processes.
\newblock {\em Adv.\ Appl.\ Prob.\/}, {\bf 4}, 151--176.

\bibitem[{Mo et~al.(1992)Mo, Deng, Xia, Schiller, and
  B{\"o}rner}]{mo:typical_scales}
Mo, H.~J., Deng, Z.~G., Xia, X.~Y., Schiller, P., and B{\"o}rner, G. (1992):
  Typical scales in the distribution of galaxies and clusters of galaxies from
  unnormalized pair counts.
\newblock {\em \providecommand{\aanda}{Astron.\ Astrophys.}{\aanda}\/}, {\bf
  257}, 1--10.

\bibitem[{Muche(1996)}]{muche:distributional}
Muche, L. (1996): Distributional properties of the three--dimensional {P}oisson
  {D}elauney cell.
\newblock {\em J.\ Stat.\ Phys.\/}, {\bf 84}, 147--167.

\bibitem[{Muche(1997)}]{muche:fragmenting}
Muche, L. (1997): Fragmenting the universe and the {V}oronoi tesselation.
\newblock { }preprint, Freiberg 1997{ }.

\bibitem[{Novikov et~al.(1999)Novikov, Feldman, and
  Shandarin}]{novikov:minkowski}
Novikov, D., Feldman, H.~A., and Shandarin, S.~F. (1999): {M}inkowski
  functionals and cluster analysis for {CMB} maps.
\newblock {\em Int. J. Mod. Phys.\/}, {\bf D8}, 291--306.

\bibitem[{Padmanabhan(1993)}]{padmanabhan:structure}
Padmanabhan, T. (1993): {\em Structure formation in the {U}niverse\/}.
\newblock Cambridge University Press, Cambridge.

\bibitem[{Padmanabhan and Subramanian(1993)}]{padmanabhan:zeldovich}
Padmanabhan, T. and Subramanian, K. (1993): {Z}el'dovich approximation and the
  probability distribution for the smoothed density field in the nonlinear
  regime.
\newblock {\em \providecommand{\apj}{Ap.\ J.}{\apj}\/}, {\bf 410}, 482--487.

\bibitem[{Peacock(1992)}]{peacock:statistics}
Peacock, J.~A. (1992): Satistics of cosmological density fields.
\newblock In V.~Martinez, M.~Portilla, and D.~Saez, eds., {\em New Insights
  into the {U}niverse\/}, number 408 in { }Lecture Notes in Physics{ }, pages
  65--126. Springer Verlag, Berlin.

\bibitem[{Peebles(1973)}]{peebles:statistical}
Peebles, P. (1973): Satistical analysis of catalogs of extragalactic objects.
  {I}. theory.
\newblock {\em \providecommand{\apj}{Ap.\ J.}{\apj}\/}, {\bf 185}, 413--440.

\bibitem[{Peebles(1980)}]{peebles:lss}
Peebles, P. J.~E. (1980): {\em The Large Scale Structure of the {U}niverse\/}.
\newblock Princeton University Press, Princeton, New Jersey.

\bibitem[{Platz{\"o}der and Buchert(1995)}]{platzoeder:ringberg}
Platz{\"o}der, M. and Buchert, T. (1995): Applications of {M}inkowski
  functionals for the statistical analysis of dark matter models.
\newblock In A.~Weiss, G.~Raffelt, W.~Hillebrandt, and F.~von Feilitzsch, eds.,
  {\em Proc. of ``1st SFB workshop on Astro-particle physics'', Ringberg,
  Tegernsee\/}, pages 251--263.
\newblock { }astro-ph/9509014{ }.

\bibitem[{Plionis and Valdarnini(1991)}]{plionis:evidence}
Plionis, M. and Valdarnini, R. (1991): Evidence for large--scale structure on
  scales about 300/h {M}pc.
\newblock {\em \providecommand{\mnras}{Mon.\ Not.\ Roy.\ Astron.\
  Soc.}{\mnras}\/}, {\bf 249}, 46--62.

\bibitem[{{Pons--Border{\'\i}a} et~al.(1999){Pons--Border{\'\i}a},
  {Mart{\'\i}nez}, Stoyan, Stoyan, and Saar}]{ponsborderia:comparing}
{Pons--Border{\'\i}a}, M.-J., {Mart{\'\i}nez}, V.~J., Stoyan, D., Stoyan, H.,
  and Saar, E. (1999): Comparing estimators of the galaxy correlation function.
\newblock {\em \providecommand{\apj}{Ap.\ J.}{\apj}\/}, {\bf 523}, 480.

\bibitem[{Retzlaff et~al.(1998)Retzlaff, Borgani, Gottl{\"o}ber, Klypin, and
  M{\"u}ller}]{retzlaff:constraining}
Retzlaff, J., Borgani, S., Gottl{\"o}ber, S., Klypin, A., and M{\"u}ller, V.
  (1998): Constraining cosmological models with cluster power spectra.
\newblock {\em New Astronomy\/}, {\bf 3}, 631--646.

\bibitem[{Sahni et~al.(1997)Sahni, Sathyaprakash, and
  Shandarin}]{sahni:probing}
Sahni, V., Sathyaprakash, B.~S., and Shandarin, S.~F. (1997): Probing
  large-scale structure using percolation and genus curves.
\newblock {\em \providecommand{\apj}{Ap.\ J.}{\apj}\/}, {\bf 476}, L1--L5.

\bibitem[{Sahni et~al.(1998)Sahni, Sathyaprakash, and
  Shandarin}]{sahni:shapefinders}
Sahni, V., Sathyaprakash, B.~S., and Shandarin, S.~F. (1998): Shapefinders: A
  new shape diagnostic for large--scale structure.
\newblock {\em \providecommand{\apj}{Ap.\ J.}{\apj}\/}, {\bf 495}, L5--L8.

\bibitem[{Sandage(1995)}]{sandage:practical}
Sandage, A. (1995): Practical cosmology: Inventing the past.
\newblock In A.~Sandage, R.~Kron, and M.~Longair, eds., {\em The Deep
  {U}niverse\/}, Saas--Fee Advanced Course Lecture Notes 1993, pages 1--232.
  Swiss society for Astrophysics and Astronomy, Springer Verlag, Berlin.

\bibitem[{Santal{\'o}(1976)}]{santalo:integralgeometry}
Santal{\'o}, L.~A. (1976): {\em Integral Geometry and Geometric Probability\/}.
\newblock Addison--Wesley, Reading, MA.

\bibitem[{Sathyaprakash et~al.(1998{\natexlab{a}})Sathyaprakash, Sahni, and
  Shandarin}]{sathyaprakash:morphology}
Sathyaprakash, B.~S., Sahni, V., and Shandarin, S.~F. (1998{\natexlab{a}}):
  Morphology of clusters and superclusters in {N}-body simulations of
  cosmological gravitational clustering.
\newblock {\em \providecommand{\apj}{Ap.\ J.}{\apj}\/}, {\bf 508}, 551--569.

\bibitem[{Sathyaprakash et~al.(1998{\natexlab{b}})Sathyaprakash, Sahni,
  Shandarin, and Fisher}]{sathyaprakash:filaments}
Sathyaprakash, B.~S., Sahni, V., Shandarin, S.~F., and Fisher, K.~B.
  (1998{\natexlab{b}}): Filaments and pancakes in the {IRAS} 1.2 {Jy} redshift
  survey.
\newblock {\em \providecommand{\apj}{Ap.\ J.}{\apj}\/}, {\bf 507}, L109--L112.

\bibitem[{Schmalzing and Buchert(1997)}]{schmalzing:beyond}
Schmalzing, J. and Buchert, T. (1997): Beyond genus statistics: A unifying
  approach to the morphology of cosmic structure.
\newblock {\em \providecommand{\apjl}{Ap.\ J.\ Lett.}{\apjl}\/}, {\bf 482},
  L1--L4.

\bibitem[{Schmalzing et~al.(1999{\natexlab{a}})Schmalzing, Buchert, Melott,
  Sahni, Sathyaprakash, and Shandarin}]{schmalzing:disentanglingI}
Schmalzing, J., Buchert, T., Melott, A., Sahni, V., Sathyaprakash, B., and
  Shandarin, S. (1999{\natexlab{a}}): Disentangling the cosmic web~{I}:
  Morphology of isodensity contours.
\newblock { }in press, astro-ph/9904384{ }.

\bibitem[{Schmalzing and Diaferio(1999)}]{schmalzing:cfa2}
Schmalzing, J. and Diaferio, A. (1999): Topology and geometry of the {CfA2}
  redshift survey.
\newblock { }in press, astro-ph/9910228{ }.

\bibitem[{Schmalzing and G{\'o}rski(1998)}]{schmalzing:minkowski_cmb}
Schmalzing, J. and G{\'o}rski, K.~M. (1998): Minkowski functionals used in the
  morphological analysis of cosmic microwave background anisotropy maps.
\newblock {\em \providecommand{\mnras}{Mon.\ Not.\ Roy.\ Astron.\
  Soc.}{\mnras}\/}, {\bf 297}, 355--365.

\bibitem[{Schmalzing et~al.(1999{\natexlab{b}})Schmalzing, Gottl\"ober,
  Kravtsov, and Klypin}]{schmalzing:quantifying}
Schmalzing, J., Gottl\"ober, S., Kravtsov, A., and Klypin, A.
  (1999{\natexlab{b}}): Quantifying the evolution of higher order clustering.
\newblock {\em \providecommand{\mnras}{Mon.\ Not.\ Roy.\ Astron.\
  Soc.}{\mnras}\/}, {\bf 309}, 1007--1016.

\bibitem[{Schmalzing et~al.(1996)Schmalzing, Kerscher, and
  Buchert}]{schmalzing:minkowski}
Schmalzing, J., Kerscher, M., and Buchert, T. (1996): {M}inkowski functionals
  in cosmology.
\newblock In S.~Bonometto, J.~Primack, and A.~Provenzale, eds., {\em
  Proceedings of the international school of physics Enrico Fermi. Course
  CXXXII: Dark matter in the {U}niverse\/}, pages 281--291. Societ{\`a}
  Italiana di Fisica, Varenna sul Lago di Como.

\bibitem[{Schneider(1993)}]{schneider:brunn}
Schneider, R. (1993): {\em Convex bodies: the {B}runn--{M}inkowski theory\/}.
\newblock Cambridge University Press, Cambridge.

\bibitem[{Schneider and Weil(1992)}]{schneider:integralgeometrie}
Schneider, R. and Weil, W. (1992): {\em Integralgeometrie\/}.
\newblock Bernd G.\ Teubner, Leipzig, Berlin.

\bibitem[{Shandarin(1983)}]{shandarin:percolation}
Shandarin, S.~F. (1983): Percolation theory and the cell--lattice structure of
  the {U}niverse.
\newblock {\em Sov.\ Astron.\ Lett.\/}, {\bf 9}, 104--106.

\bibitem[{Sharp(1981)}]{sharp:holes}
Sharp, N. (1981): Holes in the zwicky catalogue.
\newblock {\em \providecommand{\mnras}{Mon.\ Not.\ Roy.\ Astron.\
  Soc.}{\mnras}\/}, {\bf 195}, 857--867.

\bibitem[{Silverman(1986)}]{silverman:density}
Silverman, B.~W. (1986): {\em Density Estimation for Statistics and Data
  Analysis\/}.
\newblock Chapman and Hall, London.

\bibitem[{Stoyan(1998)}]{stoyan:caution}
Stoyan, D. (1998): Caution with ``fractal'' point--patterns.
\newblock {\em Statistics\/}, {\bf 25}, 267--270.

\bibitem[{Stoyan et~al.(1995)Stoyan, Kendall, and Mecke}]{stoyan:stochgeom}
Stoyan, D., Kendall, W.~S., and Mecke, J. (1995): {\em Stochastic Geometry and
  its Applications\/}.
\newblock John Wiley \& Sons, Chichester, 2nd edition.

\bibitem[{Stoyan and Stoyan(1994)}]{stoyan:fractals}
Stoyan, D. and Stoyan, H. (1994): {\em Fractals, Random Shapes and Point
  Fields\/}.
\newblock John Wiley \& Sons, Chichester.

\bibitem[{Stoyan and Stoyan(2000)}]{stoyan:improving}
Stoyan, D. and Stoyan, H. (2000): Improving ratio estimators of second order
  point process statistics.
\newblock {\em Scand.\ J.\ Satist.\/}.
\newblock { }in press{ }.

\bibitem[{Stratonovich(1963)}]{stratonovich:topicsI}
Stratonovich, R.~L. (1963): {\em Topics in the theory of random noise\/},
  volume~1.
\newblock Gordon and Breach, New York.

\bibitem[{{Sylos Labini} et~al.(1998){Sylos Labini}, Montuori, and
  Pietronero}]{labini:scale}
{Sylos Labini}, F., Montuori, M., and Pietronero, L. (1998): Scale invariance
  of galaxy clustering.
\newblock {\em Physics Rep.\/}, {\bf 293}, 61--226.

\bibitem[{Szalay(1997)}]{szalay:walls}
Szalay, A.~S. (1997): Walls and bumps in the {U}niverse.
\newblock In A.~Olinto, ed., {\em Proc. of the 18th Texas Symposium on
  Relativistic Astrophysics\/}. AIP, New York.

\bibitem[{Szapudi(1998)}]{szapudi:newmethod}
Szapudi, I. (1998): A new method for calculating counts in cells.
\newblock {\em \providecommand{\apj}{Ap.\ J.}{\apj}\/}, {\bf 497}, 16.

\bibitem[{Szapudi and Colombi(1996)}]{szapudi:cosmic}
Szapudi, I. and Colombi, S. (1996): Cosmic error and statistics of large scale
  structure.
\newblock {\em \providecommand{\apj}{Ap.\ J.}{\apj}\/}, {\bf 470}, 131.

\bibitem[{Szapudi and Gaztanaga(1998)}]{szapudi:comparison}
Szapudi, I. and Gaztanaga, E. (1998): Comparison of the large--scale clustering
  in the {APM} and the {EDSGC} galaxy surveys.
\newblock {\em \providecommand{\mnras}{Mon.\ Not.\ Roy.\ Astron.\
  Soc.}{\mnras}\/}, {\bf 300}, 493--496.

\bibitem[{Szapudi and Szalay(1998)}]{szapudi:new}
Szapudi, I. and Szalay, A.~S. (1998): A new class of estimators for the
  $n$--point correlations.
\newblock {\em \providecommand{\apj}{Ap.\ J.}{\apj}\/}, {\bf 494}, L41.

\bibitem[{Th{\"o}nnes and {van Lieshout}(1999)}]{thoennes:comparative}
Th{\"o}nnes, E. and {van Lieshout}, M.-C. (1999): A comparative study on the
  power of {van Lieshout} and {B}addeley's {J}--function.
\newblock {\em Biom.\ J.\/}, {\bf 41}, 721--734.

\bibitem[{Tomita(1986)}]{tomita:statistical}
Tomita, H. (1986): Statistical properties of random interface systems.
\newblock {\em Progr.\ Theor.\ Phys.\/}, {\bf 75}, 482--495.

\bibitem[{Totsuji and Kihara(1969)}]{totsuji:correlation}
Totsuji, H. and Kihara, T. (1969): The correlation function for the
  distribution of galaxies.
\newblock {\em \providecommand{\pasj}{Publications of the Astronomical Society
  of Japan}{\pasj}\/}, {\bf 21}, 221.

\bibitem[{{van Lieshout} and Baddeley(1996)}]{vanlieshout:j}
{van Lieshout}, M.~N.~M. and Baddeley, A.~J. (1996): A nonparametric measure of
  spatial interaction in point patterns.
\newblock {\em Statist.\ Neerlandica\/}, {\bf 50}, 344--361.

\bibitem[{Wegner et~al.(1993)Wegner, Haynes, and Giovanelli}]{wegner:survey}
Wegner, G., Haynes, M.~P., and Giovanelli, R. (1993): A survey of the
  {P}isces--{P}erseus supercluster. v -- the declination strip +33.5 deg to
  +39.5 deg and the main supercluster ridge.
\newblock {\em \providecommand{\aj}{A.\ J.}{\aj}\/}, {\bf 105}, 1251.

\bibitem[{Weil(1983)}]{weil:stereology}
Weil, W. (1983): Stereology: A survey for geometers.
\newblock In P.~M. Gruber and J.~M. Wills, eds., {\em Convexity and its
  applications\/}, pages 360--412. Birkh{\"a}user, Basel.

\bibitem[{Weinberg et~al.(1987)Weinberg, {Gott III}, and
  Melott}]{weinberg:topologyI}
Weinberg, D.~H., {Gott III}, J.~R., and Melott, A.~L. (1987): The topology of
  large--scale--structure. {I}. topology and the random phase hypothesis.
\newblock {\em \providecommand{\apj}{Ap.\ J.}{\apj}\/}, {\bf 321}, 2--27.

\bibitem[{Wei{\ss} and Buchert(1993)}]{weiss:highres}
Wei{\ss}, A.~G. and Buchert, T. (1993): High--resolution simulation of deep
  pencil beam surveys -- analysis of quasi--periodicty.
\newblock {\em \providecommand{\aanda}{Astron.\ Astrophys.}{\aanda}\/}, {\bf
  274}, 1--11.

\bibitem[{White(1979)}]{white:hierarchy}
White, S. D.~M. (1979): The hierarchy of correlation functions and its relation
  to other measures of galaxy clustering.
\newblock {\em \providecommand{\mnras}{Mon.\ Not.\ Roy.\ Astron.\
  Soc.}{\mnras}\/}, {\bf 186}, 145--154.

\bibitem[{Willmer et~al.(1998)Willmer, {da Costa}, and
  Pellegrini}]{willmer:southern}
Willmer, C., {da Costa}, L.~N., and Pellegrini, P. (1998): Southern sky
  redshift survey: Clustering of local galaxies.
\newblock {\em \providecommand{\aj}{A.\ J.}{\aj}\/}, {\bf 115}, 869.

\bibitem[{Winitzki and Kosowsky(1997)}]{winitzki:minkowski}
Winitzki, S. and Kosowsky, A. (1997): {M}inkowski functional description of
  microwave background {G}aussianity.
\newblock {\em New Astronomy\/}, {\bf 3}, 75--100.

\bibitem[{Worsley(1998)}]{worsley:testing}
Worsley, K. (1998): Testing for a signal with unknown location and scale in a
  $\chi^2$ random field, with an application to {fMRI}.
\newblock Adv.\ Appl.\ Prob. accepted.

\end{thebibliography}

\vspace{0.5cm}

%%%%%%%%%%%%%%%%%%%%%%%%%%%%
One may  find some of the more recent  articles on the  preprint servers\\
{\tt http://xxx.lanl.gov/archive/astro-ph} or \\
{\tt http://xxx.lanl.gov/archive/gr-qc}.\\
Numbers like astro-ph/9710207 refer to preprints on these servers.  An
abstract  server  for  articles  appaering  in  several  astrophysical
journals
is \\
{\tt http://ads.harvard.edu/}.  \\
Articles older than  a few years are scanned and  and may be donloaded
from there.

%%%%%%%%%%%%%%%%%%%%%%%%%%%%
\printindex

\end{document}